\documentclass[english,prd,onecolumn,superscriptaddress,nofootinbib,preprintnumbers]{revtex4}
\usepackage[latin1]{inputenc}
\usepackage{graphicx}
\usepackage{amsmath,amsthm,amssymb}
\usepackage{bm}

\makeatletter

\usepackage{amsfonts}
\usepackage{dcolumn}
\usepackage{hyperref}

\def\be{\begin{equation}}
\def\ee{\end{equation}}
\def\ba{\begin{eqnarray}}
\def\ea{\end{eqnarray}}
\def\bs{\begin{subequations}}
\def\es{\end{subequations}}

\def\tr{\tilde{r}}

\newcommand{\vp}{\varphi}
\newcommand{\rd}{{\rm d}}

\makeatother
\usepackage{babel}
\makeatother

\begin{document}

\title{Modified gravity models of dark energy}

\author{Shinji Tsujikawa}
\affiliation{Department of Physics, Faculty of Science, Tokyo University of Science, 
1-3, Kagurazaka, Shinjuku-ku, Tokyo 162-8601, Japan}

\begin{abstract}

We review recent progress of modified gravity models of 
dark energy--based on $f(R)$ gravity, scalar-tensor theories,
braneworld gravity, Galileon gravity, and other theories. 
In $f(R)$ gravity it is possible to design viable models consistent 
with local gravity constraints under a chameleon mechanism, while 
satisfying conditions for the cosmological viability.
We also construct a class of scalar-tensor dark energy models
based on Brans-Dicke theory in the presence of a scalar-field potential
with a large coupling strength $Q$ between the field and 
non-relativistic matter in the Einstein frame.
We study the evolution of matter density perturbations
in $f(R)$ and Brans-Dicke theories to place observational 
constraints on model parameters from the power spectra
of galaxy clustering and Cosmic Microwave Background (CMB). 

The Dvali-Gabadazde-Porrati braneworld model can be compatible with 
local gravity constraints through a nonlinear field self-interaction 
$\square \phi (\partial_\mu \phi \partial^\mu \phi)$ arising from 
a brane-bending mode, but the self-accelerating solution contains
a ghost mode in addition to the tension with the combined data analysis
of Supernovae Ia (SN Ia) and Baryon Acoustic Oscillations (BAO).
The extension of the field self-interaction to more general forms satisfying 
the Galilean symmetry $\partial_{\mu} \phi \to \partial_{\mu} \phi+b_{\mu}$ in the flat space-time allows a possibility to avoid the appearance of 
ghosts and instabilities, while the late-time cosmic acceleration 
can be realized by the field kinetic energy. We study observational constraints on such Galileon models by using the data of SN Ia, BAO, and 
CMB shift parameters.

We also briefly review other modified gravitational models of 
dark energy--such as those based on Gauss-Bonnet gravity and 
Lorentz-violating theories.

\end{abstract}

\date{\today}

\maketitle

\tableofcontents

\section{Introduction}

The cosmic acceleration today has been supported by independent 
observational data such as the Supernovae-type
Ia (SN Ia) \cite{Riess,Perlmutter}, 
the Cosmic Microwave Background (CMB) temperature anisotropies
measured by WMAP \cite{WMAP1,WMAP7}, and 
Baryon Acoustic Oscillations \cite{BAO1,BAO2}. 
The origin of dark energy responsible for this cosmic acceleration 
is one of the most serious problems in modern 
cosmology \cite{review1,review2,review3,review4,reviewSahni,review5,Noreview,Woodard,Durrer,Lobo,review6,Braxreview,review7,review8}.
The cosmological constant is one of the simplest candidates for dark 
energy, but it is plagued by a severe energy scale problem if it originates 
from the vacuum energy appearing in particle physics \cite{Weinberg}.

The first step toward understanding the nature of dark energy
is to clarify whether it is a simple cosmological constant
or it originates from other sources that dynamically change in time.
The dynamical models can be distinguished
from the cosmological constant by considering 
the evolution of the equation of state of dark energy ($=w_{\rm DE}$).
The scalar field models of dark energy such as 
quintessence \cite{quin1,quin2,quin3,quin4,quin5}
and k-essence \cite{kes1,kes2} predict a wide variety of the variation
of $w_{\rm DE}$, but still the current observational data
are not sufficient to distinguish those models from the 
$\Lambda$-Cold-Dark-Matter ($\Lambda$CDM) model.
Moreover it is generally difficult to construct viable scalar-field
models in the framework of particle physics because of a very 
tiny mass ($m_{\phi} \lesssim 10^{-33}$\,eV) 
required for the cosmic acceleration 
today \cite{Carroll98,Kolda}.

There exists another class of dynamical dark energy models based on 
the large-distance modification of gravity.
The models that belong to this class are 
$f(R)$ gravity \cite{fRearly1,fRearly2,fRearly3,fRearly4} 
($f$ is function of the Ricci scalar $R$), 
scalar-tensor theories \cite{stearly1,stearly2,stearly3,stearly4,stearly5,stearly6}, 
braneworld models \cite{DGP}, Galileon gravity \cite{Nicolis}, 
Gauss-Bonnet gravity \cite{NOS05,NO05}, and so on.
An attractive feature of these models is that the cosmic acceleration 
can be realized without recourse to a dark energy matter component.
If we modify gravity from General Relativity (GR), however, there 
are tight constraints coming from local gravity tests 
as well as a number of observational constraints.
Hence the restriction on modified gravity models is in general 
stringent compared to modified matter models (such as 
quintessence and k-essence).

For example, an $f(R)$ model of the form 
$f(R)=R-\mu^{2(n+1)}/R^n$ ($n>0$)
was proposed to explain the late-time cosmic 
acceleration \cite{fRearly2,fRearly3} 
(see also Refs.~\cite{Soussa,Allem,Easson,Dick04,Carloni,Brookfield} for early works).
However this model suffers from a number of problems such as 
the incompatibility with local gravity constraints \cite{Chiba,Dolgov,Olmo,Navarro}, 
the instability of density perturbations \cite{matterper1,matterper2,SongHu1,SongHu2}, 
and the absence of a matter-dominated epoch \cite{APT,APT2}.
As we will see in this review there are a number of conditions 
required for the viability of $f(R)$ dark energy 
models \cite{matterper1,matterper2,SongHu1,SongHu2,Teg,AGPT,LiBarrow,AmenTsuji07,Fay07,Pogosian,Sendouda},
which stimulated to propose viable 
models \cite{Hu07,Star07,Appleby,Tsuji08,LinderfR}. 

The simplest version of scalar-tensor theories is so-called
Brans-Dicke theory in which a scalar field $\varphi$ couples
to the Ricci scalar $R$ with the Lagrangian density 
${\cal L}=\varphi R/2-(\omega_{\rm BD}/2\varphi)\,(\nabla \varphi)^2$,
where $\omega_{\rm BD}$ is a so-called Brans-Dicke parameter \cite{BD}.
GR can be recovered by taking the limit $\omega_{\rm BD} \to \infty$.
If we allow the presence of the field potential $U(\varphi)$ in 
Brans-Dicke theory, $f(R)$ theory in the metric formalism 
is equivalent to this generalized Brans-Dicke theory with the parameter 
$\omega_{\rm BD}=0$ \cite{ohanlon,Chiba}. 
By transforming the action in generalized Brans-Dicke theory
(``Jordan frame'') to an ``Einstein frame'' action 
by a conformal transformation, the theory in the Einstein frame
is equivalent to a coupled quintessence scenario \cite{coupled} 
with a constant coupling $Q$ satisfying the relation 
$1/(2Q^2)=3+2\omega_{\rm BD}$ \cite{TUMTY}.
For example, $f(R)$ theory in the metric formalism 
corresponds to the constant coupling $Q=-1/\sqrt{6}$, i.e.
$\omega_{\rm BD}=0$.
For $|Q|$ of the order of unity it is generally difficult to satisfy 
local gravity constraints unless some mechanism can be at work 
to suppress the propagation of the fifth force 
between the field and non-relativistic matter.
It is possible for such large-coupling models to be consistent with 
local gravity constraints \cite{Cembranos,Teg,Hu07,AmenTsuji07,CapoTsuji,Van,Gannouji10} 
through the so-called chameleon mechanism \cite{KW1,KW2}, 
provided that a spherically symmetric body has a thin-shell
around its surface.

A braneworld model of dark energy was proposed by Dvali, Gabadadze, and
Porrati (DGP) by embedding a 3-brane in the 5-dimensional Minkowski bulk 
spacetime \cite{DGP}. 
In this scenario the gravitational leakage to the extra dimension
leads to a self-acceleration of the Universe on the 3-brane.
Moreover a longitudinal graviton (i.e. a brane-bending mode $\phi$)
gives rise to a nonlinear self-interaction of the form 
$(r_c^2/m_{\rm pl})\,\square \phi (\partial^{\mu} \phi 
\partial_{\mu} \phi)$ through the mixing with a transverse graviton, 
where $r_c$ is a cross-over scale (of the order of the Hubble radius 
$H_0^{-1}$ today) and $m_{\rm pl}$ is the Planck 
mass \cite{DGPnon1,DGPnon2}.
In the local region where the energy density $\rho$ is much larger
than $r_c^{-2}m_{\rm pl}^2$ the nonlinear self-interaction 
can lead to the decoupling of the field from matter through the so-called
Vainshtein mechanism \cite{Vainshtein}, which allows 
a possibility for the consistency with local gravity constraints.
However the DGP model suffers from a ghost 
problem \cite{DGPghost1,DGPghost2,DGPghost3}, 
in addition to the difficulty for satisfying the combined 
observational constraints of SN Ia and 
BAO \cite{DGPobser1,DGPobser2,DGPobser3,DGPobser4,DGPobser5,DGPobser6}.

The equations of motion following from the self-interacting Lagrangian 
$\square \phi (\partial^{\mu} \phi \partial_{\mu} \phi)$ present
in the DGP model are invariant under the Galilean shift 
$\partial_{\mu} \phi \to \partial_{\mu} \phi+b_{\mu}$ 
in the Minkowski background.
While the DGP model is plagued by the ghost problem,
the extension of the field self-interaction to more
general forms satisfying the Galilean symmetry
may allow us to avoid the appearance of ghosts.
Nicolis {\it et al.} \cite{Nicolis} showed that there are
only five field Lagrangians ${\cal L}_i$ ($i=1,\cdots, 5$)
that respect the Galilean symmetry
in the Minkowski background.
In Refs.~\cite{Deffayetga1,Deffayetga2} these terms were extended to 
covariant forms in the curved space-time.
In addition one can keep the equations of motion 
up to the second-order, while recovering the Galileon 
Lagrangian in the limit of the Minkowski space-time.
This property is welcome to avoid the appearance of 
an extra degree of freedom associated with ghosts. 
In fact, Refs.~\cite{DT2,DT3} derived the viable model parameter 
space in which the appearance of ghosts and instabilities 
associated with scalar and tensor perturbations can be avoided.
Moreover the late-time cosmic acceleration is realized by 
the existence of a stable de Sitter solution.
We shall review the cosmological dynamics of Galileon gravity 
as well as conditions for the avoidance of ghosts and instabilities.

In order to distinguish between different models of dark energy 
based on modified gravitational theories, it is important to study 
the evolution of cosmological perturbations as well as the 
background expansion history of the Universe.
In particular, the modified growth of matter perturbations 
$\delta_m$ relative to the $\Lambda$CDM model
changes the matter power spectrum of large-scale 
structures (LSS) as well as the weak lensing 
spectrum \cite{obsermo1}-\cite{Song10}.
Moreover the modification of gravity manifests itself for 
the evolution of the effective gravitational potential 
$\psi$ related with the Integrated-Sachs-Wolfe 
(ISW) effect in CMB anisotropies.
We shall review a number of observational signatures
for the modified gravitational models of dark energy.

This review is organized as follows.
In Sec.~\ref{fRsec} we construct viable dark energy models 
based on $f(R)$ theories after discussing conditions for the 
cosmological viability as well as for the consistency with 
local gravity tests.
In Sec.~\ref{scasec} we show that, in Brans-Dicke theories
with large matter couplings, it is possible to design the field 
potential consistent with both cosmological and local gravity constraints.
In Sec.~\ref{DGPsec} we derive the field equations in the DGP model 
and confront the model with observations at the background level.
In Sec.~\ref{gasec} we review the cosmological dynamics based on 
Galileon gravity as well as conditions 
for the avoidance of ghosts and Laplacian instabilities.
In Sec.~\ref{othersec} we briefly mention other modified 
gravity models of dark energy based on Gauss-Bonnet gravity
and Lorentz-violating theories. 
In Sec.~\ref{obsersec} we study 
observational signatures of dark energy models based on $f(R)$ gravity, 
Brans-Dicke theory, DGP model, and Galileon gravity, in order to 
confront them with the observations of  LSS, CMB, and 
weak lensing. Sec.~\ref{concludesec} is devoted to conclusions.

Throughout the review we use the units such that $c=\hbar=k_B=1$, 
where $c$ is the speed of light, $\hbar$ is reduced Planck's constant, 
and $k_B$ is Boltzmann's constant. 
We also adopt the metric signature $(-,+,+,+)$.

\section{$f(R)$ gravity}
\label{fRsec}

We start with the action in $f(R)$ gravity:
\begin{equation}
S=\frac{1}{2\kappa^{2}}\int{\rm d}^{4}x\sqrt{-g}f(R)
+\int {\rm d}^4 x\,{\cal L}_M(g_{\mu\nu},\Psi_{M})\,,
\label{fRaction}
\end{equation}
where $\kappa^2=8\pi G$ ($G$ is a bare gravitational constant), 
$g$ is a determinant of the metric $g_{\mu \nu}$, 
$f(R)$ is an arbitrary function in terms of the Ricci scalar $R$, 
and ${\cal L}_{M}$ 
is a matter action with matter fields $\Psi_{M}$.
Variation of the action (\ref{fRaction}) with respect to 
$g_{\mu\nu}$ leads to the following field equation
\begin{equation}
F(R)R_{\mu\nu}(g)-\frac{1}{2}f(R)g_{\mu\nu}
-\nabla_{\mu}\nabla_{\nu}F(R)+g_{\mu\nu}\square F(R)
=\kappa^{2}T_{\mu\nu}\,,\label{fREin}
\end{equation}
where $F(R) \equiv f_{,R}=\partial f/\partial R$, $R_{\mu \nu}$
is a Ricci tensor, and
$T_{\mu \nu}=-(2/\sqrt{-g}) \delta {\cal L}_M/\delta g^{\mu \nu}$ 
is an energy-momentum tensor of matter.
The trace of Eq.~(\ref{fREin}) gives
\begin{eqnarray}
3\,\square F(R)+F(R)R-2f(R)=\kappa^{2}T\,,\label{trace}
\end{eqnarray}
where $T=g^{\mu\nu}T_{\mu\nu}=-\rho+3P$. Here $\rho$ and $P$ are
the energy density and the pressure of matter, respectively.

Regarding the variation of the action (\ref{fRaction}), there is another
approach called the Palatini formalism \cite{Palatini1919} 
in which $g_{\mu\nu}$ and the affine 
connection $\Gamma_{\beta\gamma}^{\alpha}$
are treated as independent variables.\footnote{We also note that 
there is another approach for the variational principle--known as 
the metric-affine formalism--in which the matter Lagrangian ${\cal L}_M$ depends 
not only on the metric $g_{\mu \nu}$ but also on the connection 
$\Gamma_{\beta\gamma}^{\alpha}$ \cite{Hehl,Liberati,Liberati2,Capome}.}
The resulting field equations are second-order \cite{Ferraris,Vollick,Vollick2,Flanagan0,Flanagan,Flanagan2} 
and the cosmological dynamics
of dark energy models have been studied by a number of 
authors \cite{Meng,Meng2,Meng3,NOpala,Sot,Sotinf,Motapala,FayTavakol}.
However $f(R)$ theory in the Palatini formalism gives rise to a large coupling 
between a scalar field degree of freedom and ordinary 
matter \cite{Flanagan0,Flanagan,Flanagan2,Kaloper,OlmoPRL2,Olmo08,Olmo09,Barausse1}, which implies difficulty for compatibility with standard models of particle physics.
This large coupling also leads to significant growth of matter density perturbations,
unless the models are very close to the $\Lambda$CDM 
model \cite{KoivistoPala,KoivistoPala2,LiPala0,LiPala,TsujiUddin}.

In the following we focus on the variational approach (so called 
the metric formalism) given above.
The Einstein gravity without a cosmological constant corresponds to 
$f(R)=R$ and $F(R)=1$, so that the term $\square F(R)$ in Eq.~(\ref{trace})
vanishes. Since in this case $R=-\kappa^{2}T=\kappa^{2}(\rho-3P)$, 
the Ricci scalar $R$ is directly determined by matter.
In $f(R)$ gravity with a non-linear term in $R$, 
$\square F(R)$ does not vanish 
in Eq.~(\ref{trace}). Hence there is a propagating scalar
degree of freedom, $\psi\equiv F(R)$, dubbed {}
``scalaron'' in Ref.~\cite{Star80}.
The trace equation (\ref{trace})
allows the dynamics of the scalar field $\psi$.

The de Sitter point corresponds to a vacuum solution with 
constant $R$. Since $\square F(R)=0$ at this point, we obtain 
\begin{eqnarray}
F(R)R-2f(R)=0\,.
\label{fRdeSitter}
\end{eqnarray}
Since the quadratic model $f(R)=\alpha R^{2}$ satisfies this condition, 
it gives rise to an exact de Sitter solution.
In the inflation model $f(R)=R+\alpha R^{2}$ proposed by 
Starobinsky \cite{Star80}, the accelerated cosmic
expansion ends when the term $\alpha R^{2}$ 
becomes smaller than the linear term $R$. 
It is possible to construct such $f(R)$ inflation models
in the framework of supergravity \cite{Ketov1,Ketov2}.

\subsection{Cosmological dynamics in $f(R)$ gravity}

We first study cosmological dynamics for the models based 
on $f(R)$ theories in the metric formalism.
In order to derive conditions for the cosmological 
viability of $f(R)$ models we shall carry out general analysis 
without specifying the form of $f(R)$.
We consider a flat Friedmann-Lema$\hat{\rm i}$tre-Robertson-Walker (FLRW) 
background with the line element
\begin{eqnarray}
\rd s^2=-\rd t^2+a(t)^2 \rd {\bm x}^2\,,
\end{eqnarray}
where $a(t)$ is a scale factor.
For the matter Lagrangian ${\cal L}_{M}$ in Eq.~(\ref{fRaction})
we take into account non-relativistic matter and radiation, whose energy 
densities $\rho_m$ and $\rho_r$ satisfy 
the usual continuity equations $\dot{\rho}_{m}+3H\rho_{m}=0$
and $\dot{\rho}_{r}+4H\rho_{r}=0$ respectively. 
Here $H \equiv \dot{a}/a$ is the Hubble parameter and 
a dot represents a derivative with respect to cosmic time $t$.
{}From Eqs.~(\ref{fREin}) and (\ref{trace}) we obtain 
\begin{eqnarray}
3FH^{2} & = & \kappa^{2}\,(\rho_{m}+\rho_{r})+(FR-f)/2-3H\dot{F}\,,
\label{FRWfR1}\\
2F\dot{H} & = & -\kappa^{2}\left[\rho_{m}+(4/3)\rho_{r}\right]
-\ddot{F}+H\dot{F}\,,\label{FRWfR2}
\end{eqnarray}
where the Ricci scalar is given by 
\begin{eqnarray}
R=6(2H^2+\dot{H})\,.
\label{Riccisca}
\end{eqnarray}

Let us introduce the following dimensionless variables: 
\begin{equation}
x_{1}\equiv-\frac{\dot{F}}{HF}\,,\qquad x_{2}\equiv-\frac{f}{6FH^{2}}\,,
\qquad x_{3}\equiv\frac{R}{6H^{2}}\,,
\qquad x_{4}\equiv\frac{\kappa^{2}\rho_{r}}{3FH^{2}}\,,
\end{equation}
together with the density parameters
\begin{eqnarray}
\Omega_{m}\equiv\frac{\kappa^{2}\rho_{m}}{3FH^{2}}=1-x_{1}-x_{2}-x_{3}-x_{4},
\qquad\Omega_{r}\equiv x_{4}\,,\qquad\Omega_{{\rm DE}}\equiv x_{1}+x_{2}+x_{3}\,.
\label{fromedef}
\end{eqnarray}
It is straightforward to derive the following equations \cite{AGPT}: 
\begin{eqnarray}
\frac{\rd x_{1}}{\rd N} & = & -1-x_{3}-3x_{2}+x_{1}^{2}-x_{1}x_{3}+x_{4}~,\label{x1fR}\\
\frac{\rd x_{2}}{\rd N} & = & \frac{x_{1}x_{3}}{m}-x_{2}(2x_{3}-4-x_{1})~,\label{x2fR}\\
\frac{\rd x_{3}}{\rd N} & = & -\frac{x_{1}x_{3}}{m}-2x_{3}(x_{3}-2)~,\label{x3fR}\\
\frac{\rd x_{4}}{\rd N} & = & -2x_{3}x_{4}+x_{1} x_{4}\,,\label{x4fR}
\end{eqnarray}
where $N=\ln a$ and
\begin{eqnarray}
m & \equiv & \frac{\rd\ln F}{\rd\ln R}=\frac{Rf_{,RR}}{f_{,R}}\,,\\
r & \equiv & -\frac{\rd\ln f}{\rd\ln R}=-\frac{Rf_{,R}}{f}
=\frac{x_{3}}{x_{2}}\,.\label{mdef}
\end{eqnarray}
{}From Eq.~(\ref{mdef}) one can express $R$ as a function of
$x_{3}/x_{2}$. Since $m$ is a function of $R$, it follows that
$m$ is a function of $r$, i.e., $m=m(r)$. The $\Lambda$CDM model,
$f(R)=R-2\Lambda$, corresponds to $m=0$. Then the quantity $m$
characterizes the deviation from the $\Lambda$CDM model. 

The effective equation of state of the system is given by 
\begin{eqnarray}
w_{{\rm eff}} \equiv -1-\frac{2\dot{H}}{3H^2}
=-\frac13 (2x_{3}-1)\,.\label{ldef}
\end{eqnarray}
In the absence of radiation ($x_{4}=0$) the fixed points
for the dynamical system (\ref{x1fR})-(\ref{x4fR}) are 
\begin{eqnarray}
 &  & P_{1}:(x_{1},x_{2},x_{3})=(0,-1,2),\quad\Omega_{m}=0,\quad w_{{\rm eff}}=-1\,,\label{P1point} \\
 &  & P_{2}:(x_{1},x_{2},x_{3})=(-1,0,0),\quad\Omega_{m}=2,\quad w_{{\rm eff}}=1/3\,,\\
 &  & P_{3}:(x_{1},x_{2},x_{3})=(1,0,0),\quad\Omega_{m}=0,\quad w_{{\rm eff}}=1/3\,,\\
 &  & P_{4}:(x_{1},x_{2},x_{3})=(-4,5,0),\quad\Omega_{m}=0,\quad w_{{\rm eff}}=1/3\,,\\
 &  & P_{5}:(x_{1},x_{2},x_{3})=\left(\frac{3m}{1+m},-\frac{1+4m}{2(1+m)^{2}},\frac{1+4m}{2(1+m)}\right), \quad 
 \Omega_{m}=1-\frac{m(7+10m)}{2(1+m)^{2}},\quad w_{{\rm eff}}=-\frac{m}{1+m},\\
 &  & P_{6}:(x_{1},x_{2},x_{3})=\left(\frac{2(1-m)}{1+2m},\frac{1-4m}{m(1+2m)},
 -\frac{(1-4m)(1+m)}{m(1+2m)}\right), \quad 
 \Omega_{m}=0,\quad w_{{\rm eff}}=\frac{2-5m-6m^{2}}{3m(1+2m)}.
 \end{eqnarray}
The points $P_{5}$ and $P_{6}$ are on the line $m(r)=-r-1$ in
the $(r,m)$ plane.

Only the point $P_{5}$ can be responsible for 
the matter-dominated epoch ($\Omega_{m}\simeq1$
and $w_{{\rm eff}}\simeq0$). 
This is realized provided $m$ is close to 0.
In the ($r,m$) plane the matter point $P_5$ 
exists around $(r,m)=(-1,0)$. 
Either the point $P_1$ or $P_6$ can lead to the late-time
cosmic acceleration. The former corresponds to a de Sitter point 
($w_{\rm eff}=-1$) with $r=-2$, in which case the condition 
(\ref{fRdeSitter}) is satisfied.
Depending on the values of $m$,
the point $P_6$ can be responsible for the cosmic
acceleration \cite{AGPT}. 
In the following we shall focus on the case in which the 
matter point $P_5$ is followed by the de Sitter point $P_1$. 

The stability of the fixed points is known by considering small
perturbations $\delta x_{i}$ ($i=1,2,3$) around them \cite{AGPT}.
For the point $P_{5}$ the eigenvalues for the $3\times3$ Jacobian 
matrix of perturbations are 
\begin{eqnarray}
3(1+m_{5}'),~~\frac{-3m_{5}\pm\sqrt{m_{5}(256m_{5}^{3}
+160m_{5}^{2}-31m_{5}-16)}}{4m_{5}(m_{5}+1)}\,,\label{P5eig}
\end{eqnarray}
where $m_{5} \equiv m(r_5)$ and
$m_{5}'\equiv\frac{\rd m}{\rd r}(r_{5})$ with $r_{5}\approx-1$.
In the limit $|m_{5}|\ll1$ the latter two eigenvalues reduce to $-3/4\pm\sqrt{-1/m_{5}}$.
The $f(R)$ models with $m_{5}<0$ show a divergence of the eigenvalues
as $m_{5}\to-0$, in which case the system cannot remain for a long
time around the point $P_{5}$. For example the model $f(R)=R-\alpha/R^{n}$
with $n>0$ and $\alpha>0$ falls into this category. On the other
hand, if $0<m_{5}<0.327$, the latter two eigenvalues in Eq.~(\ref{P5eig})
are complex with negative real parts. Then, provided that $m_{5}'>-1$,
the point $P_{5}$ corresponds to a saddle point with a damped oscillation.
Hence the Universe can evolve toward the point $P_{5}$ from the radiation
era and leave for the late-time acceleration. 
Then the condition for the existence of the saddle matter era is 
\begin{eqnarray}
m(r)\approx+0\,,\quad\frac{\rd m}{\rd r}>-1\,,~~~{\rm at}~~~r=-1\,.
\label{m5con}
\end{eqnarray}
The first condition implies that the $f(R)$ models need to be close
to the $\Lambda$CDM model during the matter era.

The eigenvalues for the Jacobian matrix of perturbations
about the point $P_{1}$ are 
\begin{equation}
-3,~~~-\frac{3}{2}\pm\frac{\sqrt{25-16/m_{1}}}{2}\,,
\end{equation}
where $m_{1}=m(r=-2)$. This shows that the condition for the stability
of the de Sitter point $P_{1}$ is \cite{Muller,Faraonista1,Faraonista2,AGPT}
\begin{equation}
0<m(r=-2)\le1\,.\label{m1con}
\end{equation}
The trajectories that start from the saddle matter point $P_{5}$ 
with the condition (\ref{m5con}) and then approach the
stable de Sitter point $P_{1}$ with the condition (\ref{m1con})
are cosmologically viable.

\begin{figure}
\begin{centering}
\includegraphics[width=3.4in,height=3.2in]{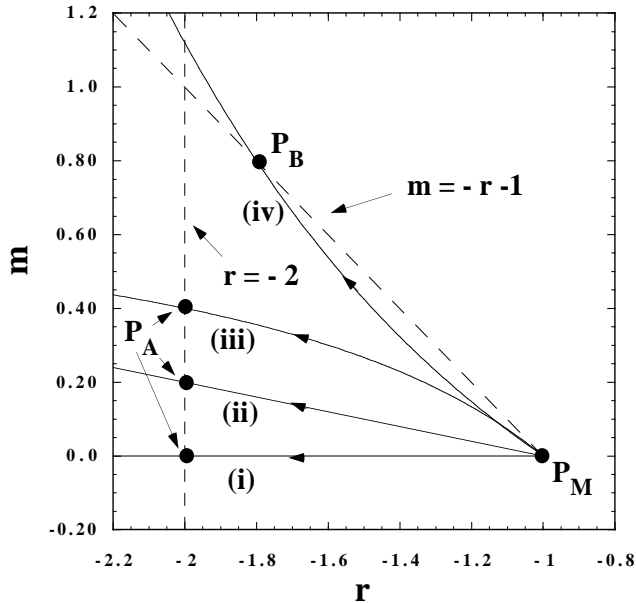} 
\par\end{centering}
\caption{Four trajectories in the $(r,m)$ plane. Each trajectory corresponds
to the models: (i) $\Lambda$CDM, (ii) $f(R)=(R^{b}-\Lambda)^{c}$,
(iii) $f(R)=R-\alpha R^{n}$ with $\alpha>0,0<n<1$, and (iv) $m(r)=-C(r+1)(r^{2}+ar+b)$.
Here $P_{M}$, $P_{A}$ and $P_{B}$ are the matter point $P_{5}$,
the de Sitter point $P_{1}$, and the accelerated point
$P_{6}$, respectively. 
{}From Ref.~\cite{AmenTsuji07}.}
\centering{}\label{mrplane} 
\end{figure}

Let us consider a couple of viable $f(R)$ models in the $(r,m)$
plane. The $\Lambda$CDM model, $f(R)=R-2\Lambda$, corresponds to
$m=0$, in which case the trajectory is a straight line from $P_{5}$:
$(r,m)=(-1,0)$ to $P_{1}$: $(r,m)=(-2,0)$. 
The trajectory (ii) in Fig.~\ref{mrplane} represents the 
model $f(R)=(R^{b}-\Lambda)^{c}$ \cite{AmenTsuji07},
which corresponds to the straight line $m(r)=[(1-c)/c]r+b-1$ in
the $(r,m)$ plane. The existence of a saddle matter epoch requires
the condition $c \ge 1$ and $bc \approx 1$. 
The trajectory (iii) represents the model \cite{AGPT,LiBarrow}
\begin{equation}
f(R)=R-\alpha R^{n} \qquad (\alpha>0,~0<n<1),
\label{powermodel}
\end{equation}
which corresponds to the curve $m=n(1+r)/r$. 
The trajectory (iv) in Fig.~\ref{mrplane}
shows the model $m(r)=-C(r+1)(r^{2}+ar+b)$, in which case
the late-time accelerated attractor is the point $P_6$
with $(\sqrt{3}-1)/2<m<1$.

In Ref.~\cite{AGPT} it was shown that the variable $m$ needs to be
close to 0 during the radiation-dominated epoch as well.
Hence the viable $f(R)$ models are close to the $\Lambda$CDM 
model, $f(R)=R-2\Lambda$,
in the region $R\gg R_{0}$ (where $R_{0}$ is the present cosmological
Ricci scalar). The Ricci scalar $R$ given in Eq.~(\ref{Riccisca}) remains positive
from the radiation era to the present epoch, as long as the it does
not oscillate.
As we will see in Sec.~\ref{fRconsec}, we require the condition 
$f_{,R}>0$ to avoid ghosts.
Then the condition $m>0$ for the presence of the matter-dominated 
epoch translates to $f_{,RR}>0$. 
The model $f(R)=R-\alpha/R^n$ ($\alpha>0$, $n>0$) is not viable
because the condition $f_{,RR}>0$ is violated.
We also note that the power-law models with $f(R)=R^n$  
do not give rise to a successful cosmological 
trajectory \cite{APT,APT2} (unlike the claims in Ref.~\cite{CNOT}).

In order to derive the equation of state of dark energy to confront
with SN Ia observations for the cosmologically viable models, 
we rewrite Eqs.~(\ref{FRWfR1}) and (\ref{FRWfR2}) 
as follows\footnote{If the field equations are written in this form, 
we can also show that the background cosmological dynamics has 
a correspondence with equilibrium thermodynamics on the apparent horizon \cite{BGT09}.}: 
\begin{eqnarray}
 &  & 3AH^{2}=\kappa^{2} \left(\rho_{m}+\rho_{r}+\rho_{{\rm DE}} 
 \right)\,,\label{mofR1}\\
 &  & -2A\dot{H}=\kappa^{2}\left[\rho_{m}+(4/3)\rho_{r}
 +\rho_{{\rm DE}}+P_{{\rm DE}}\right]\,,
 \label{mofR2}
\end{eqnarray}
where $A$ is some constant and 
\begin{eqnarray}
\kappa^{2}\rho_{{\rm DE}} & \equiv & (1/2)(FR-f)-3H\dot{F}+3H^{2}
(A-F)\,,\label{fRrhode}\\
\kappa^{2}P_{{\rm DE}} & \equiv & \ddot{F}+2H\dot{F}
-(1/2)(FR-f)-(3H^{2}+2\dot{H})(A-F)\,.\label{fRPde}
\end{eqnarray}
Defining $\rho_{{\rm DE}}$ and $P_{{\rm DE}}$ in this
way, one can show that these satisfy the usual continuity equation
\begin{eqnarray}
\dot{\rho}_{{\rm DE}}+3H(\rho_{{\rm DE}}+P_{{\rm DE}})=0\,.
\label{rhodecon}
\end{eqnarray}

The dark energy equation of state related with SN Ia observations
is given by $w_{{\rm DE}}\equiv P_{{\rm DE}}/\rho_{{\rm DE}}$.
{}From Eqs.~(\ref{mofR1}) and (\ref{mofR2}) it follows that 
\begin{eqnarray}
w_{{\rm DE}}=-\frac{2A\dot{H}+3AH^2+\kappa^2 \rho_r/3}
{3AH^2-\kappa^2 (\rho_m+\rho_r)}
\simeq\frac{w_{{\rm eff}}}{1-(F/A)\Omega_m}\,,\label{wDEfR}
\end{eqnarray}
where the last approximate equality in Eq.~(\ref{wDEfR}) 
is valid in the regime where the radiation density $\rho_r$ is negligible
relative to the matter density.
The viable $f(R)$ models approach the $\Lambda$CDM model in the past, 
i.e. $F \to 1$ as $R \to \infty$.
In order to reproduce the standard matter era for the redshifts $z \gg 1$,
one can choose $A=1$ in Eqs.~(\ref{mofR1}) and (\ref{mofR2}).
Another possible choice is $A=F_0$, where $F_0$ is the present value
of $F$. This choice is suitable if the deviation of $F_0$ from 1 is small
(as in the scalar-tensor theory with a massless scalar field \cite{Torres,Boi00,Espo}).
In both cases the equation of state $w_{\rm DE}$ can be smaller than $-1$
before reaching the de Sitter 
attractor \cite{AmenTsuji07,Tsuji08,Motohashi10,Bamba10,Bamba10d}.
Thus $f(R)$ gravity models give rise to a phantom equation of state
without violating stability conditions of the system.

As we see in Eq.~(\ref{wDEfR}), the presence of non-relativistic 
matter is important to lead to the apparent phantom behavior.
We wish to stress here that for viable $f(R)$ models constructed
to satisfy all required conditions [such as the models 
(\ref{Amodel}), (\ref{Bmodel}), and (\ref{Cmodel}) we will discuss later] 
the ghosts are not present even if $w_{\rm DE}<-1$.
A number of authors proposed some models to realize $w_{\rm DE}<-1$
without including non-relativistic matter \cite{Bamba08,Bamba09}, 
which means that $w_{\rm DE}=w_{\rm eff}$ from Eq.~(\ref{wDEfR}).
However, such $f(R)$ models usually imply the presence of 
ghosts\footnote{If a late-time de Sitter solution is a stable spiral, 
it happens that $w_{\rm DE}$ oscillates around $-1$ with 
a small amplitude, even for viable $f(R)$ models. 
Here we are discussing the real ghosts out of this regime.}, 
because $w_{\rm eff}<-1$ corresponds to 
$\dot{H}>0$.

The observational constraints on specific $f(R)$ models
have been carried out in Refs.~\cite{Dev,Mel09,Cardone09,Ali10} from 
the background expansion history of the Universe (see also 
Refs.~\cite{Caporecon,Mul06,Dobado06,Wu07,Carloni10} 
for the reconstruction of $f(R)$ models from observations).
Since the deviation of $w_{\rm DE}$ from that in the $\Lambda$CDM 
model ($w_{\rm DE}=-1$) is not so 
significant \cite{Hu07,Motohashi10}, the viable models
such as (\ref{Amodel}), (\ref{Bmodel}), and (\ref{Cmodel}) 
can be consistent with the data fairly easily.
In other words we do not obtain very tight bounds on model parameters from 
the information of the background expansion history only.
However, the models can be more strongly 
constrained at the level of perturbations, as we will see 
in Sec.~\ref{fRsecper}.

\subsection{Conditions for the avoidance of ghosts and tachyonic instabilities}
\label{fRconsec}

In this subsection we shall derive conditions for the avoidance of 
ghosts and tachyonic instabilities in $f(R)$ theories.
In doing so we expand the action (\ref{fRaction}) up to the 
second-order by considering the following perturbed metric
about the FLRW background
\begin{equation}
{\rm d}s^2=-(1+2\alpha){\rm d}t^2-2a(t) \partial_i \beta
{\rm d}t {\rm d}x^i+a^2(t) \left( \delta_{ij}+2\psi \delta_{ij}
+2\partial_i \partial_j \gamma \right) {\rm d}x^i {\rm d}x^j\,,
\end{equation}
where $\alpha, \beta, \psi, \gamma$ are scalar 
metric perturbations \cite{Bardeen}.

Introducing the perturbation $\delta F$ for the quantity 
$F=\partial f/\partial R$, one can construct 
the gauge-invariant curvature perturbation 
\begin{equation}
{\cal R} \equiv \psi-\frac{H}{\dot{F}}\delta F\,.
\end{equation}
Expanding the action (\ref{fRaction}) without the 
matter source, we obtain 
the second-order action for the curvature 
perturbation \cite{Hwang97,review7}
\begin{equation}
\label{eq:actPhid}
\delta S^{(2)}=\int {\rm d} t\,{\rm d}^3x\,a^3\,Q_s\left[\frac12\,\dot{\cal R}^2
 -\frac12\,\frac{1}{a^2}(\nabla{\cal R})^2\right]\,,
\end{equation}
where 
\begin{equation}
Q_s \equiv \frac{3\dot{F}^2}{2\kappa^2 F [H+\dot{F}/(2F)]^2}\,.
\end{equation}
The negative sign of $Q_s$ corresponds to a ghost field because 
of the negative kinetic energy. Hence the condition for the avoidance 
of ghosts is given by 
\begin{equation}
F>0\,.
\label{noghost}
\end{equation}
For the matter sector the ghost does not appear for 
$\rho_M (1+w_M)/w_M>0$ (where $w_M$ is the equation of 
state for the matter fluid) \cite{DMT}, which is satisfied for 
radiation ($w_M=1/3$) and 
non-relativistic matter ($w_M \simeq +0$).

If $Q_s$ is positive, the action~(\ref{eq:actPhid}) can be written 
in the following form by introducing the new variables 
$u=z_s {\cal R}$ and $z_s=a\sqrt{Q}_s$:
\begin{equation}
\label{eq:actPhid2}
\delta S^{(2)}=\int {\rm d} \eta\,{\rm d}^3x
\left[\frac12\,u'^2
 -\frac12 (\nabla u)^2+
 \frac12 \frac{z_s''}{z_s}u^2 \right]\,,
\end{equation}
where a prime represents a derivative with respect to 
the conformal time $\tau=\int a^{-1} \mathrm{d}t$.
Equation (\ref{eq:actPhid2}) shows that the scalar
degree of freedom has the effective mass
\begin{eqnarray}
M_s^2 \equiv -\frac{1}{a^2}\frac{z_s''}{z_s}
&=&
\frac{\dot Q_s^2}{4Q_s^2}-\frac{\ddot Q_s}{2Q_s}
-\frac{3H\dot Q_s}{2Q_s} \nonumber \\
&=&-\frac{72 F^2 H^4}{(2 F H+f_{,RR} \dot R)^2}+
\frac{1}{3} F \left(\frac{288 H^3-12 H R}
{2 F H+f_{,RR} \dot R}+\frac{1}{f_{,RR}}\right)
+\frac{f_{,RR}^2 \dot R^2}{4 F^2}
-24 H^2+\frac76 R\,,
\label{Ms2}
\end{eqnarray}
where we have eliminated the term 
$\dot{H}$ by using the background
equations.

In Fourier space the perturbation $u$ satisfies 
the equation of motion 
\begin{equation}
u''+\left( k^2+M_s^2 a^2 \right)u=0\,.
\end{equation}
For $k^2/a^2 \gg M_s^2$, the propagation speed $c_s$
of the field $u$ is equivalent to the speed of light $c$.
Hence, in $f(R)$ gravity, the gradient instability 
associated with negative $c_s^2$ is absent.
For small $k$ satisfying $k^2/a^2 \ll M_s^2$, 
we require that $M_s^2>0$ to 
avoid the tachyonic instability of perturbations.
The viable dark energy models based on $f(R)$ theories need to satisfy 
the condition $Rf_{,RR}\ll F$ (i.e. $m=Rf_{,RR}/f_{,R} \ll 1$) at early 
cosmological epochs in order to have successful cosmological 
evolution from radiation domination till matter domination. 
At these epochs the mass squared is approximately given by 
\begin{equation}
\label{eq:massfR}
M_s^2 \simeq \frac{F}{3f_{,RR}}\,.
\end{equation}
Under the no-ghost condition (\ref{noghost}) the tachyonic
instability is absent for 
\begin{equation}
f_{,RR}>0\,.
\label{notachyon}
\end{equation}

The viable $f(R)$ dark energy models have been constructed to 
satisfy the conditions (\ref{noghost}) and (\ref{notachyon})
in the regime $R \ge R_1$, where $R_1$ 
is the Ricci scalar at the late-time de Sitter point.
Moreover we require that the models are 
consistent with the conditions (\ref{m5con}) and (\ref{m1con}).
The model (\ref{powermodel}) can be consistent with all 
these conditions, but the local gravity constraints demand
that the variable $m$ is very much smaller than 1
in the regions of high density (i.e. $R \gg R_0$, where $R_0$ is the 
cosmological Ricci scalar today).
In the model (\ref{powermodel}) one has $m \simeq n(-r-1)$
around $r \simeq -1$. For the consistency with local gravity constraints
we require that $n \lesssim 10^{-10}$ \cite{CapoTsuji}, but in this case
the deviation from the $\Lambda$CDM model around the 
present epoch ($R \approx R_0$) is very small.

If the variable $m$ behaves as $m=C(-r-1)^p$ with $p>1$ in the region 
$R \gg R_0$, then it is possible to satisfy local gravity
constraints (i.e. $m \ll 0.01 \sim 0.1$ for $R \gg R_0$)
while at the same time showing deviations from the 
$\Lambda$CDM ($m \gtrsim 0.01 \sim 0.1$ for $R \simeq R_0$). 
The models constructed in this vein are
\begin{eqnarray}
 &  & {\rm (A)}~f(R)=R-\mu R_{c}\frac{(R/R_{c})^{2n}}{(R/R_{c})^{2n}+1}
 \qquad{\rm with}~~\mu>0,~~n>0~{\rm and}~R_{c}>0\,,\label{Amodel}\\
 &  & {\rm (B)}~f(R)=R-\mu R_{c}\left[1-\left(1+R^{2}/R_{c}^{2}\right)^{-n}\right]
 \qquad{\rm with}~~\mu>0,~~n>0~{\rm and}~R_{c}>0\,,\label{Bmodel}
\end{eqnarray}
which were proposed by Hu and Sawicki \cite{Hu07} and 
Starobinsky \cite{Star07}, respectively.
$R_c$ is roughly of the order of the present cosmological 
Ricci scalar $R_0$ for $\mu$ and $n$ of the order of unity.
The models (A) and (B) asymptotically behave as 
\begin{equation}
f(R) \simeq  R-\mu R_{c}[1-(R^{2}/R_{c}^{2})^{-n}]
\qquad {\rm for} \quad R\gg R_{c}\,,
\label{fRasy}
\end{equation}
which gives $m(r)=C(-r-1)^{2n+1}$.

Another viable model that leads to the even rapid decrease of 
$m$ toward the past is \cite{Tsuji08}
\begin{equation}
{\rm (C)}~f(R)=R-\mu R_{c}{\rm tanh}\,(R/R_{c})
\qquad{\rm with}~~\mu>0,~~R_{c}>0\,.
\label{Cmodel}
\end{equation}
Other similar models were proposed by Appleby and Battye \cite{Appleby}
and Linder \cite{LinderfR}.

In what follows we shall discuss local gravity constraints 
on the above models.

\subsection{Local gravity constraints on $f(R)$ gravity models}
\label{fRLGC}

Let us proceed to discuss local gravity constraints on 
$f(R)$ gravity models.
In the region of high density like Earth or Sun, 
the Ricci scalar $R$ is much larger than the background 
cosmological value $R_0$.
In this case the linear expansion of $R=R_0+\delta R$ cannot 
be justified. In such a non-linear regime the chameleon 
mechanism \cite{KW1,KW2} plays an important role 
for the $f(R)$ models to satisfy local gravity 
constraints \cite{Cembranos,Teg,Hu07,AmenTsuji07,CapoTsuji,Van,Gannouji10}
(see also Refs.~\cite{OlmoPRL,Olmo05,Fara06,Erick06,Chiba07,Kainu07,Kainu08}).

To discuss the chameleon mechanism in $f(R)$ gravity, it
is convenient to transform the action (\ref{fRaction}) to the 
so-called Einstein frame action via the 
conformal transformation \cite{Maeda}:
\begin{eqnarray}
\tilde{g}_{\mu\nu}=\Omega^{2}g_{\mu\nu}\,,\qquad
\Omega^2=F\,.
\label{ctrans}
\end{eqnarray}
The action in the Einstein frame includes a linear term in 
$\tilde{R}$, where the tilde represents quantities in the Einstein frame.
Introducing a new scalar field $\phi=\sqrt{3/2\kappa^2}\,{\rm ln}\, F$,
we obtain the action in the Einstein frame, as \cite{Maeda,review7} 
\begin{eqnarray}
S_{E}=\int\rd^{4}x\sqrt{-\tilde{g}}\left[\frac{1}{2\kappa^{2}}\tilde{R}
-\frac{1}{2}\tilde{g}^{\mu\nu}\partial_{\mu}\phi\partial_{\nu}\phi-V(\phi)\right]
+\int {\rm d}^4 x\,{\cal L}_M(F^{-1}\tilde{g}_{\mu\nu},\Psi_{M})\,,\label{Ein}
\end{eqnarray}
where 
\begin{eqnarray}
V(\phi)=\frac{RF-f}{2\kappa^{2}F^2}\,.
\label{fRpo}
\end{eqnarray}

In the Einstein frame the scalar field $\phi$ directly couples with
non-relativistic matter. The strength of this coupling depends on 
the conformal factor $\Omega=\sqrt{F}$.
We define the coupling $Q$ as 
\begin{equation}
Q \equiv -\frac{\Omega_{,\phi}}{\Omega}=-\frac{F_{,\phi}}{2F}
=-\frac{1}{\sqrt{6}}\,,
\end{equation}
which is of the order of unity in $f(R)$ gravity.
If the field potential $V(\phi)$ is absent,
the field propagates freely with a large coupling $Q$.
Since a potential (\ref{fRpo}) with a gravitational origin is present
in $f(R)$ gravity, it is possible for $f(R)$ dark energy models 
to satisfy local gravity constraints through the
chameleon mechanism \cite{Teg,Hu07,CapoTsuji,Van}.

In a spherically symmetric space-time under a weak gravitational 
background (i.e. neglecting the backreaction of gravitational potentials), 
variation of the action (\ref{Ein}) with respect to the scalar 
field $\phi$ leads to 
\begin{eqnarray}
\frac{{\rm d}^2 \phi}{{\rm d} \tr^2}+
\frac{2}{\tr} \frac{{\rm d}\phi}{{\rm d}\tr}=
\frac{{\rm d}V_{\rm eff}}{{\rm d}\phi}\,,
\label{dreq}
\end{eqnarray}
where $\tr$ is a distance from the center of symmetry, 
and $V_{\rm eff}(\phi)$ is an effective potential defined by 
\begin{eqnarray}
V_{\rm eff}(\phi)=V(\phi)+
e^{Q \phi}\rho^*\,.
\label{Veff}
\end{eqnarray}
Here $\rho^*$ is a conserved quantity in the Einstein frame,
which is related to the density $\tilde{\rho}$
in the Jordan frame via the relation $\rho^*=e^{3Q \phi}\tilde{\rho}$.
By the end of this section we use the unit $\kappa^2=1$.

We assume that a spherically symmetric body has a constant
density $\rho^*=\rho_A$ inside the body ($\tr<\tr_c$)
and that the density outside the body ($\tr>\tr_c$)
is $\rho^*=\rho_B$. 
The mass $M_c$ of the body and the gravitational potential $\Phi_c$
at the radius $\tr_c$ are given by $M_c=(4\pi/3)\tr_c^3 \rho_A$
and $\Phi_c=M_c/8\pi \tr_c$, respectively.
The effective potential $V_{\rm eff} (\phi)$ has two minima 
at the field values $\phi_A$ and $\phi_B$
satisfying $V_{\rm eff}' (\phi_A)=0$ and 
$V_{\rm eff}' (\phi_B)=0$, respectively (here a prime represents 
a derivative with respect to $\phi$). 
The former corresponds to the region with a high density that gives rise
to a large mass squared $m_A^2 \equiv V_{{\rm eff}}''(\phi_A)$,
whereas the latter to the lower density region with a smaller mass
squared $m_B^2 \equiv V_{{\rm eff}}''(\phi_B)$.
When the ``dynamics'' of the field $\phi$
with the field equation (\ref{dreq}) is studied,  
we need to consider the effective potential $(-V_{\rm eff})$ 
so that it has two maxima
at $\phi=\phi_A$ and $\phi=\phi_B$.

We impose the two boundary conditions
$({\rm d}\phi/{\rm d}\tr)(\tr=0)=0$
and $\quad\phi(\tr \to\infty)=\phi_{B}$.
The field $\phi$ is at rest at $\tr=0$ and begins to roll down the
potential when the matter-coupling term $Q\rho_{A}e^{Q\phi}$ becomes
important at a radius $\tr_{1}$ in Eq.~(\ref{dreq}).
As long as $\tr_1$ is close to $\tr_c$ such that 
$\Delta \tilde{r}_c \equiv \tr_c-\tr_1 \ll \tr_c$, the body has 
a thin-shell inside the body.
The field acquires a sufficient kinetic energy in the thin-shell regime
($\tr_1<\tr<\tr_c$) and hence the field climbs up the potential 
hill outside the body ($\tr>\tr_c$).

The field profile can be obtained by matching the solutions of 
Eq.~(\ref{dreq}) at the radius $\tr=\tr_1$ and $\tr=\tr_c$. 
Neglecting the mass term $m_B$, the thin-shell field profile 
outside the body is given by \cite{TT08} 
\begin{eqnarray}
\phi(\tr)=\phi_{B}-2Q_{{\rm eff}}\frac{GM_{c}}{\tr}\,,\label{phisothin}
\end{eqnarray}
where
\begin{eqnarray}
Q_{\rm eff} \simeq 3Q \epsilon_{\rm th}\,,\qquad
\epsilon_{\rm th} \equiv \frac{\phi_B-\phi_A}{6Q\Phi_c}\,.
\label{epdef}
\end{eqnarray}
Here $\epsilon_{\rm th}$ is called a thin-shell parameter.
Under the conditions $\Delta \tilde{r}_c/\tilde{r}_c \ll 1$ 
and $1/(m_A \tilde{r}_c) \ll 1$, the thin-shell parameter 
is approximately given by \cite{TT08} 
\begin{eqnarray}
\epsilon_{\rm th} \simeq \frac{\Delta \tilde{r}_c}{\tilde{r}_c}
+\frac{1}{m_A \tilde{r}_c}\,.
\label{thinshelles}
\end{eqnarray}
Provided that $\epsilon_{\rm th} \ll 1$, the amplitude of the 
effective coupling $Q_{\rm eff}$ can be much smaller than 1.
It is then possible for the $f(R)$ models ($|Q|=1/\sqrt{6}$) 
to be consistent with local gravity experiments.
Originally the thin-shell solution was derived by assuming that 
the field is frozen in the region 
$0<\tilde{r}<\tilde{r}_1$ \cite{KW1,KW2}. 
In this case the thin-shell parameter
is given by $\epsilon_{\rm th} \simeq \Delta \tilde{r}_c/\tilde{r}_c$,
which is different from Eq.~(\ref{thinshelles}).
However, this difference is not important because the condition 
$ \Delta \tilde{r}_c/\tilde{r}_c \gg 1/(m_A \tilde{r}_c)$
is satisfied for most of viable models \cite{TT08}.

Consider the bound on the thin-shell 
parameter from the possible violation of equivalence principle (EP). 
The tightest bound comes from the solar system tests of weak EP using
the free-fall acceleration of Moon ($a_{{\rm Moon}}$) 
and Earth ($a_{\oplus}$) toward Sun \cite{KW2}. 
The experimental bound on the difference
of two accelerations is given by \cite{Will05} 
\begin{eqnarray}
\frac{|a_{{\rm Moon}}-a_{\oplus}|}
{(a_{{\rm Moon}}+a_{\oplus})/2}<10^{-13}\,.
\label{etamoon}
\end{eqnarray}
Provided that Earth, Sun, and Moon have thin-shells,
the field profiles outside the bodies are given by Eq.~(\ref{phisothin})
with the replacement of corresponding quantities. The
acceleration induced by a fifth force with the field profile 
$\phi(\tr)$ and the effective coupling $Q_{{\rm eff}}$ is 
$a^{{\rm fifth}}=|Q_{{\rm eff}}\nabla\phi(\tr)|$.
Using the thin-shell parameter $\epsilon_{{\rm th},\oplus}$ for Earth,
the accelerations $a_{\oplus}$ and $a_{{\rm Moon}}$ 
toward Sun (mass $M_{\odot}$) are 
\begin{equation}
a_{\oplus} \simeq \frac{GM_{\odot}}{\tr^{2}}\left[1+18Q^{2}
\epsilon_{{\rm th},\oplus}^{2}\frac{\Phi_{\oplus}}{\Phi_{\odot}}
\right]\,,\qquad
a_{{\rm Moon}} \simeq  
\frac{GM_{\odot}}{\tr^{2}}\left[1+18Q^{2}\epsilon_{{\rm th},\oplus}^{2}
\frac{\Phi_{\oplus}^{2}}{\Phi_{\odot}\Phi_{{\rm Moon}}}\right]\,,
\label{etamoon2}
\end{equation}
where $\Phi_{\odot}\simeq2.1\times10^{-6}$, $\Phi_{\oplus}\simeq7.0\times10^{-10}$,
and $\Phi_{{\rm Moon}}\simeq3.1\times10^{-11}$ are the gravitational
potentials of Sun, Earth and Moon, respectively. Then the condition
(\ref{etamoon}) translates to
\begin{equation}
\epsilon_{{\rm th},\oplus} < 8.8\times10^{-7}/|Q|\,.
\label{boep}
\end{equation}
Since the condition $|\phi_B| \gg |\phi_A|$ is satisfied for viable $f(R)$ models (as we will see below), we have $\epsilon_{{\rm th},\oplus} \simeq \phi_B/(6Q \Phi_{\oplus})$ from Eq.~(\ref{epdef}). Hence the condition (\ref{boep}) corresponds to 
\begin{equation}
|\phi_{B,\oplus}|<3.7 \times 10^{-15}\,.
\label{boep2}
\end{equation}

Let us consider local gravity constraints on the $f(R)$ models given
in Eqs.~(\ref{Amodel}) and (\ref{Bmodel}). In the region of high
density where local gravity experiments are carried out, 
it is sufficient to use the asymptotic form given in Eq.~(\ref{fRasy}).
In order for these models to be responsible for the present cosmic 
acceleration, $R_{c}$ is roughly the same order as the cosmological
Ricci scalar $R_{0}$ today for $\mu$ and $n$ of the order of unity.
For the functional form (\ref{fRasy}) we have the following relations
\begin{eqnarray}
F & = & e^{2\phi/\sqrt{6}}=1-2n\mu(R/R_{c})^{-(2n+1)}\,,\label{FdeffR}\\
V_{{\rm eff}}(\phi) & \simeq & \frac{1}{2}\mu R_{c}e^{-4\phi/\sqrt{6}}
\left[1-(2n+1)\left(\frac{-\phi}{\sqrt{6}n\mu}\right)^{2n/(2n+1)}\right]
+\rho^* e^{-\phi/\sqrt{6}}\,.\label{Veff2}
\end{eqnarray}
Inside and outside the body the effective potential (\ref{Veff2}) 
has minima at 
\begin{eqnarray}
\phi_{A}\simeq-\sqrt{6}n\mu(R_{c}/\rho_{A})^{2n+1}\,,\quad
\phi_{B}\simeq-\sqrt{6}n\mu(R_{c}/\rho_{B})^{2n+1}\,.
\end{eqnarray}
If $\rho_A \gg \rho_B$, then one has $|\phi_B| \gg |\phi_A|$.

The bound (\ref{boep2}) translates into 
\begin{eqnarray}
\frac{n\mu}{x_{1}^{2n+1}}\left(\frac{R_{1}}{\rho_{B}}
\right)^{2n+1}<1.5\times10^{-15}\,.
\label{consmo1}
\end{eqnarray}
Here $x_{1}$ is defined by $x_{1}\equiv R_{1}/R_{c}$, where 
$R_1$ is the Ricci scalar at the late-time de Sitter fixed point $P_1$
given in Eq.~(\ref{P1point}).
Let us consider the model described by the Lagrangian density (\ref{fRasy})
for $R\ge R_{1}$. 
If we use the models (\ref{Amodel}) and (\ref{Bmodel}), then there are
some modifications for the estimation of $R_{1}$.
However this change is not significant when we place constraints 
on model parameters.

The de Sitter solution for the model (\ref{fRasy}) satisfies
$\mu=x_{1}^{2n+1}/[2(x_{1}^{2n}-n-1)]$.
Substituting this relation into Eq.~(\ref{consmo1}), it follows that 
\begin{equation}
\frac{n}{2(x_{1}^{2n}-n-1)}\left(\frac{R_{1}}{\rho_{B}}
\right)^{2n+1}<1.5\times10^{-15}\,.\label{cons2}
\end{equation}
{}For the stability of the de Sitter point we require that $m(R_{1})<1$,
which translates into the condition $x_{1}^{2n}>2n^{2}+3n+1$. Hence
the term $n/[2(x_{1}^{2n}-n-1)]$ in Eq.~(\ref{cons2}) is smaller
than 0.25 for $n>0$.

Let us use the simple approximation that 
$R_{1}$ and $\rho_{B}$ are of the orders of the
present cosmological density $10^{-29}$ g/cm$^{3}$ and 
the baryonic/dark
matter density $10^{-24}$ g/cm$^{3}$ in our galaxy, respectively.
{}From Eq.~(\ref{cons2}) we obtain the constraint \cite{CapoTsuji}
\begin{equation}
n>0.9\,.\label{bound3}
\end{equation}
Thus $n$ is not required to be much larger than unity. 
Under the condition (\ref{bound3}), as $R$ decreases to 
the order of $R_{c}$,  one can cosmologically see an appreciable 
deviation from the $\Lambda$CDM model.
The deviation from the $\Lambda$CDM model appears when $R$ decreases to 
the order of $R_c$. The model (\ref{Cmodel}) also shows similar behavior. 
If we consider the model (\ref{powermodel}), it was shown in Ref.~\cite{CapoTsuji} that 
the bound (\ref{boep2}) gives the constraint $n<3 \times 10^{-10}$.
Hence the deviation from the $\Lambda$CDM model is very
small. The models (\ref{Amodel}) and (\ref{Bmodel}) are carefully constructed 
to satisfy local gravity constraints, while at the same time the deviation 
from the $\Lambda$CDM model appears even for $n={\cal O}(1)$.
Note that the model (\ref{Cmodel}) can easily satisfy local gravity constraints
because of the rapid approach to the $\Lambda$CDM in the regime $R \gg R_c$. 

In the strong gravitational background (such as neutron stars), 
Kobayashi and Maeda \cite{KM1,KM2} pointed out that for the model (\ref{Bmodel}) 
it is difficult to obtain thin-shell solutions inside a spherically symmetric body with constant density. 
For chameleon models with general couplings $Q$, a thin-shell field profile was analytically 
derived in Ref.~\cite{TTT} by employing a linear expansion in terms of the gravitational 
potential $\Phi_c$ at the surface of a compact object with constant density. 
Using the boundary condition set by analytic solutions, 
Ref.~\cite{TTT} also numerically confirmed the existence of thin-shell 
solutions for $\Phi_c \lesssim 0.3$ in the case of inverse power-law potentials 
$V(\phi)=M^{4+n}\phi^{-n}$. Ref.~\cite{Upadhye} also showed that
static relativistic stars with constant density exists for the model (\ref{Bmodel}).  
The effect of the relativistic pressure is important around the center of the body,
so that the field tends to roll down the potential quickly unless the boundary condition is 
carefully chosen. Realistic stars have densities $\rho_A (r)$ that globally 
decrease as a function of $r$. The numerical simulation of 
Refs.~\cite{Babi1,Babi2} showed that thin-shell solutions are present 
for the $f(R)$ model (\ref{Bmodel}) by considering a polytropic equation of 
state even in the strong gravitational background (see also Ref.~\cite{Cooney}).

\section{Scalar-tensor gravity}
\label{scasec}

There is another class of modified gravity called scalar-tensor
theories in which the Ricci scalar $R$ is coupled to a scalar field $\varphi$.
One of the simplest examples is the so-called Brans-Dicke theory with the action 
\begin{equation}
S=\int{\rm d}^{4}x\sqrt{-g}\left[\frac{1}{2}\vp R
-\frac{\omega_{\rm BD}}{2\vp}(\nabla\vp)^{2}-U(\vp)\right]
+\int {\rm d}^4 x\,{\cal L}_M(g_{\mu\nu},\Psi_{M})\,,
\label{BDaction}
\end{equation}
where $\omega_{\rm BD}$ is a constant (called 
the Brans-Dicke parameter), $U(\vp)$
is a field potential, and ${\cal L}_M$ is a matter Lagrangian that depends 
on the metric $g_{\mu\nu}$ and matter fields $\Psi_{m}$. 
The original Brans-Dicke theory \cite{BD} does not have the field potential.
As we will see below, metric $f(R)$ gravity discussed in Sec.~\ref{fRsec}
is equivalent to the Brans-Dicke theory with $\omega_{\rm BD}=0$.

\subsection{Scalar-tensor theories and the matter coupling in the 
Einstein frame}

The general action for scalar-tensor theories can be written as 
\begin{equation}
S=\int{\rm d}^{4}x\sqrt{-g}\left[\frac{1}{2}f(\vp,R)
-\frac{1}{2}\zeta(\vp)(\nabla\vp)^{2}\right]+
\int {\rm d}^4 x\,{\cal L}_M(g_{\mu\nu},\Psi_{M})\,,
\label{stensoraction}
\end{equation}
where $f$ depends on the scalar field $\vp$ and the
Ricci scalar $R$, $\zeta$ is a function of $\vp$.
We choose the unit $\kappa^{2}=8\pi G=1$.
The action (\ref{stensoraction}) covers a wide variety of theories
such as $f(R)$ gravity ($f(\vp,R)=f(R)$, $\zeta=0$), 
Brans-Dicke theory ($f=\vp R$ and $\zeta=\omega_{{\rm BD}}/\vp$), 
and dilaton gravity ($f=e^{-\vp}R$ and $\zeta=-e^{-\vp}$).

Let us consider theories of the type 
\begin{eqnarray}
f(\vp,R)=F(\vp)R-2U(\vp)\,.
\label{stensor}
\end{eqnarray}
In order to avoid the appearance of ghosts we require that 
$F(\vp)>0$.  Under the conformal transformation (\ref{ctrans}) 
with the conformal factor $\Omega=\sqrt{F}$,
the action (\ref{stensoraction}) can be transformed to that 
in the Einstein frame:
\begin{equation}
S_{E}=\int{\rm d}^{4}x\sqrt{-\tilde{g}}\left[\frac{1}{2}\tilde{R}
-\frac{1}{2}(\tilde{\nabla}\phi)^{2}-V(\phi)\right]+
\int {\rm d}^4 x\,{\cal L}_M(\tilde{g}_{\mu\nu}F^{-1} (\phi),
\Psi_{M})\,,\label{SEframe}
\end{equation}
where 
\begin{equation}
V=U/F^{2}\,.\label{Upo}
\end{equation}
We have introduced a new scalar field $\phi$ 
in order to make the kinetic term canonical:
\begin{equation}
\phi \equiv \int{\rm d}\vp\,\sqrt{\frac{3}{2}\left(\frac{F_{,\vp}}{F}\right)^{2}
+\frac{\zeta}{F}}\,,\label{phire}
\end{equation}

We define the coupling between dark energy and non-relativistic matter
in the Einstein frame:
\begin{equation}
Q\equiv-\frac{F_{,\phi}}{2F}=-\frac{F_{,\vp}}{F}
\left[\frac{3}{2}\left(\frac{F_{,\vp}}{F}\right)^{2}+
\frac{\zeta}{F}\right]^{-1/2}\,.\label{Q}
\end{equation}
Recall that in metric $f(R)$ gravity we have that $Q=-1/\sqrt{6}$. 
If $Q$ is a constant, the following
relations hold from Eqs.~(\ref{phire}) and (\ref{Q}): 
\begin{equation}
F=e^{-2Q\phi}\,,\qquad
\zeta=(1-6Q^{2})F\left(\frac{{\rm d}\phi}
{{\rm d}\vp}\right)^{2}\,.\label{conf_factor}
\end{equation}
Then the action (\ref{stensoraction}) in the Jordan frame 
can be written as \cite{TUMTY} 
\begin{equation}
S=\int{\rm d}^{4}x\sqrt{-g}\Bigg[\frac{1}{2}F(\phi)R-
\frac{1}{2}(1-6Q^{2})F(\phi)(\nabla\phi)^{2}-U(\phi)\Bigg]
+\int {\rm d}^4 x\,{\cal L}_M(g_{\mu\nu},\Psi_{M})\,.
\label{action2}
\end{equation}
In the limit that $Q \to 0$, the action (\ref{action2}) reduces to the
one for a minimally coupled scalar field $\phi$ with the potential
$U(\phi)$. The transformation of the Jordan frame action (\ref{action2})
via a conformal transformation $\tilde{g}_{\mu\nu}=F(\phi)g_{\mu\nu}$
gives rise to the Einstein frame action (\ref{SEframe}) with a constant
coupling $Q$. The action (\ref{SEframe}) is equivalent to 
the action (\ref{Ein})
with $\tilde{g}_{\mu \nu}=e^{-2 Q \phi}g_{\mu \nu}$.

One can compare (\ref{action2}) with the action (\ref{BDaction})
in Brans-Dicke theory.
Setting $\varphi=F=e^{-2Q\phi}$, one finds that two actions
are equivalent if the parameter $\omega_{{\rm BD}}$ is related to
$Q$ via the relation \cite{KW2,TUMTY} 
\begin{equation}
3+2\omega_{{\rm BD}}=\frac{1}{2Q^{2}}\,.
\label{BD}
\end{equation}
Using this relation, we find that the General Relativistic limit 
($\omega_{{\rm BD}}\to\infty$)
corresponds to the vanishing coupling ($Q \to 0$).
Since $Q=-1/\sqrt{6}$ in metric $f(R)$ gravity, this corresponds to 
the Brans-Dicke parameter $\omega_{{\rm BD}}=0$ \cite{ohanlon,Chiba}. 
The experimental bound on $\omega_{\rm BD}$ for a {\it massless}
scalar field is given by $\omega_{\rm BD}>40000$ \cite{Will05,Iess}, 
which translates into the condition 
\begin{equation}
|Q|<2.5 \times 10^{-3}\qquad ({\rm for~a~massless~field}).
\label{Qbound}
\end{equation}
In such cases it is difficult to find a large difference relative to 
the uncoupled quintessence model.
In the presence of the field potential, however, it is possible
for large coupling models ($|Q| \sim 1$) to satisfy local gravity 
constraints via the chameleon mechanism \cite{TUMTY}.

The above Brans-Dicke theory is one of the examples in scalar-tensor theories.
In general the coupling $Q$ is field-dependent apart from 
Brans-Dicke theory.
If we consider a nonminimally coupled scalar field with
$F(\varphi)=1-\xi \varphi^2$ and $\zeta (\varphi)=1$, 
then it follows that $Q(\phi)=\xi \phi/[1-\xi \phi^2 (1-6\xi)]^{1/2}$.
The cosmological dynamics in such a theory have been studied by 
a number of authors \cite{stearly1,stearly2,stearly3,stearly4,stearly5,stearly6,Peri1,Gannouji06,Peri2,Martin,Gunzig,Verde,Agarwal,Jarv,Leach}.
If the field is nearly massless during most of the cosmological 
epochs, the coupling $Q$ needs to be suppressed to avoid the propagation 
of the fifth force.

In the following we shall study the cosmological dynamics and local 
gravity constraints on the constant coupling models based
on the action (\ref{action2}) with $F(\phi)=e^{-2Q\phi}$.

\subsection{Cosmological dynamics in Brans-Dicke theory}

We study the cosmological dynamics for the Jordan frame action
(\ref{action2}) in the presence of a non-relativistic fluid with
energy density $\rho_{m}$ and a radiation fluid with energy density
$\rho_{r}$. We regard the Jordan frame as a physical frame due to the
usual conservation of non-relativistic matter ($\rho_{m}\propto a^{-3}$).
In the flat FLRW background variation of the action (\ref{action2})
with respect to $g_{\mu\nu}$ and $\phi$ gives the following 
equations of motion 
\begin{eqnarray}
&  & 3FH^{2}=(1-6Q^{2})F\dot{\phi}^{2}/2+U-3H\dot{F}+\rho_{m}+\rho_{r}\,,
\label{scabe1}\\
&  & 2F\dot{H}=-(1-6Q^{2})F\dot{\phi}^{2}-\ddot{F}+H\dot{F}-\rho_{m}-(4/3)\rho_{r}\,,
\label{scabe2}\\
&  & (1-6Q^{2})F [\ddot{\phi}+3H\dot{\phi}+\dot{F}/(2F)\dot{\phi}]
+U_{,\phi}+QFR=0\,.
\label{scabe3}
\end{eqnarray}

Let us introduce the following variables 
\begin{equation}
x_{1}\equiv\frac{\dot{\phi}}{\sqrt{6}H}\,,\qquad 
x_{2}\equiv\frac{1}{H}\sqrt{\frac{U}{3F}}\,,\qquad 
x_{3}\equiv\frac{1}{H}\sqrt{\frac{\rho_{r}}{3F}}\,,
\end{equation}
and 
\begin{eqnarray}
\Omega_{m}\equiv\frac{\rho_{m}}{3FH^{2}}\,,\qquad
\Omega_{{\rm rad}}\equiv x_{3}^{2}\,,\qquad
\Omega_{{\rm DE}}\equiv(1-6Q^{2})x_{1}^{2}+x_{2}^{2}+2\sqrt{6}Qx_{1}\,.
\end{eqnarray}
These satisfy the relation $\Omega_{m}+\Omega_{{\rm rad}}+\Omega_{{\rm DE}}=1$
from Eq.~(\ref{scabe1}).
Using Eqs.~(\ref{scabe1})-(\ref{scabe3}), we obtain the differential
equations for $x_{1}$, $x_{2}$ and $x_{3}$: 
\begin{eqnarray}
\frac{{\rm d}x_{1}}{{\rm d}N} & = & \frac{\sqrt{6}}{2}(\lambda x_{2}^{2}-\sqrt{6}x_{1})
+\frac{\sqrt{6}Q}{2}\left[(5-6Q^{2})x_{1}^{2}+2\sqrt{6}Qx_{1}-3x_{2}^{2}+x_{3}^{2}-1\right]
-x_{1}\frac{\dot{H}}{H^{2}}\,,\label{au1}\\
\frac{{\rm d}x_{2}}{{\rm d}N} & = & 
\frac{\sqrt{6}}{2}(2Q-\lambda)x_{1}x_{2}-x_{2}\frac{\dot{H}}{H^{2}}\,,\label{au2}\\
\frac{{\rm d}x_{3}}{{\rm d}N} & = & 
\sqrt{6}Qx_{1}x_{3}-2x_{3}-x_{3}\frac{\dot{H}}{H^{2}}\,,\label{au3}
\end{eqnarray}
where $N=\ln a$, $\lambda=-U_{,\phi}/U$, and 
\begin{eqnarray}
\frac{\dot{H}}{H^{2}}=-\frac{1-6Q^{2}}{2}\left(3+3x_{1}^{2}
-3x_{2}^{2}+x_{3}^{2}-6Q^{2}x_{1}^{2}+2\sqrt{6}Qx_{1}\right)
+3Q(\lambda x_{2}^{2}-4Q)\,.
\end{eqnarray}
The effective equation of state of the system is given by 
by $w_{\rm eff}=-1-2\dot{H}/(3H^2)$.

If $\lambda$ is a constant, one can derive 
the fixed points of the system (\ref{au1})-(\ref{au3})
in the absence of radiation ($x_3=0$) \cite{TUMTY}: 
\begin{itemize}
\item (a) 
\begin{equation}
(x_{1},x_{2})=\left(\frac{\sqrt{6}Q}{3(2Q^{2}-1)},0\right),~~~
\Omega_{m}=\frac{3-2Q^{2}}{3(1-2Q^{2})^{2}},~~~w_{{\rm eff}}=\frac{4Q^{2}}{3(1-2Q^{2})}\,.\label{fp1}
\end{equation}

\item (b) 
\begin{equation}
(x_{1},x_{2})=\left(\frac{1}{\sqrt{6}Q\pm1},0\right)\,,\quad\Omega_{m}=0\,,\quad w_{{\rm eff}}=\frac{3\mp\sqrt{6}Q}{3(1\pm\sqrt{6}Q)}\,.
\end{equation}

\item (c) 
\begin{equation}
 (x_{1},x_{2})=\left(\frac{\sqrt{6}(4Q-\lambda)}{6(4Q^{2}-Q\lambda-1)},\left[\frac{6-\lambda^{2}+8Q\lambda-16Q^{2}}{6(4Q^{2}-Q\lambda-1)^{2}}\right]^{1/2}\right)\,, \quad 
 \Omega_{m}=0\,,\quad
 w_{{\rm eff}}=-\frac{20Q^{2}-9Q\lambda-3+\lambda^{2}}{3(4Q^{2}-Q\lambda-1)}\,.
 \end{equation}

\item (d) 
\begin{equation}
\hspace*{-0.5em}(x_{1},x_{2})=\left(\frac{\sqrt{6}}{2\lambda},\sqrt{\frac{3+2Q\lambda-6Q^{2}}{2\lambda^{2}}}\right),\quad\Omega_{m}=1-\frac{3-12Q^{2}+7Q\lambda}{\lambda^{2}},\quad w_{{\rm eff}}=-\frac{2Q}{\lambda}.\label{scaling}\end{equation}

\item (e) 
\begin{equation}
(x_{1},x_{2})=(0,1)\,,\quad\Omega_{m}=0\,,\quad w_{{\rm eff}}=-1\,.\label{fp5}
\end{equation}

\end{itemize}
The point (e) corresponds to the de Sitter point, which
exists only for $\lambda=4Q$ [this can be confirmed
by setting $\dot{\phi}=0$ in Eqs.~(\ref{scabe1})-(\ref{scabe3})].

We first study the case of non-zero values of $Q$ with constant $\lambda$,
i.e. for the exponential potential $U(\phi)=U_{0}e^{-\lambda\phi}$.
We do not consider the special case of $\lambda=4Q$.
The matter-dominated era can be realized
either by the point (a) or by the point (d). If the point (a) is responsible
for the matter era, the condition $Q^{2} \ll 1$ is required. We then
have $\Omega_{m}\simeq1+10Q^{2}/3>1$ and $w_{{\rm eff}}\simeq4Q^{2}/3$.
When $Q^{2}\ll1$ the scalar-field dominated point (c) yields an accelerated
expansion of the Universe provided that $-\sqrt{2}+4Q<\lambda<\sqrt{2}+4Q$.
Under these conditions the point (a) is followed by the late-time
cosmic acceleration. The scaling solution (d) can give rise to the
equation of state, $w_{{\rm eff}}\simeq0$ for $|Q|\ll|\lambda|$.
In this case, however, the condition $w_{{\rm eff}}<-1/3$ for the
point (c) gives $\lambda^{2}<2$. Then the energy fraction
of the pressureless matter for the point (d) does not satisfy the
condition $\Omega_{m}\simeq1$. From the above discussion the viable
cosmological trajectory for constant $\lambda$ corresponds to the
sequence from the point (a) to the scalar-field dominated point (c)
under the conditions $Q^{2}\ll1$ and $-\sqrt{2}+4Q<\lambda<\sqrt{2}+4Q$.

We shall proceed to the case where $\lambda$ varies with time.
The fixed points derived above for constant $\lambda$ 
can be regarded as the {}``instantaneous'' fixed points, 
provided that the time scale of the variation of $\lambda$ is 
smaller than that of the cosmic expansion.
The matter era can be realized by the point (d) with $|Q|\ll|\lambda|$.
The solutions finally approach either the de Sitter point (e) with 
$\lambda=4Q$ or the accelerated point (c).

In the following we focus on the case in which the matter era
with the point (d) is followed by the accelerated epoch
with the de Sitter solution (e). 
To study the stability of the point (e) we define 
a variable $x_{4}\equiv F$, satisfying the following equation
\begin{equation}
\frac{{\rm d}x_{4}}{{\rm d}N}=-2\sqrt{6}Qx_{1}x_{4}\,.\label{dx4}
\end{equation}
Considering the $3\times3$ matrix for perturbations $\delta x_{1}$,
$\delta x_{2}$ and $\delta x_{4}$ around the point (e), we obtain
the eigenvalues
\begin{equation}
-3\,,\quad-\frac{3}{2}\left[1\pm\sqrt{1-\frac{8}{3}F_{1}
Q\frac{{\rm d}\lambda}{{\rm d}F}(F_{1})}\right]\,,
\end{equation}
where $F_{1}\equiv F(\phi_{1})$ is the value of $F$ at the de Sitter
point with the field value $\phi_{1}$. Since $F_{1}>0$, we find
that the de Sitter point is stable under the condition
\begin{equation}
Q\frac{{\rm d}\lambda}{{\rm d}F}(F_{1})\ge0\,,\quad{\rm i.e.},
\quad\frac{{\rm d}\lambda}{{\rm d}\phi}(\phi_{1})\le0\,.
\label{decon}
\end{equation}

Let us consider the $f(R)$ model (\ref{fRasy}) in which the models 
(\ref{Amodel}) and (\ref{Bmodel}) are recovered in the
regime $R\gg R_{c}$. 
Since $e^{2\phi/\sqrt{6}}=1-2n\mu(R/R_{c})^{-(2n+1)}$, 
the potential $U=(FR-f)/2$ is given by  
\begin{equation}
U(\phi)=\frac{\mu R_{c}}{2}\left[1-\frac{2n+1}{(2n\mu)^{2n/(2n+1)}}
\left(1-e^{2\phi/\sqrt{6}}\right)^{2n/(2n+1)}\right]\,,
\label{fRpo2}
\end{equation}
In this case the slope of the potential, $\lambda=-U_{,\phi}/U$, is 
\begin{eqnarray}
\lambda = -\frac{4ne^{2\phi/\sqrt{6}}}{\sqrt{6}(2n\mu)^{2n/(2n+1)}}
\left[1-\frac{2n+1}{(2n\mu)^{2n/(2n+1)}}\left(1-e^{2\phi/\sqrt{6}}\right)
\right]^{-2n/(2n+1)} \left(1-e^{2\phi/\sqrt{6}}\right)^{-1/(2n+1)}\,.
\end{eqnarray}
In the deep matter-dominated epoch during which the condition $R/R_{c}\gg1$
is satisfied, the field $\phi$ is very close to zero. For $n$ and
$\mu$ of the order of unity, we have $|\lambda| \gg 1$ at this stage.
Hence the matter era can be realized by the instantaneous
fixed point (d). As $R/R_{c}$ gets smaller, $|\lambda|$ decreases
to the order of unity. If the solutions reach the point $\lambda=4Q=-4/\sqrt{6}$
and satisfy the stability condition ${\rm d}\lambda/{\rm d}F\le0$,
then the final attractor corresponds to the de Sitter 
fixed point (e).

For the theories with general couplings $Q$, it is possible to construct
a scalar-field potential that is the generalization of (\ref{fRpo}).
One example is \cite{TUMTY} 
\begin{equation}
U(\phi)=U_{0}\left[1-C(1-e^{-2Q\phi})^{p}\right]
\qquad
(U_{0}>0,~C>0,~0<p<1)\,.
\label{modelscalar}
\end{equation}
The $f(R)$ model (\ref{fRasy}) corresponds to $Q=-1/\sqrt{6}$ and $p=2n/(2n+1)$. 
The slope of the potential is given by 
\begin{equation}
\lambda=\frac{2Cp\, Qe^{-2Q\phi}(1-e^{-2Q\phi})^{p-1}}
{1-C(1-e^{-2Q\phi})^{p}}\,.
\label{lambdascalar}
\end{equation}
We have $U(\phi) \to U_0$ for $\phi \to 0$ and 
$U(\phi) \to U_{0}(1-C)$ in the limits $\phi\to\infty$ 
(for $Q>0$) and $\phi \to -\infty$ (for $Q<0$).

The field is nearly frozen around the value $\phi=0$ during the deep radiation
and matter epochs. In these epochs we have $R\simeq\rho_{m}/F$ from 
Eqs.~(\ref{scabe1})-(\ref{scabe3}) by noting that $U_{0}$ is negligibly 
small compared to $\rho_{m}$ or $\rho_{r}$. 
Using Eq.~(\ref{scabe3}), it follows that 
$U_{,\phi}+Q\rho_{m}\simeq0$. Hence, in the high-curvature region,
the field $\phi$ evolves along the instantaneous minima given by
\begin{equation}
\phi_{m}\simeq\frac{1}{2Q}\left(\frac{2U_{0}pC}{\rho_{m}}
\right)^{1/(1-p)}\,.
\label{phim}
\end{equation}
The field value $|\phi_{m}|$ increases for decreasing $\rho_{m}$.
As long as the condition $\rho_{m}\gg2U_{0}pC$ is satisfied, we have
$|\phi_{m}|\ll1$ from Eq.~(\ref{phim}).

For field values around $\phi=0$ one has $|\lambda|\gg1$ from
Eq.~(\ref{lambdascalar}).
Hence the instantaneous fixed point (d) can
be responsible for the matter-dominated epoch provided 
that $|Q|\ll|\lambda|$.
The variable $F=e^{-2Q\phi}$ decreases in time irrespective of the
sign of the coupling $Q$ and hence $0<F<1$. The de Sitter solution
corresponds to $\lambda=4Q$, that is 
\begin{equation}
C=\frac{2}{(1-F_{1})^{p-1}\left[2+(p-2)F_{1}\right]}\,.
\label{Cvalue}
\end{equation}
This solution is present as long as the solution of 
this equation exists in the region $0<F_{1}<1$.

{}From Eq.~(\ref{lambdascalar}) the derivative of $\lambda$
with respect to $\phi$ is 
\begin{equation}
\frac{{\rm d}\lambda}{{\rm d}\phi}=-\frac{4CpQ^{2}F(1-F)^{p-2}
[1-pF-C(1-F)^{p}]}{[1-C(1-F)^{p}]^{2}}\,.
\end{equation}
The de Sitter point is stable under the condition 
$1-pF_1>C(1-F_1)^p$. Using Eq.~(\ref{Cvalue}) this 
translates into 
\begin{equation}
F_1>1/(2-p)\,.
\label{F1con}
\end{equation}
When $0<C<1$ one can show that ${\rm d}\lambda/{\rm d}\phi<0$
is always satisfied.
Hence the solutions approach the de Sitter attractor
after the end of the matter era.
When $C>1$, the de Sitter point is stable
under the condition (\ref{F1con}).
If this condition is violated, the solutions choose another 
stable fixed point  [such as the point (c)] as an attractor.

\begin{figure}
\begin{center}
\includegraphics[height=3.4in,width=3.4in]{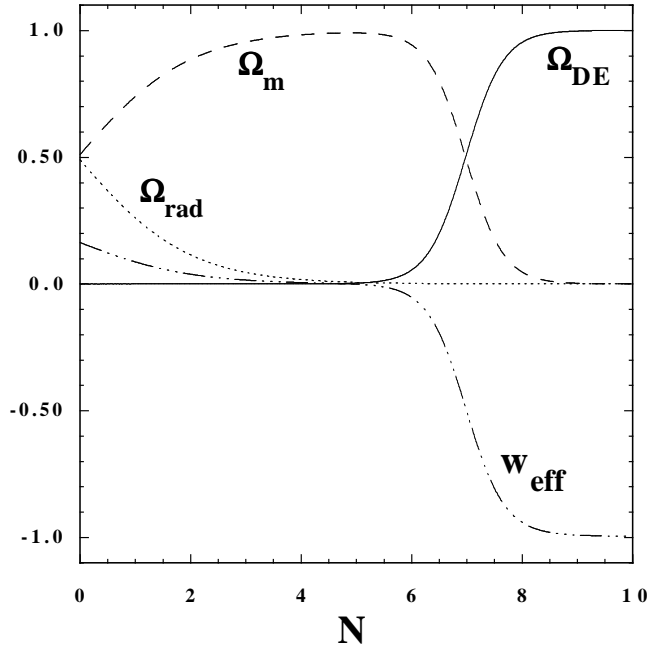}
\caption{
The evolution of $\Omega_{\rm DE}$, $\Omega_m$,
$\Omega_{\rm rad}$ and $w_{\rm eff}$
in Brans-Dicke theory with the potential (\ref{modelscalar}).
The model parameters are $Q=0.01$, $p=0.2$
and $C=0.7$ with the  initial conditions
$x_1=0$, $x_2=2.27 \times 10^{-7}$,
$x_3=0.7$, and $x_4-1=-5.0 \times 10^{-13}$.
{}From Ref.~\cite{TUMTY}.}
\label{stensorevo}
\end{center}
\end{figure}

The above discussion shows that, if $0<C<1$, the matter point (d) 
can be followed by the stable de Sitter solution (e).
In Fig.~\ref{stensorevo} we plot the evolution of $\Omega_{{\rm DE}}$,
$\Omega_{m}$, $\Omega_{{\rm rad}}$, and $w_{{\rm eff}}$ for $Q=0.01$, $p=0.2$ and $C=0.7$. 
This shows that the viable cosmological trajectory can be realized 
for the potential (\ref{modelscalar}).
In order to confront with SN Ia observations, it is possible to rewrite
Eqs.~(\ref{scabe1}) and (\ref{scabe2}) in the forms of
Eqs.~(\ref{mofR1}) and (\ref{mofR2}) by defining the dark energy
density $\rho_{{\rm DE}}$ and the pressure $P_{{\rm DE}}$ 
in the similar way.
It was shown in Ref.~\cite{TUMTY} that 
the phantom equation of state as well as the cosmological constant
boundary crossing can be realized for the field potentials $U(\phi)$ 
satisfying local gravity constraints.

\subsection{Local gravity constraints on Brans-Dicke theory}
\label{lgcstensor}

We study local gravity constraints on Brans-Dicke theory 
described by the action (\ref{action2}).
In the absence of the potential $U(\phi)$
we already mentioned that the Brans-Dicke parameter 
$\omega_{\rm BD}$ is constrained to be 
$\omega_{\rm BD}>4.0 \times 10^4$
from solar-system experiments. 
This gives the upper bound (\ref{Qbound}) on the coupling $Q$ between 
the field $\phi$ and non-relativistic matter in the Einstein frame.
This bound also applies to the case
of a nearly massless field with the potential $U(\phi)$
in which the Yukawa correction $e^{-Mr}$ is close to unity
(where $M$ is the scalar field mass and $r$ is an interaction length).

In the presence of the field-potential it is possible for large coupling
models ($|Q| \sim 1$) to satisfy local gravity constraints
provided that the mass $M$ of the field $\phi$ is sufficiently large in the
region of high density. 
In fact, the potential (\ref{modelscalar})
is designed to have a large mass in the high-density region, so that
it can be compatible with experimental tests of gravity 
through the chameleon mechanism.
In the following we study the model (\ref{modelscalar})
and derive the conditions under which local gravity constraints
can be satisfied. If we make a conformal transformation for the action
(\ref{modelscalar}), the action in the Einstein frame is given
by (\ref{action2}) with $F(\phi)=e^{-2Q\phi}$.
We can use the results obtained in Sec.~\ref{fRLGC}, because
thin-shell solutions have been derived for the general coupling $Q$.

As in the case of $f(R)$ gravity, we consider a configuration in which 
a spherically symmetric body has a constant density $\rho_{A}$ 
inside the body and that the density outside the body 
is given by $\rho=\rho_{B}~(\ll\rho_{A})$.
Under the condition $|Q\phi| \ll 1$, we have 
$V_{,\phi}\simeq-2U_{0}QpC(2Q\phi)^{p-1}$
for the potential $V=U/F^{2}$ in the Einstein frame. Then the field
values at the potential minima inside and outside the body are 
\begin{eqnarray}
\phi_{A}\simeq\frac{1}{2Q}\left(\frac{2U_{0}\, p\, C}
{\rho_{A}}\right)^{1/(1-p)}\,,
\qquad
\phi_{B}\simeq\frac{1}{2Q}\left(\frac{2U_{0}\, p\, C}
{\rho_{B}} \right)^{1/(1-p)}\,.\label{phiAB}
\end{eqnarray}
In order to realize the accelerated expansion today, the energy scale 
$U_0$ is required to be the same order as the square of the present
Hubble parameter $H_{0}$, i.e. $U_{0} \sim H_{0}^{2} \sim \rho_{0}$,
where $\rho_{0}\simeq10^{-29}$ g/cm$^{3}$ is the cosmological
density today. The baryonic/dark matter density in our galaxy corresponds
to $\rho_{B}\simeq10^{-24}$ g/cm$^{3}$. Hence the conditions
$|Q\phi_{A}|\ll1$ and $|Q\phi_{B}|\ll1$ are in fact satisfied unless
$C\gg1$. The field mass squared $m_{A}^{2}\equiv V_{,\phi\phi}$
at $\phi=\phi_{A}$ is approximately given by 
\begin{equation}
m_{A}^{2}\simeq\frac{1-p}{(2^{p}\, pC)^{1/(1-p)}}Q^{2}
\left(\frac{\rho_{A}}{U_{0}}\right)^{(2-p)/(1-p)}U_{0}\,,\label{Mphi}
\end{equation}
which means that $m_{A}$ can be much larger than $H_{0}$ because
of the condition $\rho_{A}\gg U_{0}$.
This large mass allows the chameleon mechanism to work, because  
the condition $1/(m_A \tr_c) \ll 1$ is satisfied.

The bound (\ref{boep2}) coming from the violation
of equivalence principle in the solar system translates into 
\begin{equation}
\left(2U_{0}pC/\rho_{B}\right)^{1/(1-p)}<7.4\times10^{-15}\,|Q|\,.
\label{delrcon2}
\end{equation}
Let us consider the case in which the solutions finally approach 
the de Sitter solution (e).
At the point (e), one has $3F_{1}H_{1}^{2}=U_{0}[1-C(1-F_{1})^{p}]$
with $C$ given in Eq.~(\ref{Cvalue}). 
Hence we get the following relation 
\begin{equation}
U_{0}=3H_{1}^{2}\left[2+(p-2)F_{1}\right]/p\,.
\end{equation}
Plugging this into Eq.~(\ref{delrcon2}), it follows that  
\begin{equation}
\left(R_{1}/\rho_{B}\right)^{1/(1-p)}(1-F_{1})
<7.4\times10^{-15}|Q|\,,
\end{equation}
where $R_{1}=12H_{1}^{2}$ is the Ricci scalar at the de Sitter point.
Since the term $(1-F_{1})$ is smaller than 1/2 from the condition
(\ref{F1con}), we obtain the inequality 
$(R_{1}/\rho_{B})^{1/(1-p)}<1.5\times10^{-14}|Q|$.
Using the values $R_{1}=10^{-29}$ g/cm$^{3}$ and
$\rho_{B}=10^{-24}$ g/cm$^{3}$, 
we obtain the following bound 
\begin{equation}
p>1-\frac{5}{13.8-{\rm \log}_{10}\,|Q|}\,.\label{EPsca}
\end{equation}
When $|Q|=10^{-1}$ and $|Q|=1$ we have $p>0.66$ and $p>0.64$,
respectively. 
Thus the model can be compatible with local gravity experiments
even for $|Q|={\cal O}(1)$. 

In Ref.~\cite{Gannouji10} it was shown that in order to satisfy both 
local gravity and cosmological constraints the chameleon potentials 
in the Einstein frame need to be of the form 
$V(\phi)=M^4 [1+f(\phi)]$, where the function $f(\phi)$ is 
smaller than 1 today and $M$ is a mass that corresponds to 
the dark energy scale ($M \sim 10^{-12}$\,GeV).
The potential $V(\phi)=M^4 \exp[\mu (M/\phi)^n]$ \cite{KW2,Braxchame} 
is one of those viable candidates, but the allowed model parameter space is severely 
constrained by the 2006 E\"{o}t-Wash experiment \cite{Kapner:2006si}.
Unless the parameter $\mu$ is unnaturally small ($\mu \lesssim 10^{-5}$),
this potential is incompatible with local gravity constraints 
for $\{n,Q \} = {\cal O}(1)$.

On the other hand, the chameleon potential 
$V(\phi)=V_0 [1-\mu (1-e^{-\phi})^n ]$ ($0<n<1$) 
can satisfy both local gravity and cosmological 
constraints\footnote{In the Einstein frame
the potential (\ref{modelscalar}) takes the form 
$V(\phi)=U/F^2=U_{0}e^{4Q\phi}\left[1-C(1-e^{-2Q\phi})^{p}\right]$, 
so in the region $|Q \phi| \ll 1$ this potential is similar to 
$V(\phi)=V_0 [1-\mu (1-e^{-\phi})^n ]$.}.
In Ref.~\cite{Gannouji10} this potential is consistent with 
the constraint coming from 2006 E\"{o}t-Wash experiments as well as 
the WMAP bound on the variation of the field-dependent 
mass \cite{Nagata} for natural model parameters.

\section{DGP model}
\label{DGPsec}

In this section we review braneworld models of dark energy motivated 
by string theory. In braneworlds standard
model particles are confined on a 3-dimensional (3D) brane embedded in
5-dimensional bulk with large extra dimensions \cite{Randall1,Randall2}.
Dvali, Gabadadze, and Porrati (DGP) \cite{DGP} proposed a braneworld
model in which the 3-brane is embedded in a Minkowski bulk 
with infinitely large extra dimensions.
One can recover Newton's law by adding a 4D Einstein-Hilbert 
action sourced by the brane curvature to the 5D action \cite{DGP2}. 
The presence of such a 4D term may be induced by quantum corrections 
coming from the bulk gravity and its coupling with matter on the brane. In the
DGP model the standard 4D gravity is recovered at small distances,
whereas the effect from the 5D gravity manifests itself for large
distances. Interestingly one can realize the self cosmic
acceleration without introducing 
a dark energy component \cite{Deffayet1,Deffayet2}
(see also Ref.~\cite{Shtanov}).

\subsection{Self-accelerating solution}

The action of the DGP model is given by 
\begin{equation}
S=\frac{1}{2\kappa_{(5)}^{2}}\int\rd^{5}X\sqrt{-\tilde{g}}\,
\tilde{R}+\frac{1}{2\kappa_{(4)}^{2}}\int\rd^{4}X\sqrt{-g}R-
\int\rd^{5}X\sqrt{-\tilde{g}}\,{\cal L}_{M}\,,
\label{DGPaction}
\end{equation}
where $\tilde{g}_{AB}$ is the metric in the 5D bulk and 
$g_{\mu\nu}=\partial_{\mu}X^{A}\partial_{\nu}X^{B}\tilde{g}_{AB}$
is the induced metric on the brane with $X^{A}(x^{c})$ being the
coordinates of an event on the brane labelled by $x^{c}$.
The 5D and 4D gravitational constants, $\kappa_{(5)}^{2}$ and 
$\kappa_{(4)}^{2}$, are related with the 5D and 4D Planck masses,
$M_{(5)}$ and $M_{(4)}$, via
\begin{equation}
\kappa_{(5)}^{2}=1/M_{(5)}^3\,,\qquad
\kappa_{(4)}^{2}=1/M_{(4)}^2\,.
\end{equation}
The first and second terms in Eq.~(\ref{DGPaction}) correspond to
Einstein-Hilbert actions in the 5D bulk and on the brane, respectively.
There is no contribution to the Lagrangian ${\cal L}_{M}$ from 
the bulk because we are considering a Minkowski bulk.
Then the matter action consists of a brane-localized matter 
whose action is given by 
$\int\rd^{4}x\sqrt{-g}\,(\sigma+{\cal L}_{M}^{{\rm brane}})$,
where $\sigma$ is the 3-brane tension and ${\cal L}_{M}^{{\rm brane}}$
is the Lagrangian density on the brane. 
Since the tension is unrelated to the Ricci scalar $R$, 
it can be adjusted to be zero (as we do in the following).

In order to study the cosmological dynamics on the brane 
(located at $y=0$), 
we take a metric of the form:
\begin{equation}
\rd s^{2}=-n^{2}(\tau,y)\rd\tau^{2}+a^{2}(\tau,y)\,
\gamma_{ij}\rd x^{i}\rd x^{j}+\rd y^{2}\,,
\label{branemetric}
\end{equation}
where $\gamma_{ij}$ represents a maximally symmetric space-time with
a constant curvature $K$. 
The 5D Einstein equations are 
\begin{equation}
\tilde{G}_{AB}\equiv\tilde{R}_{AB}-\frac{1}{2}\tilde{R}\,\tilde{g}_{AB}
=\kappa_{(5)}^{2}\tilde{T}_{AB}\,,
\label{GAB}
\end{equation}
where $\tilde{R}_{AB}$ is the 5D Ricci tensor, $\tilde{T}_{AB}$
is the sum of the energy momentum tensor $T_{AB}^{({\rm brane)}}$
on the brane and the contribution $\tilde{U}_{AB}$ coming from the
scalar curvature of the brane: 
\begin{equation}
\tilde{T}_{AB}=T_{AB}^{{\rm (brane)}}+\tilde{U}_{AB}\,.
\end{equation}
Since we are considering a homogeneous and isotropic Universe on
the brane, one can write ${T_{B}^{A}}^{{\rm (brane)}}$ in the form
\begin{equation}
{T_{B}^{A}}^{{\rm (brane)}}=
\delta(y)\,{\rm diag}(-\rho_{M},P_{M},P_{M},P_{M},0)\,,
\end{equation}
where $\rho_{M}$ and $P_{M}$ are functions of $\tau$ only.
The non-vanishing components coming from the Ricci scalar $R$ of
the brane are 
\begin{eqnarray}
&  & \tilde{U}_{00}=-\frac{3}{\kappa_{(4)}^{2}}
\left(\frac{\dot{a}^{2}}{a^{2}}+K\frac{n^{2}}{a^{2}}\right)\,\delta(y)\,,\\
&  & \tilde{U}_{ij}=-\frac{1}{\kappa_{(4)}^{2}}\left[\frac{a^{2}}{n^{2}}
\left(-\frac{\dot{a}^{2}}{a^{2}}+2\frac{\dot{a}}{a}\frac{\dot{n}}{n}
-2\frac{\ddot{a}}{a}\right)-K\right]\gamma_{ij}\,\delta(y)\,,
\end{eqnarray}
where a dot represents a derivative with respect to $\tau$. 
The non-vanishing components of the 5D Einstein tensor 
$\tilde{G}_{AB}$ are \cite{Langlois1,Langlois2,Deffayet1}
\begin{eqnarray}
&  & \tilde{G}_{00}=3\left[\frac{\dot{a}^{2}}{a^{2}}-n^{2}
\left(\frac{a''}{a}+\frac{a'^{2}}{a^{2}}\right)
+K\frac{n^{2}}{a^{2}}\right]\,,\label{DGPG1}\\
&  & \tilde{G}_{ij}=\left[a^{2}\left(2\frac{a''}{a}+\frac{n''}{n}
+\frac{a'^{2}}{a^{2}}+2\frac{a'n'}{an}\right)+
\frac{a^{2}}{n^{2}}\left(-2\frac{\ddot{a}}{a}
-\frac{a'^{2}}{a^{2}}+2\frac{\dot{a}\dot{n}}{an}\right)
-K\right]\gamma_{ij}\,,\label{DGPG2} \\
&  & \tilde{G}_{05}=3\left(\frac{\dot{a}n'}{an}
-\frac{\dot{a}'}{a}\right)\,,\label{DGPG3}\\
&  & \tilde{G}_{55}=3\left(\frac{a'^{2}}{a^{2}}+\frac{a'n'}{an}\right)
-\frac{3}{n^{2}}\left(\frac{\ddot{a}}{a}+\frac{\dot{a}^{2}}{a^{2}}
-\frac{\dot{a}\dot{n}}{an}\right)-3\frac{K}{a^{2}}\,,\label{DGPG4}
\end{eqnarray}
where a prime represents a derivative with respect to $y$.

Assuming no flow of matter along the 5-th dimensions, 
we have $\tilde{T}_{05}=0$ and hence $\tilde{G}_{05}=0$. 
Then Eqs.~(\ref{DGPG1}) and (\ref{DGPG4}) can be written as 
\begin{equation}
\tilde{G}_{00}=-\frac{3n^{2}}{2a^{3}a'}I'\,,\qquad
\tilde{G}_{55}=-\frac{3}{2a^{3}\dot{a}}\dot{I}\,,
\end{equation}
where 
\begin{equation}
I\equiv(a'a)^{2}-\frac{(\dot{a}a)^{2}}{n^{2}}-Ka^{2}\,.
\end{equation}
Since we are considering the Minkowski bulk, we have $\tilde{G}_{00}=0$
and $\tilde{G}_{55}=0$ locally in the bulk. This gives $I'=0$
and $\dot{I}=0$. Integrations of these equations lead to 
\begin{equation}
(a'a)^{2}-\frac{(\dot{a}a)^{2}}{n^{2}}-Ka^{2}+C=0\,,
\label{dgpaa}
\end{equation}
where $C$ is a constant independent of $\tau$ and $y$.

We shall find solutions of the Einstein equations (\ref{GAB}) in the
vicinity of $y=0$. The metric needs to be continuous across
the brane in order to have a well-defined geometry. 
However, its derivatives with respect to $y$ can be discontinuous at
$y=0$. The Einstein tensor is made of the metric up to the second
derivatives with respect to $y$, so the Einstein equations with a
distributional source are written in the 
form \cite{Langlois1,Langlois2,Deffayet1}
\begin{equation}
g''=T\,\delta(y)\,,
\end{equation}
where $\delta(y)$ is a Dirac's delta function. 
Integrating this equation across the brane gives 
\begin{equation}
[g']=T\,,\qquad{\rm where}\qquad[g']
\equiv g'(0^{+})-g'(0^{-})\,.
\label{g0}
\end{equation}
The jump of the first derivative of the metric is
equivalent to the energy-momentum tensor on the brane.

Equations (\ref{DGPG1}) and (\ref{DGPG2}) include the 
derivatives $a''$ and $n''$ of the metric. Integrating the Einstein
equations $\tilde{G}_{00}=\kappa_{(5)}^{2}\tilde{T}_{00}$ 
and $\tilde{G}_{ij}=\kappa_{(5)}^{2}\tilde{T}_{ij}$
across the brane, we obtain 
\begin{eqnarray}
&  & \frac{[a']}{a_{b}}=-\frac{\kappa_{(5)}^{2}}{3}\rho_{M}
+\frac{\kappa_{(5)}^{2}}{\kappa_{(4)}^{2}n_{b}^{2}}
\left(\frac{\dot{a}_{b}^{2}}{a_{b}^{2}}+K\frac{n_{b}^{2}}
{a_{b}^{2}}\right)\,,\label{aDGP}\\
&  & \frac{[n']}{n_{b}}=\frac{\kappa_{(5)}^{2}}{3}(3P_{M}
+2\rho_{M})-\frac{\kappa_{(5)}^{2}}{\kappa_{(4)}^{2}n_{b}^{2}}
\left(\frac{\dot{a}_{b}^{2}}{a_{b}^{2}}+2\frac{\dot{a}_{b}}{a_{b}}
\frac{\dot{n}_{b}}{n_{b}}-2\frac{\ddot{a}_{b}}{a_{b}}
+K\frac{n_{b}^{2}}{a_{b}^{2}}\right)\,,
\label{nDGP}
\end{eqnarray}
where the subscript {}``$b$'' represents the quantities on the brane.

We assume the symmetry $y\leftrightarrow-y$, in which case $[a']=2a'(0^{+})$
and $[n']=2n'(0^{+})$. Substituting Eq.~(\ref{aDGP}) into Eq.~(\ref{dgpaa}),
we obtain the modified Friedmann equation on the brane: 
\begin{equation}
\epsilon\sqrt{H^{2}+\frac{K}{a_{b}^{2}}-\frac{C}{a_{b}^{4}}}=
\frac{\kappa_{(5)}^{2}}{2\kappa_{(4)}^{2}}\left(H^{2}
+\frac{K}{a_{b}^{2}}\right)
-\frac{\kappa_{(5)}^{2}}{6}\rho_{M}\,,\label{DGPeq1}
\end{equation}
where $H\equiv\dot{a}_{b}/(a_{b}n_{b})$ is the Hubble parameter
and $\epsilon=\pm1$ is the sign of $[a']$. The constant $C$
can be interpreted as the term coming from the 5D bulk Weyl 
tensor \cite{SMS,Deffayet1,Deffayet2}. 
Since the Weyl tensor vanishes for
the Minkowski bulk, we set $C=0$ in the following discussion.
We introduce a length scale 
\begin{equation}
r_{c}\equiv\frac{\kappa_{(5)}^{2}}{2\kappa_{(4)}^{2}}=
\frac{M_{(4)}^{2}}{2M_{(5)}^{3}}\,.
\label{crossover}
\end{equation}
Then Eq.~(\ref{DGPeq1}) can be written as 
\begin{equation}
\frac{\epsilon}{r_{c}}\sqrt{H^{2}+\frac{K}{a^{2}}}
=H^{2}+\frac{K}{a^{2}}
-\frac{\kappa_{(4)}^{2}}{3}\rho_{M}\,,\label{DGBf1}
\end{equation}
where we have omitted the subscript {}``$b$'' for the 
quantities at $y=0$.

Plugging the junction conditions (\ref{aDGP}) and (\ref{nDGP}) into
the $(05)$ component of the Einstein equations, $\tilde{G}_{05}=0$,
the following matter conservation equation holds on the brane:
\begin{equation}
\frac{\rd\rho_{M}}{\rd t}+3H(\rho_{M}+P_{M})=0\,,
\label{DGBf2}
\end{equation}
where $t$ is the cosmic time related to the time $\tau$ via the
relation $\rd t=n_{b}\rd\tau$. If the equation of state, 
$w_{M}=P_{M}/\rho_{M}$, is specified, 
the cosmological evolution is known by solving Eqs.~(\ref{DGBf1})
and (\ref{DGBf2}).

For a flat geometry ($K=0$), Eq.~(\ref{DGBf1}) reduces to 
\begin{equation}
H^{2}-\frac{\epsilon}{r_{c}}H=
\frac{\kappa_{(4)}^{2}}{3}\rho_{M}\,.
\label{DGPK0eq}
\end{equation}
If the crossover scale $r_c$ is much larger than the Hubble radius $H^{-1}$,
the first term in Eq.~(\ref{DGPK0eq})
dominates over the second one. In this case the standard Friedmann
equation, $H^{2}=\kappa_{(4)}^{2}\rho_{M}/3$, is recovered. 
On the other hand, in the regime $r_{c}< H^{-1}$, 
the presence of the second term in Eq.~(\ref{DGPK0eq}) 
leads to a modification to the standard
Friedmann equation. In the Universe dominated by non-relativistic
matter ($\rho_{M}\propto a^{-3}$), the Universe approaches a de Sitter
solution for the branch $\epsilon=+1$: 
\begin{equation}
H \to H_{{\rm dS}}=1/r_{c}\,.
\end{equation}
We can realize the cosmic acceleration today provided
that $r_{c}$ is of the order of the present Hubble radius $H_{0}^{-1}$.

\subsection{Observational constraints on the DGP model and other aspects of the model}
%

\begin{figure}
\begin{centering}
\includegraphics[width=3.4in,height=3.4in]{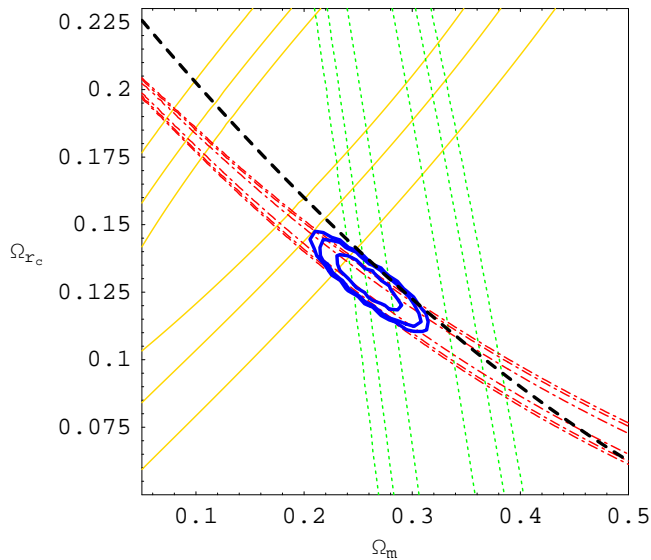} 
\par\end{centering}
\caption{Observational constraints on the DGP model from the SNLS
data \cite{Astier05} (solid thin), the BAO \cite{BAO1} (dotted), 
and the CMB shift parameter from
the WMAP 3-year data \cite{WMAP3} (dot-dashed). 
The thick line represents the curve
(\ref{curveflat}) for the flat model ($\Omega_{K}^{(0)}=0$).
The figure labels $\Omega_m$ and $\Omega_{r_c}$ correspond to 
$\Omega_m^{(0)}$ and $\Omega_{r_c}^{(0)}$, respectively.
{}From Ref.~\cite{DGPobser3}.}
\centering{}\label{dgpfig} 
\end{figure}

Equation (\ref{DGBf1}) can be written as 
\begin{equation}
H^{2}+\frac{K}{a^{2}}=\left(\sqrt{\frac{\kappa_{(4)}^{2}}{3}\rho_{M}
+\frac{1}{4r_{c}^{2}}}+\frac{1}{2r_{c}}\right)^{2}\,.\label{DGPH1}
\end{equation}
For the matter on the brane, we consider non-relativistic matter
with the energy density $\rho_{m}$ and the equation of state $w_{m}=0$.
We then have $\rho_{m}=\rho_{m}^{(0)}(1+z)^{3}$ from Eq.~(\ref{DGBf2}).
Let us introduce the following density parameters
\begin{eqnarray}
\Omega_{K}^{(0)}=-\frac{K}{a_{0}^{2}H_{0}^{2}}\,,
\qquad\Omega_{r_{c}}^{(0)}=\frac{1}{4r_{c}^{2}H_{0}^{2}}\,,
\qquad\Omega_{m}^{(0)}=
\frac{\kappa_{(4)}^{2}\rho_{m}^{(0)}}{3H_{0}^{2}}\,.
\end{eqnarray}
Then Eq.~(\ref{DGPH1}) reads 
\begin{equation}
H^{2}(z)=H_{0}^{2}\left[\Omega_{K}^{(0)}(1+z)^{2}+\left\{
\sqrt{\Omega_{m}^{(0)}(1+z)^{3}+\Omega_{r_c}^{(0)}}+
\sqrt{\Omega_{r_c}^{(0)}}\right\} ^{2}\right]\,.
\label{DGPhubble}
\end{equation}
The normalization condition at $z=0$ is given by 
\begin{equation}
\Omega_{m}^{(0)}+\Omega_{K}^{(0)}+2\sqrt{1-\Omega_{K}^{(0)}}
\sqrt{\Omega_{r_{c}}^{(0)}}=1\,.
\label{DGPhubble2}
\end{equation}
For the flat universe\index{flat universe} ($K=0$) 
this relation corresponds to
\begin{eqnarray}
\Omega_{r_{c}}^{(0)}=(1-\Omega_{m}^{(0)})^{2}/4\,.
\label{curveflat}
\end{eqnarray}

The parametrization\index{parametrization} (\ref{DGPhubble}) of 
the Hubble parameter together
with the normalization (\ref{DGPhubble2}) can be used to place observational
constraints on the DGP model at the background 
level \cite{DGPobser1,DGPobser2,DGPobser3,DGPobser4,DGPobser5}.
In Ref.~\cite{DGPobser1} the authors found a significantly worse 
fit to Supernova Ia (SN Ia) data and the distance to the last-scattering 
surface \cite{WMAP1} relative to the $\Lambda$CDM model.
In Refs.~\cite{DGPobser2} and \cite{DGPobser4} the authors showed that 
the flat DGP model is disfavored from the combined data analysis
of SN Ia \cite{Riess04,Astier05} and BAO \cite{BAO1}.
In Fig.~\ref{dgpfig} we show the joint observational constraints \cite{DGPobser3} from the data
of SNLS \cite{Astier05}, BAO \cite{BAO1}, and the CMB shift 
parameter \cite{WMAP3}. 
While the flat DGP model can be consistent with the SN Ia data, it is under
strong observational pressure by adding the data of BAO
and the CMB shift parameter. The open DGP model gives a slightly better 
fit relative to the flat model \cite{DGPobser3,DGPobser5}.
The joint analysis using the data of SN Ia, BAO, CMB, 
gamma ray bursts, and the linear growth factor of matter perturbations
show that the flat DGP model is incompatible with current 
observations \cite{DGPobser6}.

In the DGP model a brane-bending mode $\phi$ 
(i.e. longitudinal graviton) gives 
rise to a field self-interaction of the form 
$\square \phi (\partial^\mu \phi \partial_{\mu} \phi)$
through a mixing with the transverse graviton \cite{DGPnon1,DGPnon2}.
This can lead to the decoupling of the field $\phi$ from gravitational dynamics 
in the local region by the so-called Vainshtein mechanism \cite{Vainshtein}.
The General Relativistic behavior can be recovered within a radius
$r_*=(r_g r_c^2)^{1/3}$, where $r_g$ is the Schwarzschild radius of a source.
Since $r_*$ is larger than the solar-system scales, 
the DGP model can evade local gravity 
constraints \cite{DGPnon1,Gruzinov:2001hp,DGPnon2}.
However the DGP model is plagued by a strong coupling problem for 
typical distances smaller than 1000 km \cite{Luty}.
Some regularization methods have been proposed to avoid the strong coupling 
problem, such as smoothing out the delta profile on the brane \cite{Kolano1,Kolano2} 
or re-using the delta function profile but in a higher-dimensional 
brane \cite{CascoDGP,CascoDGP2,CascoDGP3}.

As we will see in \ref{DGPpersec}, the analysis of 5D cosmological 
perturbations on the scales larger than $r_*$ shows that the DGP model 
contains a ghost mode in the scalar sector
of the gravitational field \cite{DGPghost1,DGPghost2,DGPghost3}.
There are several ways of the generalization of the DGP model to avoid
the appearance of ghosts. 
One way is to consider the 6D braneworld set-up as in the 
Cascading gravity \cite{deRham}.
Another is to generalize the field self-interaction term 
$\square \phi (\partial^\mu \phi \partial_{\mu} \phi)$
to more general forms in the 4D gravity \cite{Nicolis}.
In Sec.~\ref{gasec} we shall discuss the latter approach 
(``Galileon gravity'') in detail.

\section{Galileon gravity}
\label{gasec}

In the DGP model the field derivative self-coupling
$\square \phi (\partial_\mu \phi \partial^\mu \phi)$, 
arising from a brane-bending mode, allows the decoupling of 
the field from matter within a Vainshtein radius.
In the local regions where solar-system experiments are 
carried out, the field is nearly frozen 
through the non-linear self-interaction.
This is different from the chameleon mechanism in which 
the presence of the field potential with a matter coupling 
gives rise to a minimum with a large mass in the regions of high density.

Under the Galilean shift $\partial_{\mu} \phi \to 
\partial_{\mu} \phi+b_{\mu}$, the field equation following from 
the Lagrangian $\square \phi (\partial^{\mu} \phi \partial_{\mu} \phi)$ 
is unchanged in the Minkowski space-time.
The generalization of the nonlinear field Lagrangian to more 
general cases may be useful, e.g., to overcome the ghost problem 
associated with the DGP model. 
In fact Nicolis {\it et al.} \cite{Nicolis} derived five Lagrangians 
that lead to the field equations invariant under the Galilean shift
$\partial_\mu \phi \to\partial_\mu\phi+b_\mu$
in the Minkowski space-time.
The scalar field respecting the Galilean symmetry is dubbed ``Galileon''.
Each of the five terms only leads to second-order differential equations, 
keeping the theory free from unstable spin-2 ghost degrees of freedom. 

If we extend the analysis in Ref.~\cite{Nicolis} to that in 
the curved space-time,  the Lagrangians should be promoted to 
covariant forms.
Deffayet {\it et al.} \cite{Deffayetga1,Deffayetga2} derived covariant 
Lagrangians ${\cal L}_i$ ($i=1, \cdots,5$) that keep the field equations 
up to second-order, while recovering the five Lagrangians derived by 
Nicolis {\it et al.} in the Minkowski space-time.
This can be achieved by introducing
field-derivative couplings with the Ricci scalar $R$ and
the Einstein tensor $G_{\nu \rho}$ in the expression of
${\cal L}_{4,5}$. Since the existence of those terms
affects the effective gravitational coupling, the Galileon
gravity based on the covariant Lagrangians
${\cal L}_i$ ($i=1,\cdots, 5$) can be classified as one of
modified gravitational theories.

The cosmological dynamics including the terms up to ${\cal L}_{4}$
and ${\cal L}_{5}$ have been studied by a number of 
authors \cite{Sami10,DT2}.
In particular Refs.~\cite{DT2,DT3} have shown that, for the covariant
Galileon theory having de Sitter attractors, cosmological solutions
with different initial conditions converge to a common
trajectory-- a tracker solution.
Moreover there is a viable parameter space in which the conditions 
for the avoidance of ghosts and Laplacian instabilities 
of scalar and tensor perturbations are satisfied.

The generalization of Galileon gravity, which mostly corresponds to 
the modification of the term
${\cal L}_3=\square \phi (\partial^{\mu} \phi \partial_{\mu} \phi)$, 
has been also extensively studied recently \cite{JustinGal}-\cite{Hirano}.
One application is to introduce the non-linear field self-interaction of 
the form $\xi(\phi) \square \phi (\partial_\mu \phi \partial^\mu \phi)$
in the action of (generalized) Brans-Dicke 
theories \cite{KazuyaGal,Kobayashi1,Kobayashi2,DT1,DMT}, 
where $\xi$ is a function of $\phi$.
For suitable choices of the function $\xi (\phi)$,
there exist de Sitter (dS) solutions responsible for dark energy
even in the absence of the field potential.
The cosmology based on a further general term $G(\phi,X) \square \phi$
has been discussed in the context of either dark energy and 
inflation \cite{KYY,DPSV,Mizuno,Burrage,Kimura,Kamada,Hirano}.

In the following we review the cosmological dynamics in the Galileon 
dark energy model based on the covariant Lagrangians ${\cal L}_i$ ($i=1,\cdots, 5$)
and study observational constraints on the model.
We will also discuss the modified version of Galileon gravity in which 
the term 
${\cal L}_3=\square \phi (\partial^{\mu} \phi \partial_{\mu} \phi)$ is generalized.

\subsection{Cosmology of a covariant Galileon field}
\label{galisec}

We start with the covariant Galileon gravity described
by the action \cite{Deffayetga1,Deffayetga2}
\begin{equation}
S=\int {\rm d}^4 x \sqrt{-g}\,\left[ \frac{M_{\rm pl}^2}{2}R+
\frac12 \sum_{i=1}^5 c_i {\cal L}_i \right]
+\int {\rm d}^4 x\, {\cal L}_{M}\,,
\label{action}
\end{equation}
where $M_{\rm pl}=(8\pi G)^{-1/2}$ is the reduced Planck mass, 
and $c_i$'s are constants.
The five Lagrangians ${\cal L}_i$ ($i=1, \cdots, 5$)
satisfying the Galilean symmetry in the limit
of the Minkowski space-time are given by
\begin{eqnarray}
& & {\cal L}_1=M^3 \phi\,,\quad
{\cal L}_2=(\nabla \phi)^2\,,\quad
{\cal L}_3=(\square \phi) (\nabla \phi)^2/M^3\,, \quad
{\cal L}_4=(\nabla \phi)^2 \left[2 (\square \phi)^2
-2 \phi_{;\mu \nu} \phi^{;\mu \nu}-R(\nabla \phi)^2/2 \right]/M^6,
\nonumber \\
& & {\cal L}_5=(\nabla \phi)^2 [ (\square \phi)^3
-3(\square \phi)\,\phi_{; \mu \nu} \phi^{;\mu \nu} 
+2{\phi_{;\mu}}^{\nu} {\phi_{;\nu}}^{\rho}
{\phi_{;\rho}}^{\mu}
-6 \phi_{;\mu} \phi^{;\mu \nu}\phi^{;\rho}G_{\nu \rho} ]
/M^9\,,
\label{lag}
\end{eqnarray}
where $M$ is a constant having a dimension of mass, 
and $G_{\nu \rho}$ is the Einstein tensor.
For the matter Lagrangian ${\cal L}_{M}$
we take into account perfect fluids of
non-relativistic matter (energy density $\rho_m$, equation of
state $w_m=0$) and
radiation (energy density $\rho_r$, equation of state $w_r=1/3$).

Let us consider the FLRW metric with the cosmic curvature $K$:
\begin{equation}
{\rm d}s^2=-{\rm d}t^2+a^2(t) \left[
\frac{{\rm d} r^2}{1-Kr^2}+r^2 ({\rm d}\theta^2+
\sin^2 \theta\,{\rm d}\phi^2) \right]\,.
\label{metric}
\end{equation}
Variation of the action (\ref{action}) with respect to $g_{\mu \nu}$
leads to the following equations of motion
\begin{eqnarray}
& & 3M_{\rm pl}^2 H^2=\rho_{\rm DE}+\rho_m+\rho_r+\rho_K\,,
\label{basic1} \\
& & 3M_{\rm pl}^2 H^2+2M_{\rm pl}^2 \dot{H}=-P_{\rm DE}
-\rho_r/3+\rho_K/3\,,
\label{basic2}
\end{eqnarray}
where $\rho_K \equiv -3KM_{\rm pl}^2/a^2$, and
\begin{eqnarray}
\rho_{\rm DE} &\equiv& -c_1 M^3 \phi/2-c_2 \dot{\phi}^2/2
+3c_3 H \dot{\phi}^3/M^3-45 c_4 H^2 \dot{\phi}^4/(2M^6)
+21c_5 H^3 \dot{\phi}^5/M^9,\\
P_{\rm DE} &\equiv&  c_1 M^3 \phi/2-c_2 \dot{\phi}^2/2
-c_3 \dot{\phi}^2 \ddot{\phi}/M^3
+3c_4 \dot{\phi}^3 [8H\ddot{\phi} +(3H^2+2\dot{H})
\dot{\phi}]/(2 M^6) \nonumber \\
& &-3c_5 H \dot{\phi}^4 
[5H \ddot{\phi}+2(H^2+\dot{H})\dot{\phi} ]/M^9\,.
\end{eqnarray}
The matter fluids satisfy the continuity equations
$\dot{\rho}_m+3H\rho_m=0$ and 
$\dot{\rho}_r+4H\rho_r=0$.
We define the dark energy equation of state $w_{\rm DE}$
and the effective equation of state $w_{\rm eff}$, as
\begin{equation}
w_{\rm DE} \equiv \frac{P_{\rm DE}}{\rho_{\rm DE}}\,,\qquad
w_{\rm eff} \equiv -1-\frac{2\dot{H}}{3H^2}\,.
\label{wdeweff}
\end{equation}
Using the continuity equation $\dot{\rho}_{\rm DE}+3H (\rho_{\rm DE}+P_{\rm DE})=0$, 
it follows that $w_{\rm DE}=w_{\rm eff}-\dot{\Omega}_{\rm DE}/(3H\Omega_{\rm DE})$.

Since we are interested in the case where the late-time cosmic
acceleration is realized by the field kinetic energy, we
set $c_1=0$ in the following 
discussion\footnote{In this case the only solution in the Minkowski background ($H=0$)
corresponds to $\dot{\phi}=0$ for $c_2 \neq 0$.}.
Then the de Sitter solution ($H=H_{\rm dS}={\rm constant}$) can
be present for $\dot{\phi}=\dot{\phi}_{\rm dS}={\rm constant}$.
We normalize the mass $M$ to be $M^3=M_{\rm pl}H_{\rm dS}^2$,
which gives $M \approx 10^{-40}M_{\rm pl}$ for
$H_{\rm dS} \approx 10^{-60}M_{\rm pl}$.
Defining $x_{\rm dS} \equiv \dot{\phi}_{\rm dS}/(H_{\rm dS}M_{\rm pl})$,
Eqs.~(\ref{basic1}) and (\ref{basic2}) lead to the following relations
at the de Sitter solution:
\begin{equation}
c_2 x_{\rm dS}^2=6+9\alpha-12\beta\,,\qquad
c_3 x_{\rm dS}^3=2+9\alpha-9\beta\,,
\label{dscondition}
\end{equation}
where
\begin{equation}
\alpha \equiv c_4 x_{\rm dS}^4\,,\qquad
\beta \equiv c_5 x_{\rm dS}^5\,.
\label{albe}
\end{equation}
It is convenient to use the variables $\alpha$ and $\beta$, because
the coefficients of physical quantities and dynamical equations
can be expressed by $\alpha$ and $\beta$.
The relations (\ref{dscondition}) do not change under 
the rescaling $x_{\rm dS} \to \gamma x_{\rm dS}$
and $c_i \to c_i/\gamma^i$, where $\gamma$ is a real constant.
Then the rescaled choice of $c_i$ can provide the same physics.

In order to study the cosmological dynamics, we introduce
the following dimensionless variables:
\begin{equation}
r_1 \equiv \frac{\dot{\phi}_{\rm dS}H_{\rm dS}}{\dot{\phi}H}\,,
\qquad
r_2 \equiv \frac{1}{r_1} \left( \frac{\dot{\phi}}{\dot{\phi}_{\rm dS}}
\right)^4\,.
\label{r1r2}
\end{equation}
At the de Sitter solution $r_1=1$ and $r_2=1$.
We define the dark energy density parameter
\begin{eqnarray}
\Omega_{\rm DE} &\equiv& \frac{\rho_{\rm DE}}{3M_{\rm pl}^2 H^2}
= -(2+3\alpha-4\beta)r_1^3r_2/2+(2+9\alpha-9\beta)r_1^2r_2
-15\alpha r_1 r_2/2+7\beta r_2\,.
\label{OmeDE}
\end{eqnarray}
Then Eq.~(\ref{basic1}) can be written as 
$\Omega_{\rm DE}+\Omega_m+\Omega_r+\Omega_K=1$, 
where $\Omega_m \equiv \rho_m/(3M_{\rm pl}^2 H^2)$,
$\Omega_r \equiv \rho_r/(3M_{\rm pl}^2 H^2)$, and
$\Omega_K \equiv \rho_K/(3M_{\rm pl}^2 H^2)=-K/(aH)^2$.

The autonomous equations for the variables $r_1$, $r_2$,
$\Omega_r$, and $\Omega_K$ are given by \cite{DT3,Nesseris10}:
\begin{widetext}
\begin{eqnarray}
\label{eq:DRr1}
\hspace{-0.5cm}
r_1' &=& \frac{1}{\Delta} \left(r_1-1\right)
r_1 \left[ r_1 \left(r_1 (-3 \alpha +4 \beta -2)
+6 \alpha -5 \beta \right)-5 \beta \right] \nonumber\\
&&{}\times \left[ 2 \left(\Omega _r-\Omega_K+9\right)
+3 r_2 \left( r_1^3 (-3 \alpha +4\beta -2)+
2 r_1^2 (9 \alpha -9 \beta +2)-15 r_1 \alpha
+14 \beta \right)\right]\,,\\
\label{eq:DRr2}
\hspace{-0.5cm}
r_2' &=& -\frac{1}{\Delta}
[ r_2 (6 r_1^2 (r_2 (45 \alpha ^2-4 (9 \alpha +2) \beta
+36 \beta^2)-(\Omega_r-\Omega_K-7) (9 \alpha -9 \beta +2))
+r_1^3 (-2(\Omega_r-\Omega_K+33) \nonumber \\ & &
\times (3 \alpha -4 \beta +2)
-3 r_2 (-2 (201 \alpha +89) \beta +15\alpha
(9 \alpha +2)+356 \beta ^2))
-3 r_1 \alpha (-28 \Omega _r+28\Omega_K+123 r_2
\beta +36) \nonumber \\
& &+10 \beta (-11 \Omega _r+11\Omega_K
+21 r_2 \beta -3) +3r_1^4 r_2
(9\alpha ^2-30 \alpha (4 \beta +1)+2 (2-9 \beta )^2)
+3r_1^6 r_2 (3 \alpha -4 \beta+2)^2 \nonumber \\
& &+3 r_1^5 r_2
(9 \alpha -9 \beta +2) (3 \alpha -4 \beta +2))]\,, \\
\label{eq:DRr3}
\hspace{-0.5cm}
\Omega_r' &=& \Omega_r \left( -4 -2H'/H \right)\,,\\
\hspace{-0.5cm}
\label{eq:DRr4}
\Omega_K' &=& \Omega_K \left( -2 -2H'/H \right)\,,
\end{eqnarray}
where a prime represents a derivative with respect to
$N=\ln a$, and
\begin{eqnarray}
\hspace{-0.5cm}\Delta &\equiv & 2 r_1^4 r_2 [ 72 \alpha ^2+30 \alpha
(1-5 \beta )+(2-9 \beta )^2 ]+4 r_1^2
[ 9 r_2 (5 \alpha ^2+9 \alpha \beta +(2-9 \beta ) \beta )
+2 (9 \alpha -9 \beta +2) ] \nonumber\\
\hspace{-0.5cm}&&+4 r_1^3 [ -3 r_2 \left(-2 (15 \alpha +1) \beta
+3 \alpha  (9 \alpha +2)+4 \beta ^2\right)-3 \alpha
+4 \beta -2]-24 r_1 \alpha  (16 r_2 \beta +3)
+10\beta (21 r_2 \beta +8)\,.
\end{eqnarray}
\end{widetext}
The Hubble parameter follows from the equation
$H'/H=-5r_1'/(4r_1)-r_2'/(4r_2)$.
The solutions to Eqs.~(\ref{eq:DRr3}) and (\ref{eq:DRr4})
are given by $\Omega_r (N)=\Omega_r^{(0)} e^{-4N}H_0^2/H^2(N)$
and $\Omega_K (N)=\Omega_K^{(0)} e^{-2N}H_0^2/H^2(N)$
respectively, where the subscript ``(0)'' represents the values
today ($N=0$).

{}From Eqs.~(\ref{eq:DRr1}) and (\ref{eq:DRr2}) we find that 
there are three distinct fixed points: (A) $(r_1, r_2)=(0, 0)$, 
(B) $(r_1, r_2)=(1, 0)$, and (C) $(r_1, r_2)=(1, 1)$.
As we have already mentioned, the point (C) corresponds to 
the de Sitter solution. By considering homogeneous perturbations about 
this point, we can show that the de Sitter solution (C)
is always classically stable \cite{DT3}.
The point (B) is a tracker solution found in Ref.~\cite{DT2}, 
along which the field velocity evolves as $\dot{\phi} \propto 1/H$.
During the radiation and matter eras the variable $r_2$ 
is much smaller than 1.
The fixed point (B) is followed by the stable de Sitter point (C)
once $r_2$ grows to the order of 1.
If the initial conditions of both $r_1$ and $r_2$ in the radiation era
are much smaller than 1, then the solutions are close to the 
point (A) at the initial stage. At late times the solutions 
approach the tracker at $r_1=1$.
Depending on the initial values of $r_1$, the epoch at which 
the solutions reach the tracker is different.
In the following we consider the background evolution 
in two regimes: (i) $r_1=1$ and (ii) $r_1 \ll 1$
in more detail.

\subsubsection{Tracker solution $(r_1=1)$}

Along the tracker ($r_1=1$) the dark energy density parameter 
(\ref{OmeDE}) is given by 
\begin{equation}
\Omega_{\rm DE}=r_2\,,
\label{Omedetra}
\end{equation}
which is much smaller than 1 during the radiation and matter eras.
{}From Eqs.~(\ref{eq:DRr3}) and (\ref{eq:DRr4})
we obtain $r_2'/r_2=8+2\Omega_r'/\Omega_r$.
This is integrated to give
\begin{equation}
r_2=d_1 a^8 \Omega_r^2\,,
\label{r2eq2}
\end{equation}
where $d_1$ is a constant.
{}From Eqs.~(\ref{eq:DRr2}) and (\ref{eq:DRr3}) we have
$\Omega_K'/\Omega_K-\Omega_r'/\Omega_r=2$, 
which is integrated to give
\begin{equation}
\frac{\Omega_K}{\Omega_r}=d_2 a^2\,,\quad
{\rm with} \quad
d_2=\frac{\Omega_K^{(0)}}{\Omega_r^{(0)}}\,.
\label{Omera}
\end{equation}

Substituting Eqs.~(\ref{r2eq2}) and (\ref{Omera})
into Eq.~(\ref{eq:DRr3}), we obtain the cosmologically viable solution 
to $\Omega_r$, as 
\begin{equation}
\label{eq:omsol}
\Omega_r=\frac{-1+d_3 a-d_2 a^2+\sqrt{4d_1 a^8+
(-1+d_3 a-d_2 a^2)^2}}{2d_1 a^8}\,,
\end{equation}
where $d_3$ is another constant.
Since the density parameter (\ref{eq:omsol}) evolves as
$\Omega_r \simeq 1+d_3a$ in the early time ($a \ll 1$),
this demands the condition $d_3<0$
(provided $\Omega_{\rm DE}>0$).
Using the density parameters today, the constants $d_1$ and
$d_3$ can be expressed as
$d_1=[1-\Omega_m^{(0)}-\Omega_r^{(0)}-\Omega_K^{(0)}]/
(\Omega_r^{(0)})^2$ and 
$d_3=-\Omega_m^{(0)}/\Omega_r^{(0)}$.
Since $\Omega_r \propto \rho_r/H^2 \propto 1/(a^4 H^2)$,
we have $H^2/H_0^2=(\Omega_r^{(0)}/\Omega_r)(1/a^4)$.
Using Eq.~(\ref{eq:omsol}), the Hubble parameter
can be expressed in terms of the redshift $z=1/a-1$:
\begin{eqnarray}
  \left( \frac{H(z)}{H_0} \right)^2 & = & \frac{1}{2}
  \Omega^{(0)}_K (1 + z)^2 +
  \frac{1}{2} \Omega^{(0)}_m (1 + z)^3 + \frac{1}{2} 
  \Omega^{(0)}_r (1 +
  z)^4 \nonumber \\
  & & +  \sqrt{1 - \Omega^{(0)}_m - \Omega^{(0)}_r - \Omega^{(0)}_K +
  \frac{(1 + z)^4}{4}  \left[ \Omega^{(0)}_K + \Omega^{(0)}_m (1 + z) +
  \Omega^{(0)}_r (1 + z)^2 \right]^2}\,,
  \label{Htracker1}
\end{eqnarray}
which is useful to test the viability of
the tracker solution from observations.

\begin{figure}
\begin{centering}
\includegraphics[width=3.3in,height=3.2in]{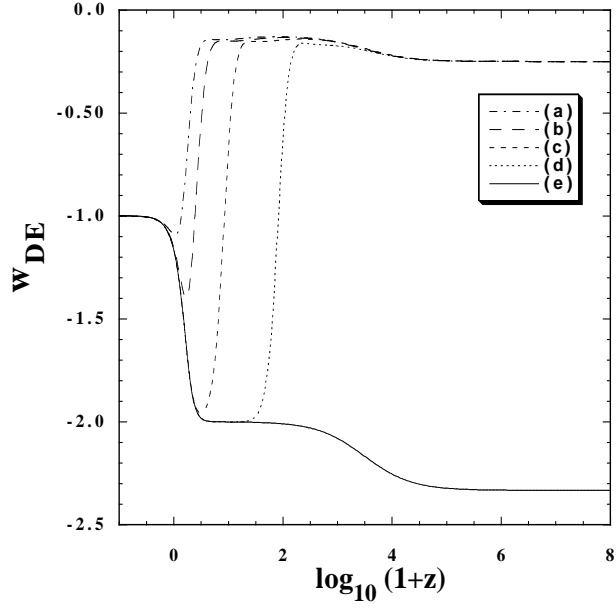} 
\par\end{centering}
\caption{Evolution of $w_{\rm DE}$ versus $z$ for $\alpha=0.3$ and 
$\beta=0.14$ [cases (a)-(d)] in the flat FLRW Universe ($K=0$).
We choose four different initial conditions: 
(a) $r_1=5.000 \times 10^{-11}$, $r_2=8.000 \times 10^{-12}$, and 
$\Omega_r=0.999995$ at $z=5.89 \times 10^8$, 
(b) $r_1=1.500 \times 10^{-10}$, $r_2=2.667 \times 10^{-12}$, and 
$\Omega_r=0.999992$ at $z=3.63 \times 10^8$,
(c) $r_1=5.000 \times 10^{-9}$, $r_2=8.000 \times 10^{-14}$, and 
$\Omega_r=0.99995$ at $z=6.72 \times 10^7$, 
(d) $r_1=5.000 \times 10^{-6}$, $r_2=8.000 \times 10^{-17}$, and 
$\Omega_r=0.9986$ at $z=2.04 \times 10^6$.
The case (e) corresponds to the tracker solution.
{}From Ref.~\cite{DT3}.}
\centering{}\label{galileonwde} 
\end{figure}

On the tracker, the equations of state defined in Eq.~(\ref{wdeweff})
are given by 
\begin{equation}
w_{\rm DE}=-\frac{\Omega_r-\Omega_K+6}{3(r_2+1)}\,,
\qquad
w_{\rm eff}=\frac{\Omega_r-\Omega_K-6r_2}{3(r_2+1)}\,.
\end{equation}
During the cosmological sequence of radiation ($\Omega_r \simeq 1$,
$|\Omega_K| \ll 1$, $r_2 \ll 1$), matter
($\Omega_r \ll 1$, $|\Omega_K| \ll 1$, $r_2 \ll 1$),
and de Sitter ($\Omega_r \ll 1$, $|\Omega_K| \ll 1$, $r_2=1$)
eras, the dark energy equation of state evolves as
$w_{\rm DE}=-7/3 \to -2 \to -1$, whereas the effective equation
of state evolves as $w_{\rm eff}= 1/3 \to 0 \to -1$.
This peculiar evolution of $w_{\rm DE}$ for the tracker corresponds to the 
case (e) in Fig.~\ref{galileonwde}. 
Although the effect of the cosmic curvature does not affect the dynamics 
of $w_{\rm DE}$ significantly, it can change
the diameter distance as well as the luminosity distance 
relative to the flat Universe.

The epoch at which the solutions reach the tracking regime 
$r_1 \simeq 1$ depends on model parameters and initial conditions.
The approach to this regime occurs later for smaller initial values of 
$r_1$, see Fig.~\ref{galileonwde}.
In Ref.~\cite{DT3} it was shown that the tracker 
is stable in the direction of $r_1$ by considering a 
homogeneous perturbation $\delta r_1$.
This means that once the solutions reach the tracker
the variable $r_1$ does not repel away from 1.
If $r_1 \lesssim 2$ initially, numerical simulations show
that the solutions approach the tracker with the 
late-time cosmic acceleration. Meanwhile, 
for the initial conditions with $r_1 \gtrsim 2$, 
the dominant contribution to $\Omega_{\rm DE}$
comes from the Lagrangian ${\cal L}_2$, so that 
the field energy density decreases rapidly as in 
the standard massless scalar field. 

\subsubsection{Solutions in the regime $r_1 \ll 1$}
\label{genesol}

There is another case in which the solutions start to evolve from
the regime $r_1 \ll 1$ (where the term ${\cal L}_5$ gives the 
dominant contribution to the field dynamics). 
In this regime, the variables $r_1$ and $r_2$ satisfy
the following approximate equations
\begin{equation}
r_1' \simeq \frac{9+\Omega_r-\Omega_K+21 \beta r_2}
{8+21 \beta r_2} r_1\,,\qquad
r_2' \simeq\frac{3+11\Omega_r-11\Omega_K-21\beta r_2}
{8+21\beta r_2}r_2\,.
\end{equation}
As long as $\{\beta r_2, |\Omega_K| \} \ll 1$, the evolution of
$r_1$ and $r_2$ during the radiation (matter) era is given by
$r_1 \propto a^{5/4}$ and $r_2 \propto a^{7/4}$
($r_1 \propto a^{9/8}$ and $r_2 \propto a^{3/8}$).
Then the field velocity grows as 
$\dot{\phi} \propto t^{3/8}$ during the radiation era 
and $\dot{\phi} \propto t^{1/4}$ during the matter era. 
The evolution of $\dot{\phi}$ is slower than 
that for the tracker (i.e. $\dot{\phi} \propto t$). 

In the regime $r_1 \ll 1$ the equations of state are
\begin{equation}
w_{\rm DE} \simeq -\frac{1+\Omega_r-\Omega_K}
{8+21\beta r_2}\,, \qquad
w_{\rm eff} \simeq \frac{8\Omega_r
-8\Omega_K-21\beta r_2}{3(8+21\beta r_2)}\,.
\end{equation}
Provided that $\{\beta r_2, |\Omega_K| \} \ll 1$, one has
$w_{\rm DE} \simeq -1/4, w_{\rm eff} \simeq 1/3$
during the radiation era and
$w_{\rm DE} \simeq -1/8, w_{\rm eff} \simeq 0$
during the matter era.
This evolution of $w_{\rm DE}$ is quite different 
from that for the tracker solution.

In Fig.~\ref{galileonwde} the variation of $w_{\rm DE}$ is plotted 
for a number of different initial conditions with $r_1 \ll 1$ [which
correspond to the cases (a)-(d)].
As expected, the solutions start to evolve from 
the value $w_{\rm DE} \simeq -1/4$ in the
radiation era. For larger initial values of $r_1$
they approach the tracker earlier.
This tracking behavior also occurs in the presence of
the cosmic curvature $K$ \cite{Nesseris10}.

\subsubsection{Conditions for the avoidance of ghosts and Laplacian instabilities}
\label{ghostcon}

Let us find a model parameter space in which
the appearance of ghosts and instabilities can be avoided 
in covariant Galileon gravity.
In doing so, we need to study a linear perturbation 
theory on the FLRW background.
For simplicity we focus on the flat Universe with $K=0$. 
Let us consider the perturbed metric 
\begin{equation}
\label{eq:perto1}
{\rm d} s^2=-[1+2\Psi(t,\bm{x})]\,{\rm d} t^2+
\partial_i\chi(t,\bm{x})\,{\rm d} t\,{\rm d} x^i
+a^2(t) [1+2\Phi(t,\bm{x})]\,{\rm d} \bm{x}^2\,,
\end{equation}
where $\Psi$, $\Phi$, and $\chi$ are scalar metric perturbations.
We have chosen the gauge $\delta\phi=0$ without 
a non-diagonal scalar perturbation in the spatial part of 
the metric, i.e. $\partial_{ij}\gamma=0$ \cite{Bardeen}.
Taking into account two perfect fluids with the equations of state 
$w_i=P_i/\rho_i$ ($i=1, 2$), there are three propagating scalar 
degrees of freedom. 
The velocity potentials $v_i$ ($i=1,2$) of perfect fluids
are related with the energy-momentum tensor ${T^0_j}^{(i)}$, as 
${T^0_j}^{(i)}=-(\rho_i+P_i)\partial_j v_i$ ($i=1,2$).

Introducing the vector $\vec{{\cal Q}}=(v_1,v_2,\Phi)$ and 
expanding the action (\ref{action}) up to the second-order,
we obtain the second-order action for scalar perturbations \cite{DT3}
(see also Refs.~\cite{Faddeev,Suyama,DMT}):
\begin{equation}
\label{eq:perto2}
\delta S^{(2)}_S=\frac12\int {\rm d} t\,{\rm d}^3x\, a^3\,
\left[ \dot{\vec{{\cal Q}}}^t\bm{A} \dot{\vec{{\cal Q}}}
-\frac{1}{a^2}\nabla{\vec{{\cal Q}}}^t\,\bm{C}\, \nabla{\vec{{\cal Q}}}
-\dot{\vec{{\cal Q}}}^t\bm{B} {\vec{{\cal Q}}}
-{\vec{{\cal Q}}}^t\bm{D} {\vec{{\cal Q}}}\,\right]\,,
\end{equation}
where the fields $\Psi$ and $\chi$ are integrated out.
$\bm{A}$, $\bm{C}$ and $\bm{D}$ are $3\times3$ symmetric 
matrices and $\bm{B}$ is an antisymmetric matrix 
(for which we do not write explicit forms). 

In order to avoid the appearance of ghosts we require that 
the matrix $\bm{A}$ is  positive definite.
This corresponds to the conditions
$(1+w_1)\rho_1/w_1>0$, $(1+w_2)\rho_2/w_2>0$, and 
\begin{equation}
 \frac{Q_S}{M_{\rm pl}^2} \equiv -\frac{6(1+\mu_1)
(\mu_1+\mu_2+\mu_1 \mu_2
-2\mu_3-\mu_3^2)}{(1+\mu_3)^2} > 0\,,
\label{gali1}
\end{equation}
where 
\begin{eqnarray}
& & \mu_1 \equiv 3\alpha r_1 r_2/2-3\beta r_2\,,\\ 
& & \mu_2 \equiv (3\alpha-4\beta+2)r_1^3 r_2/2
-2(9\alpha-9\beta+2)r_1^2 r_2 
+45\alpha r_1 r_2/2-28\beta r_2\,,\\
& & \mu_3 \equiv -(9\alpha-9\beta+2)r_1^2 r_2/2+
15\alpha r_1 r_2/2 -21 \beta r_2/2\,.
\end{eqnarray}
The propagation speeds $c_S$ of three scalar degrees of freedom 
are known by solving the equation
\begin{equation}
\label{eq:cc}
\det(c_S^2\bm{A}-\bm{C})=0\,.
\end{equation}
For the two perfect fluids we have   
$c_S^2=w_1$ and $c_S^2=w_2$, which are 
are positive for both radiation and non-relativistic matter.
The third stability condition associated with another scalar 
degree of freedom is given by 
\begin{equation}
c_S^2 \equiv \frac{(1+\mu_1)^2[2\mu_3'-(1+\mu_3)(5+3w_{\rm eff})
+4\Omega_r+3\Omega_m]-4\mu_1' (1+\mu_1)(1+\mu_2)
+2(1+\mu_3)^2(1+\mu_4)}
{6(1+\mu_1)(\mu_1+\mu_2+\mu_1 \mu_2
-2\mu_3-\mu_3^2)} > 0 \,,
\label{gali2}
\end{equation}
where 
\begin{equation}
\mu_4 \equiv -\alpha r_1 r_2/2-3\beta r_2
(r_1'/r_1+r_2'/r_2)/4\,.
\end{equation}

Let us consider tensor perturbations with $\delta g_{ij}=a^2 h_{ij}$, 
where $h_{ij}$ is traceless ($h^i{}_i=0$) and 
divergence-free ($h^{ij}{}_{,j}=0$). 
We expand the action (\ref{action}) at second-order in terms of 
the two polarization modes, 
$h_{ij}=h_{\oplus}\,\epsilon^{\oplus}_{ij}+h_{\otimes}\,\epsilon^{\otimes}_{ij}$, 
where $\epsilon^{\oplus}_{ij}$ and $\epsilon^{\otimes}_{ij}$ 
are the polarization tensors. 
For the polarization mode $h_{\oplus}$, 
the second-order action is given by 
\begin{equation}
\label{eq:parto2GW}
\delta S^{(2)}_T=\frac12\int {\rm d} t\,{\rm d}^3x\,a^3\,
Q_{T}\left[\dot h_{\oplus}^2-\frac{c_T^2}{a^2}\,
(\nabla h_{\oplus})^2\right]\,.
\end{equation}
The conditions for the avoidance of ghosts and Laplacian instabilities 
of tensor perturbations correspond, respectively, to 
\begin{eqnarray}
& & \frac{Q_T}{M_{\rm pl}^2} \equiv 
\frac12+\frac34 \alpha r_1 r_2-\frac32 \beta r_2 > 0\,,
\label{gali3}\\
& & c_T^2 \equiv \frac{2r_1 (2-\alpha r_1 r_2)
-3\beta (r_2 r_1'+r_1 r_2')}
{2r_1 (2+3\alpha r_1 r_2-6\beta r_2)} > 0\,.
\label{gali4}
\end{eqnarray}
The same conditions also follow from $h_{\otimes}$.

\begin{figure}
\begin{centering}
\includegraphics[width=3.1in,height=3.2in]{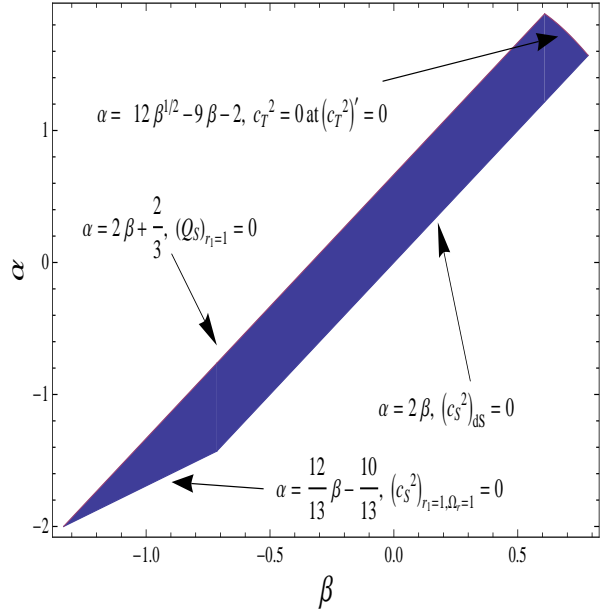} 
\par\end{centering}
\caption{The viable parameter region (blue colored region)
along the tracker solution ($r_1=1$) in the covariant 
Galileon model (where $r_2>0$). 
If the solutions start from the regime $r_1 \ll 1$, 
we also require that $\beta>0$.
{}From Ref.~\cite{DT2}.}
\centering{}\label{galileoncon} 
\end{figure}

In the regime $r_1 \ll 1$ and $r_2 \ll 1$ one has 
$Q_S/M_{\rm pl}^2 \simeq 60\beta r_2$
and $Q_T \simeq 1/2$. For the initial conditions with 
$r_2>0$ we require that $\beta>0$ to avoid the 
scalar ghost. Since $c_S^2 \simeq (1+\Omega_r)/40$
and $c_T^2 \simeq 1+3\beta r_2 (5-3\Omega_r)/8 \simeq 1$, 
there are no Laplacian instabilities of scalar and tensor 
perturbations in this regime.

In the tracking regime characterized by $r_1=1$
(either $r_2 \ll 1$ or $r_2=1$), the conditions 
(\ref{gali1}), (\ref{gali2}), (\ref{gali3}), and (\ref{gali4})
give the bounds on the parameters $\alpha$ and $\beta$.
In the regime $r_2 \ll 1$ these conditions translate to
\begin{eqnarray}
& & Q_S \simeq 3(2-3\alpha+6\beta)r_2>0\,,
\label{Qscon2} \\
& & c_S^2 \simeq \frac{8+10\alpha-9\beta+
\Omega_r (2+3\alpha-3\beta)}
{3(2-3\alpha+6\beta)}>0\,.
\label{cscon2}
\end{eqnarray}
For the branch $r_2>0$ the first condition 
reduces to $2-3\alpha+6\beta>0$. 
Since $c_T^2\simeq 1-r_2(4\alpha+3\beta+3\beta\Omega_r)/2 
\simeq 1$ and $Q_T/M_{\rm pl}^2=1/2+3(\alpha-2\beta)r_2/4>0$, 
the tensor modes do not provide additional constraints.
At the de Sitter point ($r_2=1$) we require that 
\begin{eqnarray}
& & \frac{Q_S}{M_{\rm pl}^2} =
\frac{4-9(\alpha-2\beta)^2}{3(\alpha-2\beta)^2} > 0\,,
\qquad
\frac{Q_T}{M_{\rm pl}^2} = \frac14\,(2+3\alpha-6\beta) > 0\,,
\label{at2}\\
& & c_S^2 = \frac{(\alpha-2\beta)(4+15\alpha^2-48\alpha \beta+36\beta^2)}{2[4-9(\alpha-2\beta)^2]} > 0\,, 
\qquad
c_T^2 =\frac{2-\alpha}{2+3\alpha-6\beta} > 0\,.
\label{at4}
\end{eqnarray}

If $\beta>0$, it can happen that $c_T^2$ has a minimum
during the transition from the regime 
$r_2 \ll 1$ to $r_2 \simeq 1$ \cite{DT2,DT3}. 
This value tends to decrease as $\beta$ approaches 1.
Imposing that $c_T^2>0$ at the minimum, we obtain the bound 
\begin{equation}
\alpha<12\sqrt\beta-9\beta-2\,.
\label{minimum}
\end{equation}
In Fig.~\ref{galileoncon} we plot the parameter space
in the $(\alpha, \beta)$ plane constrained by the conditions 
(\ref{Qscon2})-(\ref{minimum}).
Clearly there are viable model parameters satisfying all the 
theoretical constraints.

\subsubsection{Observational constraints on Galileon cosmology from the 
background cosmic expansion history}

Since the evolution of the dark energy equation of state in covariant 
Galileon gravity is rather peculiar, the observational data related with 
the background cosmic expansion history may place tight 
constraints on the model.
Especially the analytic formula (\ref{Htracker1}) for the tracker
is useful for such a purpose.
In Ref.~\cite{Nesseris10} the authors confronted the Galileon model 
by using the observational data of SN Ia (Constitution \cite{Hicken} 
and Union2 sets \cite{Amanullah}), 
the CMB (WMAP7) shift parameters \cite{WMAP7}, 
and BAO (SDSS7) \cite{BAO2}. 

If either of the SN Ia data (Constitution or Union2) is used in the data analysis, the $\chi^2$ for the tracker is similar to that 
in the $\Lambda$CDM model.
In the presence of the cosmic curvature $K$,
the tracker solution is compatible with the individual
observational bound constrained from either CMB or BAO.
However, the combined data analysis of Constitution+BAO+CMB shows
that the difference of $\chi^2$ between the tracker and the $\Lambda$CDM
is $\delta \chi^2 \sim 22$ (or $\sim 4.3 \sigma$).
This means that the tracker is severely disfavored with 
respect to the $\Lambda$CDM.
A similar conclusion was reached from the combined data analysis 
of Union2+BAO+CMB.
The reason for this incompatibility is that the SN Ia data favor the large values
of $\Omega_m^{(0)}$~($\gtrsim 0.32$), whereas the CMB and BAO
data constrain smaller values of $\Omega_m^{(0)} (\lesssim 0.27)$.

The general solutions starting from the regime $r_1 \ll 1$
finally approach the tracker as $r_1$ grows to 1.
In Ref.~\cite{Nesseris10} the authors carried out the likelihood 
analysis for such general solutions and found that the solutions 
approaching the tracker at late times 
(such as the case (a) in Fig.~\ref{galileonwde}) are 
favored from the combined data analysis.
In the flat FLRW background the best-fit model 
parameters are 
$\alpha=1.411 \pm 0.056$, $\beta=0.422 \pm 0.022$
(Constitution$+$CMB$+$BAO, 68\,\% CL), and 
$\alpha=1.404 \pm 0.057$, $\beta=0.419 \pm 0.023$
(Union2$+$CMB$+$BAO, 68\,\% CL).

For several fixed values of $\Omega_K^{(0)}$ it was shown that 
the late-time tracking solutions can be consistent with the data, 
apart from the models with largely negative $\Omega_K^{(0)}$ 
such as $\Omega_K^{(0)} \lesssim-0.01$.
For example, the general solutions with $\Omega_K^{(0)}=0.01$ 
and the model parameters $(\alpha, \beta)=(1.862, 0.607)$ 
give the similar value
of $\chi^2$ to that in the the $\Lambda$CDM.
In this case the Akaike-Information-Criteiron (AIC) statistics \cite{AIC}
also have the same support for the two models (see also Ref.~\cite{Liddle04})

The Bayesian-Information-Criterion (BIC) statistics \cite{BIC} show that  
the general solutions, with all 4 parameters
$(\alpha, \beta, \Omega_m^{(0)}, \Omega_K^{(0)})$ are varied, 
are not particularly favored over the $\Lambda$CDM model.
This mainly comes from the statistical property that the numbers of
model parameters are larger than those in the flat $\Lambda$CDM.
In fact the late-time tracking solutions with a non-zero cosmic curvature can 
be well consistent with the combined data analysis at the 
background level.

\subsection{Generalized Galileon gravity}

In Sec.~\ref{scasec} we showed that in Brans-Dicke theory with 
the coupling $Q$ of the order of unity the presence 
of the field potential allows a possibility for the consistency with 
local gravity constraints through the chameleon mechanism. 
Another way to recover the General Relativistic behavior in the 
regions of high density is to introduce the Galileon-like field
self-interaction. Silva and Koyama \cite{KazuyaGal} studied 
Brans-Dicke theory in the presence of the term
$\xi (\phi) \square \phi (\partial_\mu \phi \partial^\mu \phi)$
[which is the generalization of the term 
${\cal L}_3=\square \phi (\partial_\mu \phi \partial^\mu \phi)$].
The action of this theory is given by 
\begin{equation}
S=\int {\rm d}^4 x \sqrt{-g} \left[ \frac 12 \phi R
-\frac{\omega_{\rm BD}}{2\phi} (\nabla \phi)^2+
\xi (\phi) \square \phi (\partial_\mu \phi \partial^\mu \phi)
 \right]+\int {\rm d}^4 x\,{\cal L}_M\,.
 \label{BDgene}
\end{equation}
If $\xi (\phi) \propto \phi^{-2}$, there exists a de Sitter solution
that can be responsible for the late-time acceleration.
As in the Galileon model discussed in Sec.~\ref{galisec}
the field is nearly frozen during the radiation and matter eras
through the cosmological Vainshtein mechanism, but it finally 
approaches the de Sitter solution characterized by
$\dot{\phi}=$\,constant. 
Moreover, as in the DGP model, the Vainshtein radius can be 
much larger than the solar system scale, so that the 
General Relativistic behavior can be recovered 
in the local region \cite{KazuyaGal}.

We may consider more general theories described by 
the action \cite{DT1}
\begin{equation}
S=\int {\rm d}^4 x \sqrt{-g} \left[
\frac{1}{2}F(\phi)R+B(\phi)X+
\xi (\phi) \square \phi (\partial^{\mu} \phi 
\partial_{\mu} \phi) \right]
+\int {\rm d}^4 x {\cal L}_M\,,
\label{actionge}
\end{equation}
where $X =-g^{\mu \nu}\partial_{\mu}\phi \partial_{\nu} \phi/2$, 
and $F(\phi)$, $B(\phi)$, $\xi (\phi)$ are functions of $\phi$.
{}From the requirement of having de Sitter solutions responsible 
for dark energy, it is possible to restrict the functional forms 
of $F(\phi)$, $B(\phi)$, and $\xi (\phi)$.
In the presence of non-relativistic matter (energy density $\rho_m$) 
and radiation (energy density $\rho_r$), 
the field equations are given by 
\begin{eqnarray}
& & 1=\frac{B \phi^2}{6F}x^2-\frac{\phi F_{,\phi}}{F}x
+\frac{2 \xi \phi^3}{F} H^2 x^3 \left( 1-
\frac{\phi \xi_{,\phi}}{6 \xi} x \right)
+\frac{\rho_m}{3FH^2}+\frac{\rho_r}{3FH^2}\,,
\label{alg1}
\\
& & -2\frac{\dot{H}}{H^2}=\left( \frac{\phi F_{,\phi}}{F}
-\frac{2\xi \phi^3}{F}H^2 x^2 \right)
\left( \frac{\dot{x}}{H}+x \frac{\dot{H}}{H^2}
+x^2 \right)+\left[ \frac{B \phi^2}{F}x
+\frac{F_{,\phi \phi} \phi^2}{F}x-\frac{\phi F_{,\phi}}
{F}+\frac{6 \xi \phi^3}{F}H^2 x^2 \left(1-
\frac{\phi \xi_{,\phi}}{3\xi}x \right) \right]x \nonumber \\
& &~~~~~~~~~~~~~+
\frac{\rho_m}{F H^2}+\frac{4\rho_r}{3FH^2}\,,
\label{alg2}
\end{eqnarray}
where $x \equiv \dot{\phi}/(H\phi)$.
Let us search for de Sitter solutions at which $H$ and $x$
are constants.
If $F(\phi)$ and $\xi (\phi)$ are power-law 
functions of $\phi$, the quantities such as 
$\phi F_{,\phi}/F$, $F_{,\phi \phi}\phi^2/F$, and 
$\phi \xi_{,\phi}/\xi$ remain constants.
Provided that $B/F \propto \phi^{-2}$ and $\xi/F \propto \phi^{-3}$, 
we can solve Eqs.~(\ref{alg1}) and (\ref{alg2})
for $x$ and $H$ at the de Sitter point.
These conditions are satisfied for the following functions
\begin{equation}
F(\phi)=M_{\rm pl}^2 (\phi/M_{\rm pl})^{3-n}\,,\qquad
B(\phi)=\omega ( \phi/M_{\rm pl})^{1-n}\,,\qquad
\xi (\phi)=(\lambda/\mu^3) (\phi/M_{\rm pl})^{-n}\,,
\label{funchoice}
\end{equation}
where $M_{\rm pl} \simeq 10^{18}$\,GeV is 
the reduced Planck mass, $\mu~(>0)$ 
is a constant having a dimension of mass, and $\omega$
and $\lambda$ are dimensionless constants.
One can show that  the coupling $\lambda$ must be positive
for the consistency of theories \cite{DT1}.
The Brans-Dicke theory described by the action (\ref{BDgene}) 
corresponds to $n=2$ with the Brans-Dicke parameter 
$\omega_{\rm BD}=\omega$.
Since $F(\phi)$ is constant for $n=3$, the theory with 
$n=3$ corresponds to k-essence 
minimally coupled gravity.

{}From Eqs.~(\ref{alg1}) and (\ref{alg2}) we obtain 
the following algebraic equations at the de Sitter fixed point:
\begin{eqnarray}
\omega &=& -\frac{n(n-3)^2x_{\rm dS}^3
+(n-3)(n-12)x_{\rm dS}^2-6(n-5)x_{\rm dS}
+18}{x_{\rm dS}^2 (x_{\rm dS}+3)}\, ,\label{xdS}\\
\lambda &=&\frac{\mu^3}{M_{\rm pl}H_{\rm dS}^2} 
\frac{[(n-3) x_{\rm dS}-2] 
[(n-3) x_{\rm dS}-3] }{2x_{\rm dS}^3 (x_{\rm dS}+3)}\,,
\label{HdS}
\end{eqnarray}
where $x_{\rm dS}$ and $H_{\rm dS}$ are the values 
of $x$ and $H$ at the de Sitter point, respectively.
We fix the mass scale $\mu$ to be 
$\mu=(M_{\rm pl} H_{\rm dS}^2)^{1/3}$, 
where we have used $H_{\rm dS} \simeq 10^{-60}M_{\rm pl}$.
For given $\omega$ and $n$, the quantity $x_{\rm dS}$
is determined by solving Eq.~(\ref{xdS}).
Then the dimensionless constant $\lambda$ is known from 
Eq.~(\ref{HdS}). 

In order to recover the General Relativistic behavior in the early 
cosmological epoch we require that the field initial value $\phi_i$
is close to $M_{\rm pl}$ from Eq.~(\ref{funchoice}).
The quantity $x$ is much smaller than 1
in the early cosmological epoch, so that the field is nearly 
frozen during the radiation and matter eras.
The field starts to evolve at the late cosmological 
epoch in which $x$ grows to the order of unity. 
Introducing the dimensionless quantities 
$y \equiv \lambda x^2 H^2/H_{\rm dS}^2$
and $\Omega_r \equiv \rho_r/(3FH^2)$, one 
can show that the fixed point corresponding to 
the matter era corresponds to $(x,y,\Omega_r)=(0,(3-n)/6,0)$ \cite{DT1}.
Since $y$ is positive definite, it follows that $n \le 3$.

The conditions for the avoidance of ghosts and Laplacian 
instabilities are known by employing the method presented
in Sec.~\ref{ghostcon}.
Provided $F(\phi)>0$, we require that $x>0$ to avoid
ghosts during the cosmological evolution from the
radiation era to the epoch of cosmic acceleration \cite{DT1}. 
The stability of the de Sitter point is automatically ensured for 
$n \le 3$ and $x_{\rm dS}>0$.
{}From Eq.~(\ref{xdS}) the parameter $\omega$ 
is restricted in the range
\begin{equation}
\omega<-n(n-3)^2\,.
\label{wcon}
\end{equation}
The field propagation speed squared during the radiation and matter 
eras is given by $c_S^2 \simeq 6/5$ and $c_S^2 \simeq 2/3$, 
respectively in which
case no instabilities of linear perturbations are present.
Meanwhile, at the de Sitter solution, we have \cite{DT1}
\begin{equation}
c_s^2 =-\frac{(n-2) [(n-3)(n-4)x_{\rm dS}^2-8(n-3)x_{\rm dS}+6]
x_{\rm dS}}{(n-3)(3n^2-10n+12)x_{\rm dS}^3
+12(n-5)x_{\rm dS}^2+18(n-8)x_{\rm dS}-108}\,,
\end{equation}
which is positive for $n \ge 2$.
Hence the parameter $n$ is restricted in the range
\begin{equation}
2 \le n \le 3\,,
\end{equation}
which includes Brans-Dicke theory with the action (\ref{BDgene})
as a specific case ($n=2$).

In Ref.~\cite{Kobayashi1} the authors studied the evolution 
of matter density perturbations and showed that, for the model with 
$n=2$, there is an anti-correlation between the cross-correlation of 
large scale structure and the integrated Sachs-Wolfe effect
in CMB anisotropies. We shall discuss the main reason of this anti-correlation 
in Sec.~\ref{galiper}. This property will be useful to distinguish the
above model from the $\Lambda$CDM in future observations.

\section{Other modified gravity models of dark energy}
\label{othersec}

In this section we briefly discuss other classes of modified gravity models 
of dark energy. These include (i) Gauss-Bonnet gravity with 
a scalar coupling $f(\phi){\cal G}$, 
(ii) $R/2+f({\cal G})$ gravity, and (iii) Lorentz-violating models.

\subsection{Gauss-Bonnet gravity with a scalar coupling}

In addition to the Ricci scalar $R$, we can construct other 
scalar quantities coming from the Ricci tensor $R_{\mu\nu}$
and the Riemann tensor $R_{\mu\nu\alpha\beta}$, i.e.
$P\equiv R_{\mu\nu}R^{\mu\nu}$ and 
$Q\equiv R_{\mu\nu\alpha\beta}R^{\mu\nu\alpha\beta}$ \cite{Kret}.
It is possible to avoid the appearance of spurious spin-2 ghosts
by taking a Gauss-Bonnet (GB) combination \cite{Stelle,Barth,ANunez}, defined by 
\begin{equation}
{\cal G} \equiv R^2-4R_{\mu \nu}R^{\mu \nu}+
R_{\mu \nu \alpha \beta}R^{\mu \nu \alpha \beta}\,.
\label{GBterm}
\end{equation}
A simple model that can be responsible for the cosmic 
acceleration today is \cite{NOS05}
\begin{equation}
S=\int {\rm d}^4 x \sqrt{-g} \left[ \frac12 R-\frac12 (\nabla \phi)^2
-V(\phi)-f(\phi){\cal G} \right]
+\int {\rm d}^4 x\,{\cal L}_M\,,
\label{actiongau1}
\end{equation}
where $V(\phi)$ and $f(\phi)$ are functions of 
a scalar field $\phi$. The coupling of the field with the GB
term appears in low-energy effective string theory \cite{Gasreview}.
For the exponential potential $V(\phi)=V_0e^{-\lambda \phi}$ and 
the coupling $f(\phi)=(f_0/\mu)e^{\mu \phi}$, 
the cosmological dynamics were studied in 
Refs.~\cite{NOS05,Koivisto1,TsujiSami07,Koivisto2,Neupane06,Neupane,Sanyal}.
In this model there exists a de Sitter solution due 
to the presence of the GB term.
In Refs.~\cite{Koivisto1,TsujiSami07} it was also found that  
the late-time de-Sitter solution is preceded by a scaling matter era.

Koivisto and Mota \cite{Koivisto1} placed observational constraints
on the model (\ref{actiongau1}) with the exponential potential 
$V(\phi)=V_0e^{-\lambda \phi}$, by using the Gold data set of SN Ia 
together with the CMB shift parameter data of WMAP.
The parameter $\lambda$ is constrained to be $3.5<\lambda<4.5$
at the 95\% confidence level. In Ref.~\cite{Koivisto2},
they also included the constraints coming from the BAO, LSS, 
big bang nucleosynthesis, and solar system data.
This joint analysis showed that the model is 
strongly disfavored by the data.
In Ref.~\cite{TsujiSami07} it was also found that tensor perturbations
are subject to negative instabilities when the
GB term dominates the dynamics 
(see also Refs.~\cite{Kawai,Ohta} for related works). 

Amendola {\it et al.} \cite{AmenDavis} studied local gravity constraints 
on the above model and showed that the density parameter 
$\Omega_{\rm GB}$ coming from the GB term 
is required to be strongly suppressed for the compatibility  
with solar-system experiments (which is typically of the order 
of $\Omega_{{\rm GB}}<10^{-30}$).
The above discussion indicates that the GB term with 
the scalar-field coupling $f(\phi){\cal G}$ can hardly be the 
source for dark energy.

\subsection{$R/2+f({\cal G})$ gravity}

The general Lagrangian including the scalar quantities  
constructed from the Ricci scalar $R$, the Ricci tensor $R_{\mu\nu}$
and the Riemann tensor $R_{\mu\nu\alpha\beta}$ is given by 
${\cal L}=f(R,P,Q)$. 
The dark energy models based on these theories
have been studied in Refs.~\cite{CaDe,Calcagni05,Sami05,Chibaghost,Mena,Hervik,Cognola,Sokolowski}.
In order to avoid spurious spin-2 ghosts we need to choose
the GB combination (\ref{GBterm}), i.e.
${\cal L}=f(R,Q-4P)$ \cite{Defelice1,Defelice2}.

The cosmological dynamics based on the action,
\begin{equation}
S=\int {\rm d}^4 x \sqrt{-g} \left[ \frac12 R+f({\cal G}) \right]
+\int {\rm d}^4 x\,{\cal L}_M\,,
\label{actiongau2}
\end{equation}
have been studied by a number of authors \cite{NO05,Cognola06,LBM,DeTsu,Cope09,Mohseni,Ishak}. 
In order to ensure the stability of radiation/matter solutions, 
we need to satisfy the condition $f_{,{\cal G}{\cal G}}>0$ for 
all ${\cal G}$. We also require the regularities of the functions
$f$, $f_{,{\cal G}}$, and $f_{,{\cal G}{\cal G}}$ \cite{DeTsu}.  
There exists a de Sitter solution responsible for dark energy, 
whose stability requires the
condition $0<H_{\rm dS}f_{,{\cal G}{\cal G}} (H_{\rm dS})<1/384$.
Moreover $f_{,{\cal G G}}$ must approach $+0$
in the limit $|{\cal G}| \to \infty$.

In Ref.~\cite{DeTsu} the authors proposed
a number of $f({\cal G})$ models satisfying these conditions 
(see also Ref.~\cite{Cope09}). 
One of such models is given by 
\begin{equation}
f({\cal G})=\lambda \frac{{\cal G}}{\sqrt{{\cal G}_*}}
{\rm arctan} \left( \frac{{\cal G}}{{\cal G}_*} \right)
-\alpha \lambda \sqrt{{\cal G}_*}\,,
\label{fGmodel}
\end{equation}
where $\alpha$, $\lambda$ and ${\cal G}_*$ are constants.
The numerical simulation of Ref.~\cite{DeTsu} shows that the 
model (\ref{fGmodel}) is cosmologically viable at least 
at the background level.
Moreover it can be consistent with solar-system constraints
for a wide range of model parameters \cite{DeFelicesolar}.

In order to study the viability of the theories described by the action
(\ref{actiongau2}) further, let us consider the evolution of matter density 
perturbations in the presence of a perfect fluid with the barotropic 
equation of state $w_M=P_M/\rho_M$. 
In Ref.~\cite{DMT10} it was shown that, for small scales 
(i.e. for large momenta $k$), there are two different scalar propagation speeds.
One of them corresponds to the mode for the fluid, i.e. $c_1^2=w_M$, 
whereas another is given by 
\begin{align}
c_2^2= 1+\frac{2\dot H}{H^2}+\frac{1+w_M}{1+4\mu}
\frac{\kappa^2\rho_M}{3H^2}\,,
\label{eq:eqcs2w}
\end{align}
where 
\begin{align}
\mu=H\dot{{\cal G}}f_{,{\cal G}{\cal G}}\,.
\end{align}
The parameter $\mu$ characterizes the deviation from the 
$\Lambda$CDM model (note that the linear term 
$f=c\,{\cal G}$ does not give rise to any contribution to the field equation).
For viable $f({\cal G})$ models we have $|\mu| \ll 1$ at high 
redshifts \cite{DeTsu}.
Since the background evolution during the radiation/matter domination 
is given by $3H^2 \simeq \kappa^2 \rho_M$ 
and $\dot H/H^2 \simeq -(3/2)(1+w_M)$, it follows that 
\begin{equation}
\label{eq:c2neg}
c_2^2 \simeq -1-2w_M\,.
\end{equation}
The Laplacian instability at small scales is absent only for $w_M<-1/2$.
Since $w_M=1/3$ and $w_M=0$ during the radiation and matter eras, respectively, the perturbations with large momentum modes 
are unstable. This leads to violent growth of matter density 
perturbations incompatible with the observations of large-scale 
structure \cite{LBM,DMT10}.

By considering the full perturbation equations, one can show that 
the onset of the negative instability corresponds to \cite{DMT10}
\begin{equation}
\mu \approx (aH/k)^2\,.
\label{mucon}
\end{equation}
Even when $\mu$ is much smaller than 1, we can always find a wave number 
$k~(\gg aH)$ satisfying the condition (\ref{mucon}).
For the scales smaller than that determined by the wave number 
in Eq.~(\ref{mucon}), the linear perturbation theory breaks down.
Hence the background solutions cannot be trusted for those scales, 
which makes the theory unpredictable. 
The Laplacian instability can be avoided only for $\mu=0$, 
which corresponds to the $\Lambda$CDM model.
The above property persists irrespective of
the forms of $f({\cal G})$.

For more general theories described by the Lagrangian density 
$f(R, {\cal G})$ it is possible to avoid such Laplacian 
instabilities \cite{Gerrard}, depending on the models \cite{Suyama10}
(see also Refs.~\cite{Alimo08,Alimo09,Elizalde10,Tanaka}).
It may be of interest to construct some viable dark energy models 
in such theories.

\subsection{Lorentz violating models}

The modified gravity models such as $f(R)$ gravity and Galileon gravity
can give rise to the phantom dark energy equation of state $w_{\rm DE}<-1$
without violating the conditions for the appearance of ghosts and instabilities.
In the models with a broken Lorentz invariance it is also possible to 
realize $w_{\rm DE}<-1$ without pathological behavior 
in the Ultra-Violet (UV) region \cite{gcondensate,Rubakov,Libanov}
(see also Ref.~\cite{Rubakovreview} for a review).
In order to construct Lorentz violating models without pathological behavior
of phantoms, one may start with a field theory consistent at energy 
scales from zero to the UV cutoff scale ${\cal M}$ and then deform the theory in the Infra-Red (IR) in such a way that its behavior at high energies 
remains healthy. Although the weak energy condition is violated in the 
homogenous background, pathological states are present below a
certain low scale $\epsilon$ only.
Provided that $\epsilon$ is close to the Hubble scale,
a theory of this sort should be acceptable.

Let us consider a Lorentz violating model with two-derivative kinetic 
terms with healthy behavior below the scale ${\cal M}$ \cite{Ben} plus
one-derivative term suppressed by the small parameter $\epsilon$ \cite{MLVR}.
The model has a vector field $B_{\mu}$ and a scalar field $\Phi$ 
with a potential $V(B, \Phi)$. 
The Lagrangian is given by \cite{Rubakov,Libanov}
\begin{eqnarray}
\label{lorentzaction}
& & {\cal L}={\cal L}^{(2)}+{\cal L}^{(1)}+{\cal L}^{(0)}+{\cal L}_M\,,\\
& & {\cal L}^{(2)}=-\frac12 \alpha (\Xi) g^{\nu \lambda} D_\mu B_\nu
D^\mu B_\lambda + \frac{1}{2}\beta(\Xi) D_\mu B_\nu D^\mu B_\lambda
\frac{B^\nu B^\lambda}{{\cal M}^2}
+\frac12 \partial_{\mu} \Phi \partial^{\mu} \Phi\,, \\
& & {\cal L}^{(1)}=\epsilon \partial_\mu \Phi B^\mu\,,\qquad 
{\cal L}^{(0)}=-V(B, \Phi)\,,
\end{eqnarray}
where $\Xi=B_\mu B^{\mu}/{\cal M}^2$ 
(${\cal M}$ is the UV cut-off scale).
The dimensionless parameters $\alpha$ and
$\beta$ are the functions of $\Xi$, and $\epsilon$ is a free positive
parameter that characterizes an IR scale. 
The Lorentz invariance is broken for $\Xi \neq 0$.
 
\begin{figure}
\begin{centering}
\includegraphics[width=3.1in,height=3.2in]{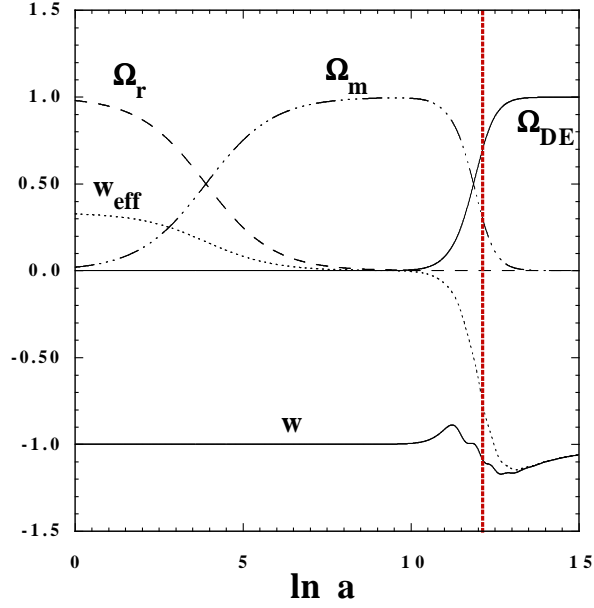} 
\par\end{centering}
\caption{Evolution of $w_{\rm DE}$ (denoted as 
$w$ in the figures) and $w_{\rm eff}=1-2\dot{H}/(3H^2)$ together 
with the density parameters of dark energy ($\Omega_{\rm DE}$), 
non-relativistic matter ($\Omega_m$), and radiation ($\Omega_r$)
for the Lorentz-violating model (\ref{lorentzaction}) 
with $V=m^2 \phi^2/2-M^2 X^2/2$.
The model parameters are chosen to be 
$\alpha=1$, $\gamma=1/2$, $\epsilon/m=3$ and $M/m=1$ with the 
initial conditions $X=\dot{X}=\dot{\phi}=0$,  $\phi = 0.5 \phi_A$, 
and $\Omega_{r}=0.99$, $\Omega_{m} = 0.01$
[where $\phi_A=Mm_{\rm pl}/(\sqrt{4\pi \alpha}m)$].
The present epoch corresponds to $\Omega_m = 0.3$
and $\Omega_{\rm DE}=0.7$, which is denoted by a vertical line.
After the cosmological constant boundary crossing, 
$w_{\rm DE}$ reaches a minimum 
$-1.19$ and then it increases toward 
the de Sitter value $-1$ from the phantom side.
{}From Ref.~\cite{Libanov}.}
\centering{}\label{lorentzfig} 
\end{figure}
 
In spatially homogeneous background we have 
\begin{equation}
B_0=X\,,\qquad
B_i=0\,,\qquad
\Phi=\phi\,,
\end{equation}
where $B_0$ and $B_i$ are time and space components 
of $B_{\mu}$. 
In the flat FLRW background with non-relativistic matter 
and radiation for the matter Lagrangian ${\cal L}_M$, 
we obtain the following equations of motion
\begin{eqnarray}
& & H^2= \frac{8\pi G}{3}
\left[\frac12 \gamma \dot{X}^2-\frac{3\alpha}{2}H^2 X^2 +\frac12
\dot{\phi}^2+W(\phi)+U(X)+ \rho_m+\rho_r \right]\,,\\
& & \gamma \left(\ddot{X}  + 3H\dot{X} \right) + \frac{1}{2}\gamma_{,X}
\dot{X}^2 +\frac{3}{2}\alpha_{,X} H^2 X^2 +3\alpha H^2 X  - \epsilon
\dot{\phi}+V_{,X}=0\,,
\label{eqX}\\
& & \ddot{\phi} + 3H \dot{\phi}+\epsilon (\dot{X} + 3HX) 
+V_{,\phi}=0\,,
\end{eqnarray}
where $\gamma(X)=X^2 \beta(X)/{\cal M}^2-\alpha (X)$.

For the separable potential $V=m^2 \phi^2/2-M^2 X^2/2$ 
with constant $X$ the cosmological dynamics of the above system 
have been studied in detail in Refs.~\cite{Rubakov,Libanov}.
In what follows we assume that both $\alpha$ and $\gamma$
are constants.
Provided $\epsilon/m>\sqrt{2\alpha/3}$ one can show that 
there is a de Sitter solution with $H=M/\sqrt{3\alpha}$, 
at which $\phi$ and $X$ are frozen.
In the early cosmological epoch the field $X$ is close to 0, whereas
the field $\phi$ slowly rolls down its potential (with the energy density 
dominating over that of $X$).

Prior to the epoch of de Sitter cosmic acceleration
there is a transient phantom regime ($w_{\rm DE}<-1$) characterized 
by $H<M/\sqrt{3\alpha}$ in which the field $\phi$ rolls up its 
potential. The Hubble parameter slowly increases toward 
the de Sitter value $H_{\rm dS}=M/\sqrt{3\alpha}$.
In Fig.~\ref{lorentzfig} we plot the evolution of the dark energy 
equation of state $w_{\rm DE}=P_{\rm DE}/\rho_{\rm DE}$
and the effective equation of state 
$w_{\rm eff}=1-2\dot{H}/(3H^2)$ together 
with the density parameters of dark energy,  
non-relativistic matter, and radiation\footnote{Here the definition of 
$\rho_{\rm DE}$ and $P_{\rm DE}$ is 
$\rho_{\rm DE}=\gamma \dot{X}^2/2+\dot{\phi}^2/2+V-3\alpha H^2 X^2/2$ and 
$P_{\rm DE}=\gamma \dot{X}^2/2+\dot{\phi}^2/2-V+\epsilon \dot{\phi}X+\alpha \dot{H} X^2+2\alpha H X \dot{X}+3\alpha H^2 X^2/2$.}.
Clearly the solution undergoes the period with $w_{\rm DE}<-1$
before reaching the de Sitter attractor.

The phantom equation of state can be realized without having ghosts, 
tachyons or superluminal modes in the UV region.
In the IR region characterized by $p \lesssim \epsilon$, 
either tachyons or ghosts appear for the spatial momenta $p$ smaller than
$\sqrt{(\epsilon^2-M^2)/\alpha-m^2}$ \cite{Rubakov,Libanov}.
The presence of tachyons at IR scales leads to the amplification 
of large-scale field
perturbations whose wavelengths are roughly comparable to the present
Hubble radius. There are two tachyonic regions of spatial momenta in this
model: (a) one is sub-horizon and its momenta are characterized by
$M^2/\gamma<p^2<(\epsilon^2-M^2)/\alpha-m^2$;  (b) another is
super-horizon with the momenta 
$0<p^2<m^2M^2/\epsilon^2$. In the region (a)
there is a parameter space in which the perturbations always 
remain smaller than the homogenous fields. 
In the region (b) the growth of the perturbations is suppressed by 
the factor $m^2/\epsilon^2$.

There are other classes of Lorentz violating models such as 
ghost condensate \cite{gcondensate} and 
Horava-Lifshitz gravity \cite{Horava} (see also 
Refs.~\cite{LVmodels1}-\cite{LVmodelsf}).
The application of such scenarios to dark energy has been 
studied by a number of authors, see e.g., \cite{Piazza,Sari,Park}.
It remains to to see whether such Lorentz-violating models can be 
observationally distinguished from other dark energy models.

\section{Observational signatures of modified gravity}
\label{obsersec}

In order to confront modified gravity models with the observations 
of large-scale structure and CMB, 
we discuss the evolution of density perturbations 
in four modified gravity models: (i) $f(R)$ gravity, (ii)  scalar-tensor gravity, 
(iii) DGP braneworld model, and (iv) Galileon gravity.
We also discuss observables to confront with weak lensing observations.

\subsection{$f(R)$ gravity}
\label{fRsecper}

Let us first consider metric $f(R)$ gravity in the presence
of non-relativistic matter. 
We take the following perturbed metric in a longitudinal gauge about 
the flat FLRW background with scalar metric 
perturbations $\Phi$ and $\Psi$ \cite{Bardeen}
\begin{equation}
\rd s^2=- (1+2\Psi)\rd t^2 
+a^2 (1+2\Phi)\delta_{ij}\rd x^i \rd x^j\,.
\label{permet}
\end{equation}
The energy momentum tensors of a non-relativistic perfect 
fluid are decomposed into background and perturbed parts, 
as $T^{0}_{0}=-(\rho_m+\delta \rho_m)$
and $T^0_{\alpha}=-\rho_m v_{m, \alpha}$
($v_m$ is a velocity potential).

The equations for matter perturbations, in the Fourier space, 
are given by \cite{Kodama,Robert,Bassett} 
\begin{eqnarray}
& & \delta \dot{\rho}_m+3H\delta \rho_m
=-\rho_m \left[ 3\dot{\Phi}+(k^2/a)v_m \right] \,,
\label{per4} \\
& & \dot{v}_m+Hv_m=\Psi/a\,,\label{per5}
\end{eqnarray}
where $k$ is a comoving wave number.
We define the gauge-invariant matter density perturbation
$\delta_m$, as 
\begin{eqnarray}
\delta_m \equiv \delta \rho_m/\rho_m
+3Hv\,,\quad {\rm where}
\quad
v \equiv a v_m\,.
\end{eqnarray}
Then Eqs.~(\ref{per4}) and (\ref{per5}) yield
\begin{eqnarray}
& & \dot{\delta}_m=-(k^2/a^2)v
-3(\Phi-Hv)^{\cdot}\,,\\
& & \dot{v}=\Psi\,,
\end{eqnarray}
from which we obtain
\begin{eqnarray}
\ddot{\delta}_m+2H\dot{\delta}_m+(k^2/a^2)\Psi
=3\ddot{B}+6H\dot{B}\,,
\label{delright}
\end{eqnarray}
where $B \equiv -\Phi+Hv$.

In $f(R)$ gravity the quantity $F(R)=\partial f/\partial R$ has
a perturbation $\delta F$. 
In the following we use the unit $\kappa^2=8\pi G=1$, but we restore
gravitational constant $G$ when it is required.
For the action given in Eq.~(\ref{fREin}), we obtain
the linearized perturbation equations in Fourier space \cite{Kofman87,Hwang02,Hwang} 
\begin{eqnarray}
&  & -\frac{k^{2}}{a^{2}}\Phi+3H(H\Psi-\dot{\Phi})
=\frac{1}{2F}\left[3H\dot{\delta F}-\left(3\dot{H}+3H^{2}
-\frac{k^{2}}{a^{2}}\right)\delta F-3H\dot{F}\Psi-3\dot{F}(H\Psi-\dot{\Phi})
-\delta\rho_{m}\right], 
\label{frper1}\\
&  & \ddot{\delta F}+3H\dot{\delta F}+\left(\frac{k^{2}}{a^{2}}
+M^2\right)\delta F
 =\frac{1}{3}\delta\rho_{m}+\dot{F}(3H\Psi+\dot{\Psi}-3\dot{\Phi})
 +(2\ddot{F}+3H\dot{F})
 \Psi\,,\label{frper2}\\
 &  & \Psi+\Phi=-\delta F/F\,.\label{frper3}
\end{eqnarray}
In Eq.~(\ref{frper2}) we have introduced the mass term
\begin{equation}
M^2 \equiv \frac13 \left( \frac{F}{f_{,RR}}-R \right)\,.
\end{equation}
For viable dark energy models the condition $F/f_{,RR} \gg R$ is satisfied
during most of the cosmological epoch, 
so that $M^2 \simeq F/(3f_{,RR})$ \cite{Star07,Tsuji08}.
This is equivalent to the mass squared $M_s^2$ 
introduced in Eq.~(\ref{eq:massfR}), which is 
required to be positive to avoid the tachyonic instability.

For the observations of large-scale structure and weak lensing 
we are interested in the modes deep inside the Hubble radius ($k \gg aH$).
In the following we employ the quasi-static approximation under 
which the dominant terms in Eqs.~(\ref{delright})-(\ref{frper3}) correspond to 
those including $k^2/a^2$, $\delta \rho_m$ (or $\delta_m$) 
and $M^2$ \cite{Star98,Boi00,review5,Tsujimatterper,TsujiUddin,Hwang10}.
We then obtain the following approximate relations from 
Eqs.~(\ref{delright})-(\ref{frper3}) :
\begin{eqnarray}
& & \ddot{\delta}_m+2H\dot{\delta}_m+(k^2/a^2)\Psi=0\,, 
\label{delmfRap}\\
& & \Phi=\frac{1}{2F}\left(\frac{a^{2}}{k^{2}}\delta\rho_{m}-\delta F\right)\,,\quad
\Psi=-\frac{1}{2F}\left(\frac{a^{2}}{k^{2}}\delta\rho_{m}+\delta F\right)\,,
\label{PhifR2}\\
& & \ddot{\delta F}+3H\dot{\delta F}+\left( k^{2}/a^{2}+M^{2}\right)\delta F
=\delta\rho_{m}/3\,.
\label{delFeq}
\end{eqnarray}
The evolution of perturbations is different depending on whether $M^2$ is larger or smaller 
than $k^2/a^2$.
We shall discuss two cases: (A) $M^{2}\gg k^{2}/a^{2}$ and (B) $M^{2}\ll k^{2}/a^{2}$,
separately.  For viable $f(R)$ models the mass squared $M^{2}$
is large in the past and it gradually decreases with time. Hence
the transition from the regime (A) to the regime (B) can occur 
in the past, depending on the wave numbers $k$.

\subsubsection{Evolution of perturbations in the regime: $M^{2}\gg k^{2}/a^{2}$}

The solutions to Eq.~(\ref{delFeq}) are given by the sum
of the oscillating solution $\delta F_{{\rm osc}}$ obtained by setting
$\delta\rho_{m}=0$ and the special solution $\delta F_{{\rm ind}}$
of Eq.~(\ref{delFeq}) induced by the presence of matter perturbations
$\delta\rho_{m}$. The oscillating part $\delta F_{{\rm osc}}$ satisfies
the equation $(a^{3/2}\delta F_{{\rm osc}})^{\cdot\cdot}
+M^{2}(a^{3/2}\delta F_{{\rm osc}})\simeq0$.
Using the WKB approximation, we obtain the following 
solution \cite{Star07}
\begin{eqnarray}
\delta F_{{\rm osc}}\propto a^{-3/2}\, f_{,RR}{}^{1/4}\,
\cos\left(\int\frac{1}{\sqrt{3f_{,RR}}}{\rm d}t\right)\,,
\end{eqnarray}
where we have used the approximation $F \simeq 1$.

For the analytic estimation of the oscillating mode 
we take the model (\ref{fRasy}), which corresponds to 
the asymptotic form of the models (\ref{Amodel}) and (\ref{Bmodel})
in the region $R\gg R_{c}$.
During the matter era in which the background
Ricci scalar evolves as $R^{(0)}=4/(3t^{2})$, the quantity $f_{,RR}$
has a dependence $f_{,RR}\propto R^{-2(n+1)}\propto t^{4(n+1)}$ 
for the model (\ref{fRasy}).
Then the evolution of the perturbation, 
$\delta R_{{\rm osc}}=\delta F_{{\rm osc}}/f_{,RR}$, is given by 
\begin{eqnarray}
\delta R_{{\rm osc}}\simeq c\, t^{-(3n+4)}\,\cos(c_{0}\, t^{-2(n+1)})\,,
\label{delos}
\end{eqnarray}
where $c$ and $c_{0}$ are constants. 
Unless the coefficient $c$ is chosen to be very small, as we go back to the past,
the perturbation $\delta R_{{\rm osc}}$ dominates over the background 
value $R^{(0)} (\propto t^{-2})$.
Since the Ricci scalar can be negative, this leads to 
the violation of the stability conditions ($f_{,RR}>0$ and $F>0$).

The special solution $\delta F_{{\rm ind}}$ to Eq.~(\ref{delFeq})
can be derived by neglecting the first and second terms 
relative to others, giving 
\begin{equation}
\delta F_{{\rm ind}}\simeq \delta\rho_{m}/(3M^{2})\,,\qquad
\delta R_{{\rm ind}}\simeq\delta\rho_{m}\,.\label{Find}
\end{equation}
Under the condition $|\delta F_{{\rm osc}}|\ll|\delta F_{{\rm ind}}|$,
one has $\delta F\simeq\delta\rho_{m}/(3M^{2})$ and hence
$\Psi=-\Phi=-(a^2/k^2)\delta\rho_{m}/(2F)$.
Then the matter perturbation equation (\ref{delmfRap}) reduces to 
\begin{equation}
\ddot{\delta}_{m}+2H \dot{\delta}_{m}
-4\pi G \rho_m \delta_m/F=0\,,
\label{matteq}
\end{equation}
where we have reproduced the gravitational constant $G$.
In Refs.~\cite{delaCruz08,Motohashi} the perturbation equations have been 
derived without neglecting the oscillating mode.

During the matter-dominated epoch ($\Omega_m=\rho_m/(3FH^2) \simeq 1$), Eq.~(\ref{matteq}) has the growing-mode solution 
\begin{equation}
\delta_{m}\propto t^{2/3}\,.\label{delmnor}
\end{equation}
This is the same evolution as that in standard General Relativity.
{}From Eq.~(\ref{Find}) the matter-induced mode evolves as 
$\delta F_{{\rm ind}}\propto t^{4(n+2/3)}$ and 
$\delta R_{{\rm ind}}\propto t^{-4/3}$.
Compared to the oscillating mode (\ref{delos}), $\delta R_{{\rm ind}}$ 
decreases more slowly and hence it dominates over $\delta R_{{\rm osc}}$
at late times. 
The evolution of the perturbation $\delta R=\delta R_{{\rm osc}}+\delta R_{{\rm ind}}$ 
relative to the background value $R^{(0)}$ is given by 
\begin{equation}
\delta R/R^{(0)} \simeq c_{1}\, t^{-(3n+2)}
\cos(c_{0}t^{-p})+c_{2}\, t^{2/3}\,,
\label{Rratio}
\end{equation}
where $c_{1}$ and $c_{2}$ are constants. In order to avoid the
dominance of the oscillating mode at the early cosmological epoch, 
we require that the coefficient $c_{1}$ is suppressed relative
to $c_{2}$ \cite{Star07,Tsuji08,Appleos1}.

This fine-tuning of initial conditions is related with the singularity 
problem raised by Frolov \cite{Frolov}.
The field $\phi=\sqrt{3/2\kappa^2}\ln F$ in the Einstein frame 
has a weak singularity at $\phi=0$ (at which the curvature $R$ 
and the mass $M$ go to infinity with a finite potential $V$).
Unless the oscillating mode of the field perturbation $\delta \phi$
is strongly suppressed relative to the background field $\phi^{(0)}$, 
the system can access the curvature singularity.
This past singularity can be cured by taking into account 
the $R^2$ term \cite{Appleos2}.
Note that the $f(R)$ models proposed in Ref.~\cite{Miranda}
[e.g., $f(R)=R-\alpha R_c \ln (1+R/R_c)$] to cure the singularity 
problem satisfy neither the local gravity 
constraints \cite{Thongkool} nor observational constraints 
of large-scale structure \cite{delaCruz}.
There are some works for 
the construction of unified models of inflation and dark energy based on 
$f(R)$ theories \cite{fRearly4,NOuni1,NOuni2}, 
but the smooth transition between two accelerated epochs
without crossing the point $f_{,RR}=0$ is not easy unless the forms of $f(R)$
are carefully chosen \cite{Appleos2}.

\subsubsection{Evolution of perturbations in the regime: 
$M^{2}\ll k^{2}/a^{2}$}

Since the mass $M$ decreases as $M\propto t^{-2(n+1)}$, the
modes initially in the region $M^{2}\gg k^{2}/a^{2}$ can
enter the regime $M^{2}\ll k^{2}/a^{2}$ during the matter-dominated
epoch. It is sufficient to consider the matter-induced mode because
the oscillating mode is already suppressed during the evolution in
the regime $M^{2}\gg k^{2}/a^{2}$. The matter-induced special solution
of Eq.~(\ref{delFeq}) in the regime $M^{2}\ll k^{2}/a^{2}$ 
is approximately given by 
\begin{eqnarray}
\delta F_{{\rm ind}}\simeq\frac{a^{2}}{3k^{2}}\delta\rho_{m}\,.
\label{FindfR}
\end{eqnarray}
{}From Eq.~(\ref{PhifR2}) the gravitational potentials satisfy
\begin{eqnarray}
\Psi=-\frac{4}{3}\cdot\frac{1}{2F}\frac{a^{2}}{k^{2}}\delta\rho_{m}\,,
\quad\Phi=\frac{2}{3}\cdot\frac{1}{2F}\frac{a^{2}}{k^{2}}\delta\rho_{m}\,.
\label{PsiPhifR}
\end{eqnarray}
Plugging Eq.~(\ref{PsiPhifR}) into Eq.~(\ref{delmfRap}), 
the matter perturbation obeys the following equation
\begin{equation}
\ddot{\delta}_m+2H \dot{\delta}_m
-\frac43 \cdot 4\pi G \rho_m \delta_m/F=0\,.
\end{equation}
During the matter-dominated epoch ($\Omega_{m} \simeq 1$ 
and $a\propto t^{2/3}$), we obtain the following evolution
\begin{equation}
\delta_{m}\propto t^{(\sqrt{33}-1)/6}\,.
\label{delmapsca}
\end{equation}
The growth rate of $\delta_{m}$ gets larger compared to 
that in the regime $M^2 \gg k^2/a^2$.

\subsubsection{Matter power spectra and the ISW effect}

If the transition from the regime $M^{2}\gg k^{2}/a^{2}$ to the regime 
$M^{2}\ll k^{2}/a^{2}$ occurs during the matter era, 
the evolution of matter perturbations changes from 
$\delta_{m}\propto t^{2/3}$ to $\delta_{m}\propto t^{(\sqrt{33}-1)/6}$.
We use the subscript {}``$k$''
for the quantities at which $k$ is equal to $aM$, whereas the subscript
{}``$\Lambda$'' is used at which the accelerated expansion starts
($\ddot{a}=0$). While the redshift $z_{\Lambda}$ is independent
of $k$, $z_{k}$ depend on $k$ and also on the mass $M$.

\begin{figure}
\begin{centering}
\includegraphics[width=3.3in,height=3.0in]{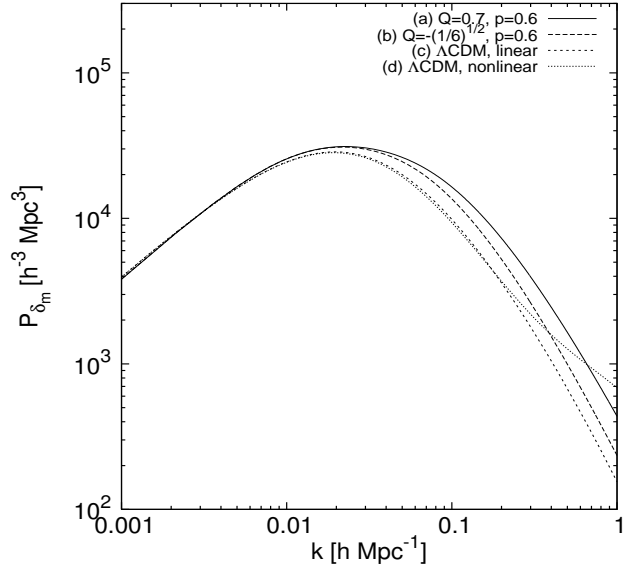} 
\par\end{centering}
\caption{The matter power spectra $P_{\delta_m}(k)$
in Brans-Dicke theory with the potential (\ref{modelscalar}).
This analysis covers the $f(R)$ model (\ref{fRasy})
with the correspondence $Q=-1/\sqrt{6}$ and 
$p=2n/(2n+1)$.
Each case corresponds to (a) $Q=0.7$, $p=0.6$, $C=0.9$, 
(b) $Q=-1/\sqrt{6}$, $p=0.6$, $C=0.9$,
(c) the $\Lambda$CDM model, and
(d) the $\Lambda$CDM model with a nonlinear halo-fitting
($\sigma_8=0.78$ and shape parameter $\Gamma=0.2$).
The model parameters are $\Omega_{m}^{(0)}=0.28$, 
$H_0=3.34 \times 10^{-4}\,h$\,Mpc$^{-1}$, 
$n_s=1$ and $\delta_H^2=3.2 \times 10^{-10}$.
{}From Ref.~\cite{TsujiTate}.}
\centering{}\label{powerspe} 
\end{figure}

For the model (\ref{fRasy}) the variable $m=Rf_{,RR}/f_{,R}$ can
grow fast from the regime $m \ll (aH/k)^2$ (i.e., $M^{2}\gg k^{2}/a^{2})$
to the regime $m \gg (aH/k)^2$ (i.e., $M^{2}\ll k^{2}/a^{2})$.
In fact, $m$ can grow to the order of $0.1$
even if $m$ is much smaller than $10^{-6}$ in the deep matter era.
For the sub-horizon modes relevant to the galaxy power spectrum, the
transition at $M^{2}=k^{2}/a^{2}$ typically occurs at the redshift
$z_{k}$ larger than 1 (provided that $n={\cal O}(1)$). 
For the mode $k/(a_{0}H_{0})=300$ one has 
$z_{k}=4.83$ for $n=1$ and $z_{k}=2.49$
for $n=2$. As $n$ gets larger, the period of non-standard evolution
of $\delta_{m}$ becomes shorter because $z_{k}$ tends to be smaller.
Since the scalaron mass evolves as $M\propto t^{-2(n+1)}$ for the
model (\ref{fRasy}), the time $t_{k}$ has a scale-dependence 
$t_{k}\propto k^{-3/(6n+4)}$.
This means that the smaller-scale modes cross the transition point
earlier. The matter power spectrum $P_{\delta_{m}}=|\delta_{m}|^{2}$
at the time $t_{\Lambda}$ shows a difference compared to the case
of the $\Lambda$CDM model: 
\begin{equation}
\frac{P_{\delta_{m}}(t_{\Lambda})}{P_{\delta_{m}}{}^{\Lambda{\rm CDM}}
(t_{\Lambda})}=\left(\frac{t_{\Lambda}}{t_{k}}\right)^{2\left(
\frac{\sqrt{33}-1}{6}-\frac{2}{3}\right)}\propto 
k^{\frac{\sqrt{33}-5}{6n+4}}\,.
\end{equation}
{}From Fig.~\ref{powerspe} we find that the matter power spectrum in 
the $f(R)$ model (\ref{fRasy}) with $n=3/4$ is in fact larger than that 
in the $\Lambda$CDM model on smaller scales. 

The galaxy matter power spectrum is modified by this effect.
Meanwhile the CMB spectrum is hardly affected except for very large 
scales (for the multipoles $\ell={\cal O}(1)$) at which the 
Integrated Sachs-Wolfe (ISW) effect becomes important.
Hence there is a difference for the spectral indices of 
two power spectra, i.e.
\begin{equation}
\Delta n(t_{\Lambda})=\frac{\sqrt{33}-5}{6n+4}\,.\label{delnesti}
\end{equation}
For larger $n$ the redshift $z_{k}$ can be as close as $z_{\Lambda}$,
which means that the estimation (\ref{delnesti}) is not necessarily
valid in such cases. Moreover the estimation (\ref{delnesti}) does
not take into account the evolution of $\delta_{m}$ after $z=z_{\Lambda}$
to the present epoch ($z=0$). It was found in Ref.~\cite{Tsuji08} that 
the estimation (\ref{delnesti}) agrees well with the numerically obtained 
$\Delta n(t_{\Lambda})$ for $n \le 2$.

In order to discuss the growth rate of matter perturbations, it is customary
to introduce the growth index \cite{Peebles} 
\begin{equation}
f_\delta \equiv \frac{\dot{\delta}_m}{H \delta_m}=
(\tilde{\Omega}_m)^{\gamma}\,,
\label{fdelta}
\end{equation}
where $\tilde{\Omega}_m=\kappa^2 \rho_m/(3H^2)$.
In the $\Lambda$CDM model $\gamma$ is nearly constant 
for the redshifts $0<z<1$, i.e. $\gamma \simeq 0.55$ \cite{Wang98,Linder05}.
In $f(R)$ gravity, if the perturbations are in the GR regime
($M^2 \gg k^2/a^2$) today, $\gamma$ is close to 0.55.
On the other hand, if the transition to the scalar-tensor regime occurs
at the redshift $z_k$ larger than 1, the growth index tends to be 
smaller than 0.55 \cite{Moraes08,Moraes09}.
Since $0<\tilde{\Omega}_m<1$, the smaller $\gamma$ implies 
a larger growth rate.

For the wave numbers relevant to the linear regime of the matter power spectra
($0.01\,h\,{\rm Mpc}^{-1} \lesssim k \lesssim 0.2\,h\,{\rm Mpc}^{-1}$, 
where $h \approx 0.7$) the viable $f(R)$ models (\ref{Amodel})-(\ref{Cmodel}) can give rise to 
the growth index $\gamma_0=0.40$ today.
Depending on the wave numbers $k$, $\gamma_0$ can be 
dispersed in the regime  $0.40 \lesssim \gamma_0 \lesssim 0.55$, 
or $\gamma_0$ can show the convergence in the regime 
$0.40 \lesssim \gamma_0 \lesssim 0.43$ \cite{Moraes08,Moraes09}. 
Moreover, when $\gamma_0$ is small, the growth index exhibits
a large variation even for low redshifts ($0<z<1$).
Although the present observational constraints 
on $\gamma$ are quite weak, the unusual evolution of $\gamma$
can be useful to distinguish the $f(R)$ models from 
the $\Lambda$CDM in future observations.
 
After the system enters the epoch of cosmic acceleration, the 
wave number $k$ can again become smaller than $aM$. 
Hence the $k$-dependence
is not necessarily negligible even for $z<z_{\Lambda}$. 
However numerical simulations show that $\Delta n(t_{0})$ is not much different from $\Delta n(t_{\Lambda})$
derived by Eq.~(\ref{delnesti}) \cite{Tsuji08}. 
Thus the analytic estimation (\ref{delnesti})
is certainly reliable to place constraints on model parameters except
for $n \gg 1$. Observationally we do not find any strong difference
for the slopes of the spectra of LSS and CMB. If we take the mild
bound $\Delta n (t_{\Lambda})<0.05$, we obtain the constraint 
$n \ge 2$ \cite{Star07}. In this case the local gravity constraint 
(\ref{bound3}) is also satisfied.

For the wave numbers $k \gtrsim 0.2\,h\,{\rm Mpc}^{-1}$
we need to take into account the non-linear effect of density perturbations.
In Refs.~\cite{Oyaizu1,Oyaizu2,Oyaizu3,ZhaoLi} the authors carried out 
$N$-body simulations for the $f(R)$ model (\ref{Bmodel}) 
(see also Refs.~\cite{Stabenau,Laszlo}).
Hu and Sawicki (HS) \cite{Huparametrization} proposed a fitting formula 
to describe the non-linear power spectrum 
based on the halo model.
Koyama {\it et al.}\ \cite{Koyama09} studied the validity of 
the HS fitting formula by comparing it with 
the results of $N$-body simulations and showed that 
the HS fitting formula can reproduce the power spectrum 
in $N$-body simulations for the scales $k<0.5h$\,Mpc$^{-1}$. 
In the quasi non-linear regime a normalized skewness, 
$S_3=\langle \delta_m^3 \rangle/\langle \delta_m^2 \rangle^2$, 
has been evaluated in $f(R)$ gravity and in Brans-Dicke 
theories \cite{Tateskew}.
The skewness in $f(R)$ dark energy models differs only by 
a few percent relative to the value $S_3=34/7$ in the 
$\Lambda$CDM model.

The modified growth of matter perturbations also affects the evolution
of the gravitational potentials $\Psi$ and $\Phi$. 
The effective potential $\psi \equiv \Phi-\Psi$ is important to discuss
the ISW effect on the CMB as well as the weak lensing 
observations \cite{Kunz}.
For the modes deep inside the Hubble radius, 
Eq.~(\ref{PhifR2}) gives
\begin{equation}
\psi=3\Omega_{m}\delta_{m} \left( \frac{aH}{k} \right)^2\,.
\label{psieq}
\end{equation}
Note that for the large-scale modes relevant to the ISW effect in CMB
we need to solve the full perturbation equations without using 
the quasi-static approximation on sub-horizon scales.
For viable $f(R)$ models, however, numerical simulations show that 
the result (\ref{psieq}) can be trustable even for the wave numbers
close to the Hubble radius today (i.e. $k \gtrsim a_0H_0$) \cite{TsujiUddin}.
In the $\Lambda$CDM model the potential $\psi$ remains constant
during the standard matter era, but it decays after the system enters
the accelerated epoch, producing the ISW
contribution for low multipoles on the CMB power spectrum. In $f(R)$
gravity the additional growth of matter perturbations in the region
$z<z_{k}$ changes the evolution of $\psi$.

{}From the observation of the CMB angular power spectrum, 
the constraint on the deviation parameter 
$B \equiv (\dot{R}H/R\dot{H})m$ from the $\Lambda$CDM is weak. 
The value of $B$ today is constrained to be $B_0<4.3$ \cite{Peiris}.
There is another observational constraint coming from the angular correlation
between the CMB temperature field and the galaxy number density field 
induced by the ISW effect.
The avoidance of a large anti-correlation between the observational data of CMB and LSS places an upper bound of $B_0 \lesssim 1$.
Since this roughy corresponds to $m (z=0) \lesssim 1$, 
the CMB observations do not provide tight constraints 
on $f(R)$ models relative to the matter power spectrum of LSS.
In weak lensing observations, the
modified evolution of the lensing potential $\psi$ directly leads
to the change even for the small-scale shear power 
spectrum \cite{Kunz,TsujiTate,Schmidt}.
Hence this can be a powerful tool to constrain $f(R)$ gravity models
from future observations.

\subsection{Brans-Dicke theory}
\label{scatensec}

Let us proceed to discuss the evolution of matter perturbations 
in Bran-Dicke theory described by the action (\ref{action2}) with 
the potential $U(\phi)$ and the coupling $F(\phi)=e^{-2Q\phi}$.
We are mainly interested in large coupling models with $|Q|$
of the order of unity \cite{KW1,KW2,TUMTY}, 
as this gives rise to significant deviation from the $\Lambda$CDM. 
We define the field mass squared to be 
$M^2 \equiv U_{,\phi\phi}$.
If the scalar field is light such that the condition $M<H_{0}$
is always satisfied irrespective of high or low density regions, the
coupling $Q$ is constrained to be $|Q| \lesssim 10^{-3}$ from local
gravity constraints. Meanwhile, if the mass $M$ in the region of high density
is much larger than that on cosmological scales, the model can 
satisfy local gravity constraints under the chameleon mechanism even
if $|Q|$ is of the order of unity. Cosmologically the mass $M$ can
decrease from the past to the present, which can allow the transition
from the {}``GR regime'' to the {}``scalar-tensor regime'' as
it happens in $f(R)$ gravity. An example of the field potential showing
this behavior is given by Eq.~(\ref{modelscalar}).

As in $f(R)$ gravity, the matter perturbation $\delta_{m}$ obeys 
Eq.~(\ref{delright}). The difference appears in the expression
of the gravitational potential $\Psi$. In Fourier space the scalar
metric perturbations obey the following equations \cite{Hwang02,Hwang,TUMTY}
\begin{eqnarray}
 &  & -\frac{k^{2}}{a^{2}}\Phi+3H(H\Psi-\dot{\Phi})=
 -\frac{1}{2F}\Biggl[\omega\dot{\phi}\delta \dot{\phi}+
 \frac{1}{2}(\omega_{,\phi}\dot{\phi}^{2}-F_{,\phi}R+2V_{,\phi})
 \delta \phi \nonumber \\
& & +\left(3\dot{H}+3H^{2}-\frac{k^{2}}{a^{2}}\right)\delta F
-3H\delta\dot{F}+(3H\dot{F}-\omega\dot{\phi}^{2})\Psi
+3\dot{F}(H\Psi-\dot{\Phi})+\delta\rho_{m}\Biggr],
\label{persca1}\\
 &  & \delta \ddot{\phi}+\left(3H+\frac{\omega_{,\phi}}{\omega}\dot{\phi}\right)
 \delta \dot{\phi}+\left[\frac{k^{2}}{a^{2}}+\left(\frac{\omega_{,\phi}}
 {\omega}\right)_{,\phi}\frac{\dot{\phi}^{2}}{2}+
 \left(\frac{2U_{,\phi}-F_{,\phi}R}{2\omega}\right)_{,\phi}\right]\delta \phi\nonumber \\
 &  & =\dot{\phi}\dot{\Psi}+\left(2\ddot{\phi}+3H\dot{\phi}
 +\frac{\omega_{,\phi}}{\omega}\dot{\phi}^{2}\right)\Psi+3\dot{\phi}
 (H\Psi-\dot{\Phi})+\frac{1}{2\omega}F_{,\phi}\delta R\,,\label{persca2}\\
 &  & \Psi+\Phi=-\frac{\delta F}{F}=-\frac{F_{,\phi}}{F}\delta \phi\,,
\label{persca3}
\end{eqnarray}
where $\delta\phi$ is the perturbation of the 
field $\phi$, $\omega=(1-6Q^{2})F$, and
\begin{equation}
\delta R=2\left[3(\dot{\Phi}-H\Psi)^{\cdot}-12H(H\Psi-\dot{\Phi})
+\left(k^2/a^2-3\dot{H}\right)\Psi+2(k^2/a^2)\Phi\right]\,.
\label{persca5}
\end{equation}
Provided that the field is sufficiently heavy to satisfy the conditions $M^{2}\gg R$, one can employ the approximation
$[(2U_{,\phi}-F_{,\phi}R)/2\omega]_{,\phi}\simeq M^{2}/\omega$ in
Eq.~(\ref{persca2}). The solution to Eq.~(\ref{persca2}) consists
of the sum of the matter-induced mode $\delta \phi_{{\rm ind}}$ sourced
by the matter perturbation and the oscillating mode $\delta \phi_{{\rm osc}}$, i.e. $\delta \phi=\delta \phi_{{\rm ind}}+ \delta \phi_{{\rm osc}}$
(as in the case of $f(R)$ gravity).

In order to know the evolution of the matter-induced mode 
we employ the quasi-static approximation on sub-horizon scales.
Under this approximation,
we have $\delta R_{{\rm ind}}\simeq2(k^{2}/a^{2})[\Phi-(F_{,\phi}/F)
\delta \phi_{{\rm ind}}]$
from Eqs.~(\ref{persca1}) and (\ref{persca5}), where the subscript
{}``ind'' represents the matter-induced mode. 
Then from Eq.~(\ref{persca2}) we find 
\begin{equation}
\delta \phi_{{\rm ind}}\simeq-\frac{2QF}{(k^{2}/a^{2})(1-2Q^{2})F+M^{2}}
\frac{k^{2}}{a^{2}}\Phi\,.\label{delphi}
\end{equation}
Using Eqs.~(\ref{persca1}) and (\ref{persca3}) we obtain 
\begin{equation}
\frac{k^{2}}{a^{2}}\Psi\simeq-\frac{\delta\rho_{m}}{2F}
\frac{(k^{2}/a^{2})(1+2Q^{2})F+M^{2}}{(k^{2}/a^{2})F+M^{2}}\,,
\qquad
\frac{k^{2}}{a^{2}}\Phi\simeq\frac{\delta\rho_{m}}{2F}
\frac{(k^{2}/a^{2})(1-2Q^{2})F+M^{2}}{(k^{2}/a^{2})F+M^{2}}\,.
\label{PsiPhisca}
\end{equation}
In the massive limit $M^{2}/F\gg k^{2}/a^{2}$, we recover the standard
result of General Relativity. In the massless limit $M^{2}/F\ll k^{2}/a^{2}$,
it follows that $(k^{2}/a^{2})\Psi\simeq-(\delta\rho_{m}/2F)(1+2Q^{2})$ and
$(k^{2}/a^{2})\Phi\simeq(\delta\rho_{m}/2F)(1-2Q^{2})$.

Plugging Eq.~(\ref{PsiPhisca}) into Eq.~(\ref{delmfRap}),
we obtain the equation for matter perturbations \cite{TUMTY} 
\begin{equation}
\ddot{\delta}_m+2H\dot{\delta}_m-4\pi G_{{\rm eff}}\rho_{m}\delta_{m}=0\,,
\label{mattereqsca2a}
\end{equation}
where the effective gravitational coupling is
\begin{equation}
G_{{\rm eff}}=\frac{G}{F}\frac{(k^{2}/a^{2})(1+2Q^{2})F
+M^{2}}{(k^{2}/a^{2})F+M^{2}}\,.\label{eq:jfg}
\end{equation}
We have recovered the bare gravitational constant $G$.
In the massless limit ($M^{2}\ll k^{2}/a^{2}$) this reduces to 
\begin{equation}
G_{{\rm eff}}\simeq\frac{G}{F}(1+2Q^{2})=\frac{G}{F}
\frac{4+2\omega_{{\rm BD}}}{3+2\omega_{{\rm BD}}}\,,
\label{Geffmassless}
\end{equation}
where in the last line we have used the relation (\ref{BD}) between
the coupling $Q$ and the Brans-Dicke parameter $\omega_{{\rm BD}}$.
In $f(R)$ gravity we have $\omega_{{\rm BD}}=0$ and hence 
$G_{\rm eff}=4G/(3F)$.

Let us consider the evolution of the oscillating mode of perturbations.
Using Eqs.~(\ref{persca1}) and (\ref{persca2}) for sub-horizon modes
($k^{2}/a^{2}\gg H^{2}$), the gravitational potentials 
can be expressed by $\delta \phi_{\rm osc}$ 
(note that $\delta \rho_m=0$ for the oscillating mode). 
{}From Eq.~(\ref{persca5}) the perturbation of $R$ corresponding to the oscillating mode is given by
\begin{eqnarray}
\delta R_{{\rm osc}}\simeq6Q\left[ \delta \ddot{\phi}_{{\rm osc}}
+3H \delta \dot{\phi}_{{\rm osc}}+(k^2/a^2)
\delta \phi_{{\rm osc}} \right]\,.
\label{delRscal}
\end{eqnarray}
Substituting Eq.~(\ref{delRscal}) into Eq.~(\ref{persca2}), it follows that 
\begin{equation}
\delta \ddot{\phi} _{{\rm osc}}+3H\delta \dot{\phi}_{{\rm osc}}+
\left( k^2/a^2+M^2/F \right) \delta \phi_{{\rm osc}}
\simeq 0\,,
\label{ddotphi}
\end{equation}
 which is valid in the regime $M^{2}\gg R$.

When $|Q|={\cal O}(1)$ the field potential $U(\phi)$ is required to be
heavy in the region of high density for the consistency with local
gravity constraints. We take the potential (\ref{modelscalar})
as an example of a viable model. During the matter era the field $\phi$
settles down at the instantaneous minima characterized by the condition (\ref{phim}).
Then we have that $\phi\propto\rho_{m}^{\frac{1}{p-1}}$
and $M^{2}\propto \rho_{m}^{\frac{2-p}{1-p}}$ during the matter-dominated
epoch. The field $\phi$ can initially be heavy to satisfy the condition
$M^{2}/F\gg k^{2}/a^{2}$ for the modes relevant to the galaxy power
spectrum. Depending upon the model parameters and the mode $k$, the
mass squared $M^{2}$ can be smaller than $k^{2}/a^{2}$ during the
matter era \cite{TUMTY}.

In the regime $M^{2}/F\gg k^{2}/a^{2}$ the matter perturbation equation
(\ref{mattereqsca2a}) reduces to the standard one in Einstein gravity,
which gives the evolution $\delta_{m}\propto t^{2/3}$. For the model
(\ref{modelscalar}) the matter-induced mode of the field perturbation
evolves as $\delta \phi_{{\rm ind}}\propto\delta\rho_{m}/M^{2}\propto t^{\frac{2(4-p)}{3(1-p)}}$.
Meanwhile, the WKB solution to Eq.~(\ref{ddotphi}) is given by 
$\delta \phi_{{\rm osc}}\propto t^{\frac{p}{2(1-p)}}\cos\left(ct^{-\frac{1}{1-p}}\right)$,
where $c$ is a constant. Since the background field $\phi$ during
the matter era evolves as $\phi\propto t^{\frac{2}{1-p}}$, we find
\begin{equation}
\delta \phi/\phi=(\delta \phi_{{\rm ind}}+\delta \phi_{{\rm osc}})
/\phi\simeq c_{1}t^{2/3}
+c_{2}t^{-\frac{4-p}{2(1-p)}}\cos\left(ct^{-\frac{1}{1-p}}\right)\,.
\label{delphiso}
\end{equation}
As long as the oscillating mode is initially suppressed relative
to the matter-induced mode, the latter remains 
the dominant contribution in the subsequent cosmic 
expansion history.

In the regime $M^{2}/F\ll k^{2}/a^{2}$ the effective gravitational
coupling is given by Eq.~(\ref{Geffmassless}).
Solving Eq.~(\ref{mattereqsca2a})
in this case, we obtain the solution
\begin{equation}
\delta_{m}\propto t^{(\sqrt{25+48Q^{2}}-1)/6}\,.\label{persol2}
\end{equation}
Setting $Q=-1/\sqrt{6}$, this recovers the solution 
$\delta_{m}\propto t^{(\sqrt{33}-1)/6}$
in metric $f(R)$ gravity.

The potential (\ref{modelscalar}) has a heavy mass $M$ 
much larger than $H$ in the deep matter-dominated epoch, 
but it gradually decreases with time.
Depending on the modes $k$, the system crosses the point 
$M^{2}/F=k^{2}/a^{2}$ at $t=t_{k}$ during the matter-dominated epoch. Since the field mass
evolves as $M\propto t^{-\frac{2-p}{1-p}}$ during the matter
era, the time $t_{k}$ has a scale-dependence 
given by $t_{k}\propto k^{-\frac{3(1-p)}{4-p}}$.
Since $0<p<1$ the smaller scale modes (i.e. larger $k$) 
cross the transition point earlier.
During the matter era the mass squared is approximately given by 
\begin{equation}
M^{2}\simeq\frac{1-p}{(2^{p}\, p\, C)^{1/(1-p)}}Q^{2}
\left(\frac{\rho_{m}}{V_{0}}\right)^{\frac{2-p}{1-p}}
U_{0}\,.\label{Mphi2}
\end{equation}
Using the relation 
$\rho_{m}=3F_{0}\Omega_{m}^{(0)}H_{0}^{2}(1+z)^{3}$,
the critical redshift $z_{k}$ at time $t_{k}$ can be
estimated as 
\begin{eqnarray}
z_{k}\simeq\left[\left(\frac{k}{a_{0}H_{0}}\frac{1}{Q}\right)^{2(1-p)}
\frac{2^{p}pC}{(1-p)^{1-p}}\frac{1}{(3F_{0}\Omega_{m}^{(0)})^{2-p}}
\frac{U_{0}}{H_{0}^{2}}\right]^{\frac{1}{4-p}}-1\,,\label{zk}
\end{eqnarray}
where $a_{0}$ is the scale factor today. The critical redshift
increases for larger $k/(a_{0}H_{0})$ and for smaller $p$.
If $k/(a_{0}H_{0})=600$ and $p=0.7$, Eq.~(\ref{zk}) gives $z_{k}=3.9$.

Defining the growth rate of matter perturbations as in Eq.~(\ref{fdelta}), 
it follows that the asymptotic values of $f_\delta$ 
in the regions $t\ll t_{k}$ and $t\gg t_{k}$ are given by 
$f_\delta=1$ and $f_\delta=(\sqrt{25+48Q^{2}}-1)/4$,
respectively. Numerical simulations show that the growth rate reaches a maximum value around the end of the matter era and then it starts to decrease during the epoch of cosmic acceleration \cite{TUMTY}.
The observational constraint on $f_\delta$ reported by 
McDonald \textit{et al.} \cite{Mc} is $f_\delta=1.46 \pm 0.49$ around 
the redshift $z=3$, whereas the data reported
by Viel and Haehnelt \cite{Viel} in the redshift range $2<z<4$ show
that even the value $f_\delta=2$ can be allowed in some of the observations.
If we use the criterion $f_\delta<2$ with the analytic estimation 
$f_\delta=(\sqrt{25+48Q^{2}}-1)/4$,
we obtain the bound $Q<1.08$. 
Note that the growth index today can be smaller than 0.4 for $|Q|$ larger than 
0.4, so this will be also useful to place tight bounds on $Q$ 
in future observations.

The relative difference of the matter power 
spectrum $P_{\delta_{m}}$ at time $t=t_{\Lambda}$ (at which $\ddot{a}=0$) 
from that in the $\Lambda$CDM is given by  
\begin{equation}
\frac{P_{\delta_{m}}(t_{\Lambda})}{P_{\delta_{m}}^{\Lambda{\rm CDM}}
(t_{\Lambda})}=\left(\frac{t_{\Lambda}}{t_{k}}\right)^{2\left(
\frac{\sqrt{25+48Q^{2}}-1}{6}-\frac{2}{3}\right)}\propto 
k^{\frac{(1-p)(\sqrt{25+48Q^{2}}-5)}{4-p}}\,.
\label{relative}
\end{equation}
In Fig.~\ref{powerspe} we plot the matter power spectrum for $Q=0.7$
and $p=0.6$, which deviates from that in the $\Lambda$CDM
model on small scales. The estimation (\ref{relative}) shows fairly good 
agreement with numerical results \cite{TsujiTate}.

{}From Eqs.~(\ref{PsiPhisca}) we find that 
the effective gravitational potential $\psi=\Phi-\Psi$ satisfies 
the same equation as (\ref{psieq}).
Since the ISW effect induced by the modified evolution of $\psi$
is limited on large-scale CMB perturbations irrelevant to 
the galaxy power spectrum, 
there is a difference between the spectral indices of the matter power
spectrum and of the CMB spectrum on the scales, $k>0.01h$\,Mpc$^{-1}$:
\begin{equation}
\Delta n(t_{\Lambda})=\frac{(1-p)(\sqrt{25+48Q^{2}}-5)}{4-p}\,.
\label{deln}
\end{equation}
This reproduces the result (\ref{delnesti}) in $f(R)$ gravity by
setting $Q=-1/\sqrt{6}$ and $p=2n+1$. 
If we use the criterion $\Delta n(t_{\Lambda})<0.05$,
as in the case of the $f(R)$ gravity, we obtain the bounds $p>0.957$
for $Q=1$ and $p>0.855$ for $Q=0.5$. As long as $p$ is close to 1, 
it is possible to satisfy both cosmological and local gravity constraints
for $|Q| \lesssim 1$.

\subsection{DGP model}
\label{DGPpersec}

In this section we study the evolution of linear matter perturbations in the
DGP braneworld model. 
The discussion below is valid for the wavelengths larger than the 
Vainshtein radius $r_*$. For the radius $r$ smaller than $r_*$
the non-linear effect coming from the brane-bending mode
becomes crucially important. 
The perturbed metric in the 5-dimensional longitudinal gauge
with four scalar metric perturbations $\Psi,\Phi,B,E$ is 
given by \cite{DGPghost2,KoyamaSilva} 
\begin{equation}
\rd s^{2}=-(1+2\Psi)n(t,y)^{2}\rd t^{2}+(1+2\Phi)A(t,y)^{2}
\delta_{ij}\rd x^{i}\rd x^{j}+2r_{c}B_{,i}\rd x^{i}\rd y+(1+2E)\rd y^{2}\,,
\end{equation}
where the brane is located at $y=0$ in the 5-th dimension characterized
by the coordinate $y$ (we are considering a flat FLRW spacetime on
the brane). Note that $B$ can be identified as a brane bending mode
describing a perturbation of the brane location and that $r_{c}$
is the crossover scale defined in Eq.~(\ref{crossover}). 
The background solution describing the self-accelerating 
Universe is \cite{Deffayet1} 
\begin{equation}
n(t,y)=1+H(1+\dot{H}/H^{2})y\,,\qquad A(t,y)=a(t)(1+Hy)\,.
\end{equation}
The Hubble parameter $H=\dot{a}/a$ satisfies Eq.~(\ref{DGPK0eq})
with $\epsilon=+1$.

In what follows we neglect the terms suppressed by the factor
$aH/k\ll1$ because we are considering sub-horizon perturbations.
We also ignore the terms such as $(A'/A)\Phi'$, where a prime represents
a derivative with respect to $y$. This comes from the fact that $\Phi'$
is of the order of $(k/a)\Phi$, as we will show later. The time-derivative
terms can be also dropped under the quasi-static approximation on 
sub-horizon scales. Then the perturbed 5-dimensional Einstein tensors 
$\delta\tilde{G}_{B}^{A}$
obey the following equations locally in the bulk \cite{KoyamaSilva}:
\begin{eqnarray}
\delta\tilde{G}_{0}^{0} & = & 3\Phi''+\frac{2}{A^{2}}\nabla^{2}\Phi+
\frac{\nabla^{2}}{A^{2}}(E-r_{c}B')-2\frac{r_{c}}{A^{2}}
\left(\frac{A'}{A}\right)\nabla^{2}B=0\,,\label{eq:gtt}\\
\delta\tilde{G}_{j}^{i} & = & -\frac{1}{A^{2}}(\nabla^{i}\nabla_{j}
-\delta_{j}^{i}\nabla^{2})(\Phi+\Psi+E-r_{c}B')+\delta_{j}^{i}(\Psi''+2\Phi'')
+\frac{r_{c}}{A^{2}}(\nabla^{i}\nabla_{j}-\delta_{j}^{i}\nabla^{2})
\left(\frac{A'}{A}+\frac{n'}{n}\right)B=0\,,\label{eq:gij}\\
\delta\tilde{G}_{i}^{5} & = & -(\Psi'+2\Phi')_{,i}=0\,,\label{eq:gyi}\\
\delta\tilde{G}_{5}^{5} & = & \frac{1}{A^{2}}\nabla^{2}(\Psi+2\Phi)
-\frac{r_{c}}{A^{2}}\left(2\frac{A'}{A}+\frac{n'}{n}\right)
\nabla^{2}B=0\,.\label{eq:gyy}
\end{eqnarray}

Taking the divergence of the traceless part of Eq.~(\ref{eq:gij}), we obtain 
\begin{equation}
\frac{\nabla^{2}}{A^{2}}(\Phi+\Psi+E-r_{c}B')
-\frac{r_{c}}{A^{2}}\left(\frac{A'}{A}+\frac{n'}{n}\right)\nabla^{2}B=0\,.
\label{eq:traceless}
\end{equation}
The consistency between Eqs.~(\ref{eq:gyi}) and (\ref{eq:gyy})
requires that 
\begin{equation}
B'=0,\qquad\Psi'+2\Phi'=0\,.\label{varphip}
\end{equation}
{}From Eqs.~(\ref{eq:gyy}) and (\ref{eq:traceless}) we find
\begin{equation}
\frac{\nabla^{2}}{A^{2}}(E-r_{c}B')=-\frac{1}{2}\frac{\nabla^{2}}
{A^{2}}\Psi+\frac{r_{c}}{2A^{2}}\frac{n'}{n}\nabla^{2}B\,.\label{Geqe}
\end{equation}
Substituting Eqs~(\ref{eq:gyy}) and (\ref{Geqe}) into Eq.~(\ref{eq:gtt})
together with the use of Eq.~(\ref{varphip}), we obtain 
\begin{equation}
\Psi''+\frac{\nabla^{2}}{A^{2}}\Psi
-\frac{n'}{n}\frac{r_{c}}{A^{2}}\nabla^{2}B=0\,.
\label{wave}
\end{equation}
Under the sub-horizon approximation ($k/aH \gg 1$) the solution 
to Eq.~(\ref{wave}), upon the Fourier transformation, is given by 
\begin{equation}
\Psi-\frac{n'}{n}r_{c}B=\left[c_{1}(1+Hy)^{-k/aH}
+c_{2}(1+Hy)^{k/aH}\right]\,,
\label{solPsi}
\end{equation}
where $c_{1}$ and $c_{2}$ are integration constants. 
In order to avoid the divergence of the perturbation 
in the limit $y \to \infty$
we choose $c_{2}=0$.

The junction condition at the brane can be expressed in terms of an
extrinsic curvature $K_{\mu\nu}$ and an energy-momentum tensor 
on the brane \cite{Nico09}: 
\begin{equation}
K_{\mu\nu}-Kg_{\mu\nu}=
-\kappa_{(5)}^{2}\,T_{\mu\nu}/2+r_{c}G_{\mu\nu}\,,
\label{junction}
\end{equation}
where $K\equiv K_{\mu}^{\mu}$. 
The extrinsic curvature is defined as 
$K_{\mu\nu}=h_{\mu}^{\lambda}\nabla_{\lambda}\, n_{\nu}$,
where $n_{\nu}$ is the unit vector normal to the 
brane and $h_{\mu\nu}=g_{\mu\nu}-n_{\mu}n_{\nu}$
is the induced metric on the brane. 
The $(0,0)$ and spatial components of the junction 
condition (\ref{junction}) give 
\begin{eqnarray}
&  & \frac{2}{a^{2}}\nabla^{2}\Phi=
-\kappa_{(4)}^{2}\delta\rho_{m}+\frac{1}{a^{2}}
\nabla^{2}B-\frac{3}{r_{c}}\Phi'\,,\label{jun:tt}\\
&  & \Phi+\Psi=B,\label{jun:traceless}\\
&  & \Psi'+2\Phi'=0\,,\label{jun:trace}
\end{eqnarray}
where $\delta\rho_{m}$ is the matter perturbation on the brane.
Equation (\ref{jun:trace}) is consistent with the latter of
Eq.~(\ref{varphip}).

From Eq.~(\ref{solPsi}) it follows that $\Phi'\sim(k/a)\Phi$ in Fourier space. 
For the perturbations whose wavelengths are much smaller than the cross-over
scale $r_{c}$, i.e., $r_{c}\, k/a\gg1$, the term $(3/r_{c})\Phi'$
in Eq.~(\ref{jun:tt}) is much smaller than $(k^{2}/a^{2})\Phi$.
In Fourier space Eq.~(\ref{jun:tt}) is approximately given by
\begin{equation}
\frac{2k^{2}}{a^{2}}\Phi=\kappa_{(4)}^{2}\delta\rho_{m}
+\frac{k^{2}}{a^{2}}B\,.\label{einstein:tt}
\end{equation}
Using the projection of Eq.~(\ref{eq:gyy}) as well as Eqs.~(\ref{jun:traceless})
and (\ref{einstein:tt}), we find that metric perturbations $\Psi$
and $\Phi$ obey the following equations 
\begin{equation}
\frac{k^{2}}{a^{2}}\Psi=-\frac{\kappa_{(4)}^{2}}{2}\left(1+
\frac{1}{3\beta}\right)\delta\rho_{m}\,,\qquad\frac{k^{2}}{a^{2}}\Phi
=\frac{\kappa_{(4)}^{2}}{2}\left(1-\frac{1}{3\beta}\right)\delta\rho_{m}\,,
\label{DGPpsiphi}
\end{equation}
where 
\begin{equation}
\beta(t)\equiv1-\frac{2r_{c}}{3}\left(2\frac{A'}{A}+\frac{n'}{n}\right)
=1-2Hr_{c}\left(1+\frac{\dot{H}}{3H^{2}}\right)\,.\label{betadef}
\end{equation}
The matter perturbation $\delta_{m}$ satisfies the same form of
equation as given in (\ref{delmfRap}) for the modes deep inside the
horizon \cite{Lue}. Substituting the former of Eq.~(\ref{DGPpsiphi})
into Eq.~(\ref{delmfRap}), we find that the matter perturbation
obeys the following equation \cite{Lue,DGPghost2}
\begin{equation}
\ddot{\delta}_m+2H\dot{\delta}_m-4\pi G_{{\rm eff}}\rho_{m}\delta_{m}=0\,,
\label{DGPeqdel}
\end{equation}
where 
\begin{equation}
G_{{\rm eff}}=\left( 1+\frac{1}{3\beta} \right)G\,.
\label{DGPeff}
\end{equation}
{}From Eq.~(\ref{DGPpsiphi}) the effective gravitational potential $\psi=\Phi-\Psi$
obeys the same equation as (\ref{psieq}).

In the deep matter era one has $Hr_{c}\gg1$ and hence $\beta\simeq-Hr_{c}$,
so that $\beta$ is largely negative ($|\beta|\gg1$). In this regime
the evolution of the matter perturbation is similar to that in General
Relativity ($\delta_{m}\propto t^{2/3}$). The solutions finally approach
the de Sitter attractor characterized by $H_{{\rm dS}}=1/r_{c}$. 
At the e Sitter solution one has $\beta\simeq1-2Hr_{c}\simeq-1$.
Since $1+1/(3\beta) \simeq 2/3 $, the growth rate in this regime is
smaller than that in GR. 
The growth index is approximately given by 
$\gamma\approx0.68$ \cite{Linder05}, 
which is different from the value $\gamma \simeq 0.55$
in the $\Lambda$CDM model. 
If the future imaging survey of galaxies can constrain 
$\gamma$ within 20 \%, it may be possible to distinguish
the DGP model from the $\Lambda$CDM model \cite{Yamamoto}.

Comparing Eq.~(\ref{DGPeff}) with the effective gravitational coupling 
(\ref{Geffmassless}) in Brans-Dicke theory with a massless limit
(or the absence of the field potential), we find that the Brans-Dicke
parameter $\omega_{{\rm BD}}$ has the following relation with $\beta$:
\begin{eqnarray}
\omega_{{\rm BD}}=\frac{3}{2}(\beta-1)\,.
\end{eqnarray}
Since $\beta<0$ for the self-accelerating DGP solution, this implies
that $\omega_{{\rm BD}}<-3/2$. 
Since in this case the kinetic energy of a scalar field degree of freedom is
negative in the Einstein frame, the DGP model contains a ghost mode.
The solution in another branch of the DGP model is not plagued by  
this problem, because the minus sign of Eq.~(\ref{betadef})
is replaced by the plus sign. The self accelerating solution in the original 
DGP model can be realized at the expense of an appearance
of the ghost state.

\subsection{Galileon gravity}
\label{galiper}

\subsubsection{Covariant Galileon gravity}

In covariant Galileon gravity described by the action (\ref{action}) the evolution of matter density perturbations was studied in Ref.~\cite{Kase}.
In spite of the complexities of full perturbation equations, they are simplified 
under the quasi-static approximation on sub-horizon scales.
Under this approximation the matter perturbation obeys the same 
equation as (\ref{DGPeqdel}) with a different effective 
gravitational coupling $G_{\rm eff}$.
In Ref.~\cite{Kase} it was found that $G_{\rm eff}$ is independent of 
the wave number $k$, as in the DGP model.
$G_{\rm eff}$ is close to the gravitational constant $G$ in the asymptotic past, but it starts to deviate at the late cosmological epoch.

We define the effective gravitational potential $\psi=\Phi-\Psi$ as well as the 
anisotropic parameter $\eta=-\Phi/\Psi$.
Under the quasi-static approximation on sub-horizon scales 
we obtain \cite{Kase}
\begin{equation}
\psi \simeq 3 \frac{G_{\rm eff}}{G} \frac{1+\eta}{2}
\Omega_m \delta_m \left( \frac{aH}{k} \right)^2\,,
\label{psiga}
\end{equation}
where $\Omega_m$ and $\delta_m$ and the density parameter and 
the perturbation of non-relativistic matter, respectively.
In $f(R)$ gravity, Brans-Dicke theory, and the DGP model
the effective gravitational potential obeys Eq.~(\ref{psieq})
for the modes deep inside the Hubble radius.
In Galileon gravity the combination $(G_{\rm eff}/G)(1+\eta)/2$
is different from 1. This means that the effective gravitational potential 
may acquire some additional growth compared to other models. 
In Galileon cosmology there are three different regimes characterized by 
(i) $r_1 \ll 1$, $r_2 \ll 1$, 
(ii) $r_1=1$, $r_2 \ll 1$, and 
(iii) $r_1=1$, $r_2=1$, where $r_1$ and $r_2$ are defined in Eq.~(\ref{r1r2}).
In these regimes we can estimate 
$G_{\rm eff}$ and $\eta$ as follows \cite{Kase}.
We stress that these analytic results are valid for the modes deep inside 
the Hubble radius.
\begin{itemize}
\item (i) $r_1 \ll 1$, $r_2 \ll 1$

Expanding $G_{\rm eff}$ and $\eta$ about $r_1=0, r_2=0$, it follows that 
\begin{equation}
\label{Geffap}
\frac{G_{\rm eff}}{G} = 1+\left( \frac{255}{8}\beta+\frac{211}{16}
\alpha r_1 \right) r_2+{\cal O} (r_2^2)\,,\qquad
\eta = 1+\left( \frac{129}{8}\beta+\frac{589}{16}
\alpha r_1 \right) r_2+{\cal O} (r_2^2)\,,
\end{equation}
where $\alpha$ and $\beta$ are defined by Eq.~(\ref{albe}).
Since $\beta>0$ to avoid ghosts (for the branch $r_2>0$), 
we have 
$G_{\rm eff}>G$ and $\eta>1$ in this regime.
This means that the growth rates of $\delta_m$ and $\psi$ 
are larger than those in the $\Lambda$CDM model.

\item (ii) $r_1=1$, $r_2 \ll 1$

Expansion of $G_{\rm eff}$ and $\eta$ about $r_2=0$ gives 
\begin{eqnarray}
\frac{G_{\rm eff}}{G} &=& 1+\frac{291\alpha^2+702\beta^2
-933\alpha \beta+20\alpha-84\beta+4}
{2(10\alpha-9\beta+8)}r_2+{\cal O}(r_2^2) \,,
\label{Gefftracker}\\
\eta &=& 1-\frac{3(126\alpha^2+306 \beta^2-405\alpha \beta
+4\alpha-30\beta)}{2(10\alpha-9\beta+8)}r_2+{\cal O}(r_2^2) \,.
\label{etatracker}
\end{eqnarray}
The evolution of $G_{\rm eff}$ and $\eta$ depends on both 
$\alpha$ and $\beta$.
If $\alpha=1.4$ and $\beta=0.4$, for example, 
we have $G_{\rm eff}/G \simeq 1+4.31r_2$ and 
$\eta \simeq 1-5.11r_2$, respectively.
In this case $G_{\rm eff}>G$, but $\eta$ is smaller 
than 1.

\item (iii) $r_1=1$, $r_2=1$

At the dS point we have
\begin{equation}
\label{Geffde}
\frac{G_{\rm eff}}{G} = \frac{1}{3(\alpha-2\beta)} \,,\qquad
\eta = 1 \,,
\end{equation}
which means that there is no anisotropic stress.
\end{itemize}

\begin{figure}
\begin{centering}
\includegraphics[width=3.3in,height=3.2in]{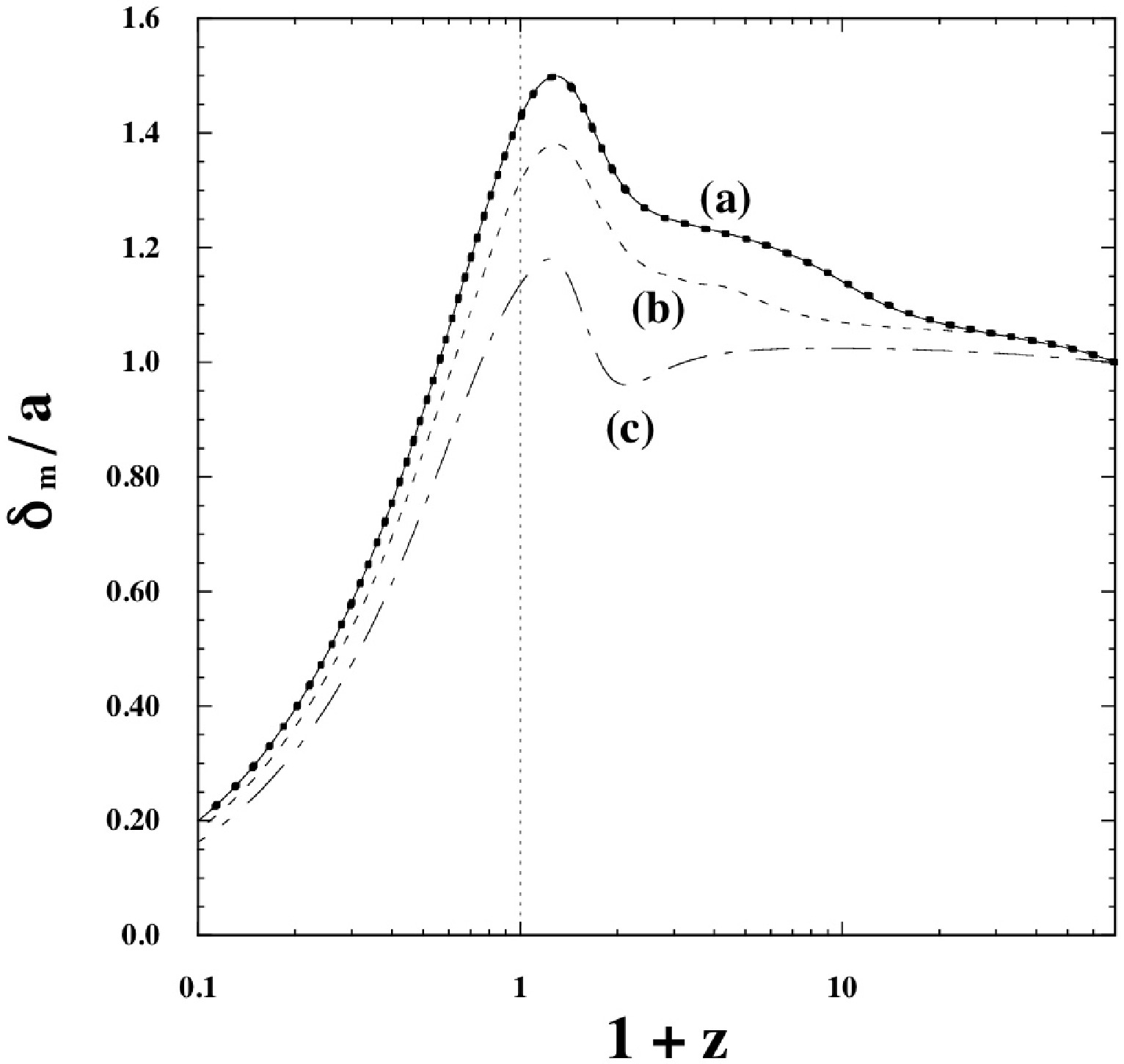} 
\includegraphics[width=3.3in,height=3.2in]{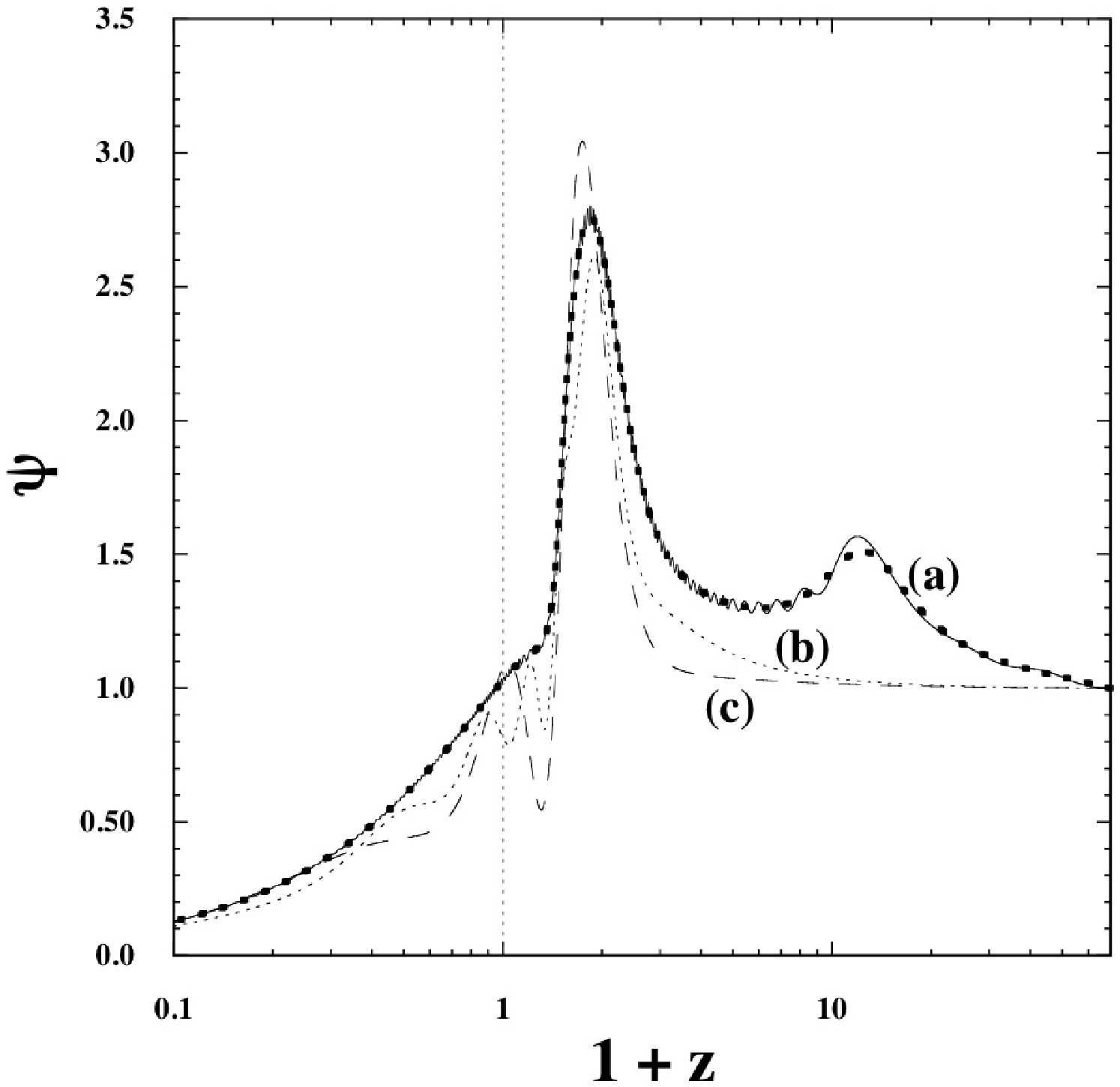} 
\par\end{centering}
\caption{Evolution of the perturbations
for $\alpha=1.37$ and $\beta=0.44$ with the 
background initial conditions $r_1=0.03$ and $r_2=0.003$.
(Left) $\delta_m/a$ versus $z$ for the wave numbers
(a) $k=300a_0H_0$, (b) $k=30a_0H_0$, and (c) $k=5a_0H_0$.
(Right) $\psi$ versus $z$ for the wave numbers
(a) $k=300a_0H_0$, (b) $k=10a_0H_0$, and (c) $k=5a_0H_0$.
Note that $\delta_m/a$ and $\psi$ are divided by 
their initial amplitudes $\delta_m (t_i)/a(t_i)$ and 
$\psi(t_i)$, respectively, so that their initial 
values are normalized to be 1.
The bold dotted lines show the results obtained under the 
quasi-static approximation on sub-horizon scales.
{}From Ref.~\cite{Kase}.}
\centering{}\label{galiperfig} 
\end{figure}

Recall that the late-time tracking solutions are favored from 
observational constraints at the background level.
In this case the solutions start from the regime (i) and finally
approach the de Sitter fixed point with a short period 
of the regime (ii).
In Fig.~\ref{galiperfig} we plot $\delta_m/a$ and 
$\psi$ versus the redshift $z$ for 
$\alpha=1.37$ and $\beta=0.44$ with the 
background initial conditions $r_1=0.03$ and $r_2=0.003$.
In this case the solutions approach the tracker at late times.
The initial conditions of perturbations are chosen to recover 
the GR behavior in the asymptotic past.
Note that these results are obtained by numerical integration 
of the full perturbation equations without quasi-static approximations.
For the mode $k=300a_0H_0$ the numerical result shows 
excellent agreement with that obtained
under the quasi-static approximation on sub-horizon scales.
The difference starts to appear for the modes 
$k/(a_0H_0)<{\cal O}(10)$.
{}From the left panel of Fig.~\ref{galiperfig} we find that, 
on larger scales, the growth of $\delta_m$
tends to be less significant.
For the modes $k \gg a_0H_0$ the matter perturbation 
evolves faster than $a$ during the matter era.

{}From the right panel of Fig.~\ref{galiperfig} we find 
that, unlike the $\Lambda$CDM model, $\psi$ 
changes in time even during the matter era
for the modes $k \gg aH$. 
Before reaching the tracker we have 
$G_{\rm eff}/G \simeq 1+255\beta r_2/8>1$ and 
$\eta \simeq 1+129\beta r_2/8>1$ from Eq.~(\ref{Geffap}).
Then the growth rates of $\psi$ and 
$\delta_m$ get larger than those in GR.
In particular the term $(G_{\rm eff}/G)(1+\eta)/2$
in Eq.~(\ref{psiga}) is larger than 1, which leads to 
the additional growth of $\psi$ to that 
coming from $\delta_m$.
In Galileon gravity the unusual behavior of the 
anisotropic parameter $\eta$ leads to the non-trivial 
evolution of perturbations.
For the model parameters $\alpha=1.37$ and $\beta=0.44$, 
Eq.~(\ref{Geffde}) gives $G_{\rm eff} \simeq 0.68 G$
at the de Sitter fixed point.
Since in this case $G_{\rm eff}$ is smaller than $G$, $\psi$ 
begins to decrease at some point after the matter era.

For the large-scale modes relevant to the ISW effect in 
CMB anisotropies ($k/(a_0H_0) \lesssim 10$),
$\psi$ is nearly 
constant in the early matter-dominated epoch.
However, as we see in Fig.~\ref{galiperfig}, 
$\psi$ exhibits temporal growth 
during the transition from the matter era 
to the epoch of cosmic acceleration.
The characteristic variation of $\psi$ in 
the Galileon model may leave interesting observational signatures
on the large-scale CMB anisotropies. 

For the model parameters constrained by SN Ia (Union2)$+$CMB$+$BAO
data sets, i.e. $\alpha=1.404 \pm 0.057$ and 
$\beta=0.419 \pm 0.023$ \cite{Nesseris10}, 
the effective gravitational coupling at the de Sitter solution 
is restricted in the range $0.5G<G_{\rm eff}<0.72G$. 
If the model parameters are close to the upper limit 
$\alpha=2\beta+2/3$ of the allowed parameter space
at the background level (i.e. $G_{\rm eff}$ is close to 
$0.5G$ at the de Sitter point), the parameter $\eta$ tends to show 
a divergence during the transition from the matter 
era to the epoch of cosmic acceleration. 
If $G_{\rm eff}$ is larger than $0.66G$, we find that 
such divergent behavior is typically avoided.
For the viable model parameters the evolution of $\delta_m$ and 
$\psi$ exhibits more or less the similar property 
to that shown in Fig.~\ref{galiperfig}.

\subsubsection{Modified Galileon gravity}

Finally we study modified Galileon theories in which the term
$\square \phi (\partial_\mu \phi \partial^\mu \phi)$ is 
generalized to $\xi (\phi) \square \phi 
(\partial_\mu \phi \partial^\mu \phi)$.
Let us consider general theories described by the action 
\begin{equation}
S=\int {\rm d}^4 x \sqrt{-g} 
\left[ \frac 12 f (R, \phi, X)
+\xi (\phi) \square \phi (\partial_\mu \phi \partial^\mu \phi)
\right]+\int {\rm d}^4 x\,{\cal L}_M\,,
\label{moaction}
\end{equation}
where 
\begin{equation}
f(R,\phi, X)=f_1(R,\phi)+f_2(\phi,X)\,.
\end{equation}
We introduce two mass scales associates with the field $\phi$ 
and the scalar gravitational degree of freedom, respectively, as
\begin{equation}
M_{\phi}^2 \equiv -f_{,\phi \phi}/2\,,\qquad
M_{R}^2 \equiv F/(3F_{,R})\,,
\end{equation}
where $F \equiv \partial f/\partial R$.

The full perturbation equations for the perturbed metric (\ref{permet}) are 
given in Ref.~\cite{DMT}. Under the quasi-static approximation 
on sub-horizon scales the matter perturbation satisfies 
Eq.~(\ref{DGPeqdel}) with the effective gravitational 
coupling \cite{DMT}
\begin{equation}
\label{Geff}
G_{\rm eff} = \frac{1}{8\pi F}\frac{1+4s_1}{1+3s_1}
\left\{ 1 + \frac{[F_{,\phi}+2(1+4s_1)\xi\dot{\phi}^2]^2}
{(1+4s_1)\mu F} \right\}\,,
\end{equation}
where 
\begin{eqnarray}
& & \mu \equiv (1+3s_1)(f_{,X}+2s_2)+3F_{,\phi}^2/F
+2\xi [4(1+3s_1)(\ddot{\phi}+2H\dot{\phi})
-2F_{,\phi}\dot{\phi}^2/F-2\xi \dot{\phi}^4 (1+4s_1)/F]\,,\\
& & s_1 \equiv k^2/(3a^2 M_R^2)\,,\qquad s_2 \equiv 
a^2M_{\phi}^2/k^2\,.
\end{eqnarray}
The anisotropic parameter $\eta=-\Phi/\Psi$ is given by 
\begin{equation}
\label{zeta}
\eta=\frac{(1+2s_1)(f_{,X}+2s_2)+2F_{,\phi}^2/F
+4 \xi [2(1+2s_1)(\ddot{\phi}+2H \dot{\phi})
-F_{,\phi}\dot{\phi}^2/F]}
{(1+4s_1)(f_{,X}+2s_2)+4F_{,\phi}^2/F
+8 \xi (1+4s_1)(\ddot{\phi}+2H\dot{\phi})}\,.
\end{equation}
The effective gravitational potential $\psi=\Phi-\Psi$ obeys 
the equation of the form (\ref{psiga}), whose explicit form is
\begin{equation}
\label{psigega} 
\psi=3\Omega_m \delta_m \left( \frac{aH}{k} \right)^2
\frac{(1+3s_1)[f_{,X}+2s_2+8\xi (\ddot{\phi}+2H\dot{\phi})]
+3F_{,\phi}^2/F-2\xi F_{,\phi} \dot{\phi}^2/F}
{(1+3s_1)[f_{,X}+2s_2+8\xi (\ddot{\phi}+2H\dot{\phi})]+3F_{,\phi}^2/F
-4\xi [F_{,\phi} \dot{\phi}^2/F+\xi \dot{\phi}^4 (1+4s_1)/F]}\,,
\end{equation}
where $\Omega_m=\rho_m/(3FH^2)$.

In the massive limits $M_{\phi}^2 \to \infty$ and $M_R^2 \to \infty$, 
i.e. $s_1 \to 0$ and $s_2 \to \infty$ we recover the standard results
in GR: $G_{\rm eff} \simeq 1/(8\pi F)$, $\eta \simeq 1$, and 
$\psi \simeq 3\Omega_m \delta_m  \left( aH/k \right)^2$.
The difference appears in the regimes characterized by the conditions 
$s_1 \gtrsim 1$ and $s_2 \lesssim 1$, as it happens for the late-time 
evolution of perturbations in $f(R)$ gravity and in Brans-Dicke theory.

For the theories with $\xi=0$, Eq.~(\ref{psigega}) gives the 
standard relation $\psi= 3\Omega_m \delta_m \left( aH/k \right)^2$.
If $\xi \neq 0$, then this relation no longer holds.
In this case $\psi$ is not directly related with $\delta_m$ due to 
the additional contribution from the $\xi$-dependent term.
This can be regarded as the main reason for the anti-correlation between 
the ISW effect in CMB and the large-scale structure found for the model 
(\ref{BDgene}) \cite{Kobayashi1}.

In Ref.~\cite{DMT} the authors derived conditions for avoiding
the appearance of ghosts and Laplacian instabilities.
Provided $F>0$ the tensor ghosts do not appear.
Since $c_T^2=1$ for the theories (\ref{moaction}), 
tensor perturbations have no Laplacian instabilities.
For the theories with $f_{,RR} \neq 0$ the conditions for 
the avoidance of ghosts and Laplacian instabilities of 
scalar perturbations are given, respectively, by 
\begin{eqnarray}
& & 24\xi H\dot{\phi}-8\xi_{,\phi}\dot{\phi}^2
+f_{,X}+f_{,XX}\dot{\phi}^2
- F_{,X}^2\dot{\phi}^2/F_{,R}> 0\,, \\
& & f_{,X} + 8(\ddot{\phi}+2H\dot{\phi})\xi-
16\xi^2\dot{\phi}^4/(3F)>0\,.
\end{eqnarray}
Similar conditions have been also derived for the theories with 
$f_{,RR} \neq 0$. The dark energy models based on 
the action (\ref{moaction}) 
need to be constructed to satisfy these conditions.

\subsection{Observables in weak lensing}

We have shown that modified gravity models of dark energy 
generally lead to changes for the growth rate of matter perturbations 
compared to the $\Lambda$CDM model. 
Since there are two free functions that determine the 
first-order metrics $\Psi$ and $\Phi$, dark energy models can be 
classified according to how the gravitational potentials are 
linked to $\delta_{m}$. 
In order to quantify this, we introduce two quantities 
$q(k,t)$ and $\zeta(k,t)$ defined by
\begin{eqnarray}
& & (k^{2}/a^{2})\Phi = 4\pi Gq\,\delta_{m}\,\rho_{m}\,,
\label{Phiwe1}\\
& & (\Phi+\Psi)/\Phi = \zeta\,,\label{eq:qeta}
\end{eqnarray}
where $G$ is the 4-dimensional bare gravitational constant. 
The $\Lambda$CDM model corresponds to $q=1$ and $\zeta=0$
(note that the cosmological constant does not cluster).
Modified gravity models give rise to different values of $q$ and $\zeta$
relative to those in the $\Lambda$CDM model. 
Therefore the functions $q$ and $\zeta$ characterize gravitational 
theories for first-order scalar perturbations on small scales.

In Brans-Dicke theory discussed in Sec.~\ref{scatensec}, the
gravitational potentials are given by Eq.~(\ref{PsiPhisca}) on sub-horizon
scales. In this case we have 
\begin{equation}
q=\frac{1}{F}\frac{(k^{2}/a^{2})(1-2Q^{2})F+M^{2}}{(k^{2}/a^{2})F
+M^{2}}\,,
\qquad
\zeta=-\frac{4F(k^{2}/a^{2})Q^{2}}{(k^{2}/a^{2})(1-2Q^{2})F+M^{2}}\,,
\label{qzeta}
\end{equation}
where we have used the unit $8\pi G=1$. In the regime 
$M^{2}/F\gg k^{2}/a^{2}$
(and $F\simeq1$) it follows that $q\simeq1$ and $\zeta\simeq0$.
In the regime $M^{2}/F\ll k^{2}/a^{2}$ we have $q\simeq(1-2Q^{2})/F$
and $\zeta\simeq-4Q^{2}/(1-2Q^{2})$, so that the deviation from the
$\Lambda$CDM model appears. The expression (\ref{qzeta}) 
covers the case of $f(R)$ gravity by setting $Q=-1/\sqrt{6}$.

In the DGP model the gravitational potentials obey Eq.~(\ref{DGPpsiphi})
and hence
\begin{equation}
q=1-1/(3\beta)\,,\qquad
\zeta=2/(1-3\beta)\,.
\end{equation}
In the deep matter era one has $|\beta|\gg1$, so that $q\simeq1$
and $\zeta\simeq0$. The deviation from $(q,\zeta)=(1,0)$ appears
when $|\beta|$ decreases to the order of unity, i.e., when the Universe
enters the epoch of cosmic acceleration.

In order to confront dark energy models with the observations of weak
lensing, it may be convenient to introduce the 
following quantity \cite{Kunz} 
\begin{equation}
\Sigma\equiv q(1-\zeta/2)\,.\label{eq:sigmamg}
\end{equation}
{}From Eqs.~(\ref{Phiwe1}) and (\ref{eq:qeta}) we find that the
effective gravitational potential $\psi=\Phi-\Psi$ associated with 
weak lensing observations can be expressed as 
\begin{equation}
\psi=8\pi G (a^2/k^2) \rho_{m}\delta_{m}\Sigma\,.
\label{psisigma}
\end{equation}
In the DGP model and in Brans-Dicke theory we have 
$\Sigma=1$ and $\Sigma=1/F$, respectively.
In (modified) Galileon theories the term $\Sigma$
is of more complicated forms, see e.g., Eq.~(\ref{psigega}).

The effect of modified gravity theories manifests itself in weak lensing
observations in at least two ways. One is the multiplication of the
term $\Sigma$ on the r.h.s. of Eq.~(\ref{psisigma}). Another is
the modification of the evolution of $\delta_{m}$. The latter depends
on two parameters $q$ and $\zeta$, or equivalently, $\Sigma$ and
$\zeta$. Thus two parameters ($\Sigma,\zeta$) will be useful to
detect signatures of modified gravity theories from future surveys
of weak lensing.

\section{Conclusions}
\label{concludesec}

We have reviewed modified gravitational models of
dark energy responsible for the cosmic acceleration today.
In addition to cosmological constraints such as the presence of 
a matter era followed by a stable de Sitter solution, we require 
that the models satisfy local gravity constraints.
There are two mechanisms for the recovery of 
GR behavior in the regions of high density.
The first one is the chameleon mechanism in which the mass of 
a scalar-field degree of freedom depends on the matter density 
in the surrounding environment. 
The chameleon mechanism can be at work in $f(R)$ gravity and 
Brans-Dicke theory, as long as the field potential is designed to 
have a large mass in the regions of high density.  
The second one is the Vainshtein mechanism 
in which the nonlinear effect of scalar-field self
interactions leads to the recovery of GR at small distances.
This can be applied to the DGP braneworld model and Galileon gravity.

The modified gravity models can give rise to the phantom equation of 
state of dark energy ($w_{\rm DE}<-1$) without having
ghosts, tachyons, and Laplacian instabilities.
The deviation of $w_{\rm DE}$ from that in the $\Lambda$CDM model
($w_{\rm DE}=-1$) is not so significant in the models based on $f(R)$ gravity and Brans-Dicke theory, so these models can be compatible with the 
observational constraints at the background cosmology fairly easily.
On the other hand, the tracker solution in covariant Galileon gravity, 
which has $w_{\rm DE}=-2$ during the matter era, is disfavored from 
the joint data analysis of SN Ia, BAO, and CMB shift parameters. 
However the late-time tracking without a significant 
deviation from $w_{\rm DE}=-1$ is allowed observationally. 

In order to confront the modified gravity models with the observations 
of large-scale structure, CMB, and weak lensing, 
we have also discussed the evolution of matter density perturbations.
In the models based on $f(R)$ and Brans-Dicke theories
there is a ``General Relativistic'' regime 
in which the field is heavy such that $M^2 \gg k^2/a^2$.
At late times this is followed by a ``scalar-tensor'' regime 
($M^2 \ll k^2/a^2$) in which the gravitational law is modified
from that in General Relativity.
In Brans-Dicke theory the evolution of matter perturbations 
during the matter era changes from $\delta_m \propto t^{2/3}$ to 
$\delta_m \propto  t^{(\sqrt{25+48Q^{2}}-1)/6}$, 
where $Q$ is related to the Brans-Dicke parameter via 
the relation $3+2\omega_{\rm BD}=1/(2Q^2)$.
The effective gravitational couplings $G_{\rm eff}$ in the DGP and 
Galileon models are independent of the wave numbers $k$.
This reflects the fact that the field is massless in those models.
In the DGP model the growth rate of $\delta_m$ is smaller than 
that in the $\Lambda$CDM, which is associated with the 
appearance of ghosts.
In (modified) Galileon gravity, a non-trivial relation between the effective 
gravitational potential $\psi$ and the matter perturbation $\delta_m$ leads to 
the extra growth of $\psi$. 

We summarize the current status of each modified gravity model of 
dark energy.
\begin{itemize}
\item (i) In $f(R)$ gravity there are some viable dark energy models such as
(\ref{Amodel}), (\ref{Bmodel}), and (\ref{Cmodel}), which can be 
consistent with both cosmological and local gravity constraints.
For the models (\ref{Amodel}) and (\ref{Bmodel}) 
the local gravity constraints can be 
satisfied for $n>0.9$ under the chameleon mechanism.
If we use the criterion that the difference between the spectral indices 
between the matter power spectrum and the CMB spectrum is smaller than 0.05, 
then we obtain the bound $n \ge 2$.
In these models the initial conditions of perturbations need to be chosen 
such that the oscillating mode does not dominate over the matter-induced
mode in the early Universe. This is associated with a weak singularity 
problem about the divergence of the mass squared $M^2 \simeq F/(3f_{,RR})$
for $R \to \infty$, but it can be circumvented by including higher-curvature 
terms such as $R^2$ to the Lagrangian.
\item (ii) In Brans-Dicke theory, even if the coupling $Q$ is of the order of unity, 
it is possible to design field potentials consistent 
with both cosmological and local gravity constraints.
One of the examples is the potential (\ref{modelscalar}), which is 
motivated by the $f(R)$ models (\ref{Amodel}) and (\ref{Bmodel}).
Depending on the couplings $Q$ the growth of matter perturbations 
is different ($Q=-1/\sqrt{6}$ in metric $f(R)$ gravity).
{}From the observational constraints on the growth rate of 
$\delta_m$, $|Q|$ is required to be smaller than the order of 1.
\item (iii) In the DGP model the self-acceleration is realized 
through the gravitational leakage to the extra dimension, but 
the joint analysis using the data of SN Ia, BAO, and CMB shift parameters
shows that the model is in tension with observations.
Moreover the linear perturbation theory beyond the Vainshtein radius 
shows that the model contains a ghost mode with the effective 
Brans-Dicke parameter $\omega_{\rm BD}$ smaller than $-3/2$.
However, the modification of the DGP model like Cascading gravity 
can alleviate this problem.
\item (iv) In covariant Galileon gravity there is a tracker that attracts solutions 
with different initial conditions to a common trajectory. The joint observational 
constraints at the background cosmology shows that the late-time tracking solutions 
are favored from the data.
Cosmological perturbations in (modified) Galileon theories exhibit peculiar features
because of non-trivial relations between the effective gravitational 
potential $\psi$ and the matter perturbation $\delta_m$. 
\item (v) The dark energy models described by a Gauss-Bonnet term with a
scalar coupling $F(\phi){\cal G}$ do not satisfy both cosmological and 
local gravity constraints. The generalized Gauss-Bonnet model 
in which the Lagrangian density is given by $R/2+f({\cal G})$ is plagued 
by a serious problem of the Laplacian instability in the presence of matter fluids.
There are some viable Lorentz-violating models of dark energy in which 
the phantom equation of state can be realized without having ghosts, 
tachyons, and Laplacian instabilities.
\end{itemize}
We hope to find some signatures for the modification of 
gravity in future high-precision observations.
This will shed new light on the nature of dark energy.

\section*{Acknowledgments}

The author thanks the organizers of Second TRR33 Winter School held in 
Passo del Tonale (Italy) for inviting him to have a series of lectures 
about modified gravity models of dark energy.
He is also grateful to Georg Wolschin for providing him an opportunity 
to write a lecture note in a book published by Springer.
This note is one of the five chapters in the book
``Lectures on Cosmology: Accelerated Expansion of the Universe 
(Lecture Notes in Physics)''.



\begin{thebibliography}{10}

\bibitem{Riess}
A.~G.~Riess {\it et al.},
Astron.\ J.\  {\bf 116}, 1009 (1998);
Astron.\ J.\  {\bf 117}, 707 (1999).

\bibitem{Perlmutter}
S.~Perlmutter {\it et al.},
Astrophys.\ J.\  {\bf 517}, 565 (1999).

\bibitem{WMAP1}
D.~N.~Spergel {\it et al.},
Astrophys.\ J.\ Suppl.\  {\bf 148}, 175 (2003).

\bibitem{WMAP7}
E.~Komatsu {\it et al.},
arXiv:1001.4538 [astro-ph.CO].

\bibitem{BAO1}
D.~J.~Eisenstein {\it et al.}  [SDSS Collaboration],
Astrophys.\ J.\  {\bf 633}, 560 (2005).

\bibitem{BAO2}
W.~J.~Percival {\it et al.},
Mon.\ Not.\ Roy.\ Astron.\ Soc.\  {\bf 401}, 2148 (2010).

\bibitem{review1}
V.~Sahni and A.~A.~Starobinsky,
Int.\ J.\ Mod.\ Phys.\  D {\bf 9}, 373 (2000).

\bibitem{review2}
S.~M.~Carroll,
Living Rev.\ Rel.\  {\bf 4}, 1 (2001).

\bibitem{review3}
T.~Padmanabhan,
Phys.\ Rept.\  {\bf 380}, 235 (2003).

\bibitem{review4}
P.~J.~E.~Peebles and B.~Ratra,
Rev.\ Mod.\ Phys.\  {\bf 75}, 559 (2003).

\bibitem{reviewSahni}
V.~Sahni,
Lect.\ Notes Phys.\  {\bf 653}, 141 (2004).

\bibitem{review5}
E.~J.~Copeland, M.~Sami and S.~Tsujikawa,
Int.\ J.\ Mod.\ Phys.\  D {\bf 15}, 1753 (2006).

\bibitem{Noreview}
S.~Nojiri and S.~D.~Odintsov,
eConf {\bf C0602061}, 06 (2006)
[Int.\ J.\ Geom.\ Meth.\ Mod.\ Phys.\  {\bf 4}, 115 (2007)].

\bibitem{Woodard}
R.~P.~Woodard,
Lect.\ Notes Phys.\  {\bf 720}, 403 (2007).

\bibitem{Durrer}
R.~Durrer and R.~Maartens,
Gen.\ Rel.\ Grav.\  {\bf 40}, 301 (2008).

\bibitem{Lobo}
F.~S.~N.~Lobo,
arXiv:0807.1640 [gr-qc].

\bibitem{review6}
T.~P.~Sotiriou and V.~Faraoni,
Rev.\ Mod.\ Phys.\ {\bf 82}, 451 (2010).

\bibitem{Braxreview}
P.~Brax,
arXiv:0912.3610 [astro-ph.CO].

\bibitem{review7}
A.~De Felice and S.~Tsujikawa,
Living Rev.\ Rel.\  {\bf 13}, 3 (2010).

\bibitem{review8}
S.~Tsujikawa,
arXiv:1004.1493 [astro-ph.CO]
(Published in 'Dark Matter and Dark Energy: A Challenge for Modern Cosmology' Series: 
Astrophysics and Space Science Library, Vol. 370, ISBN: 978-90-481-8684-6, 2011).

\bibitem{Weinberg}
S.~Weinberg,
Rev.\ Mod.\ Phys.\  {\bf 61}, 1 (1989).

\bibitem{quin1}
Y.~Fujii, Phys.\ Rev.\ D {\bf 26}, 2580 (1982).

\bibitem{quin2}
L.~H.~Ford,
Phys.\ Rev.\ D {\bf 35}, 2339 (1987).

\bibitem{quin3}
C.~Wetterich, Nucl. \ Phys \ B. {\bf 302},
668 (1988).

\bibitem{quin4}
B.~Ratra and J.~Peebles,
Phys. \ Rev \ D {\bf 37}, 321 (1988).

\bibitem{quin5}
R.~R.~Caldwell, R.~Dave and P.~J.~Steinhardt,
Phys.\ Rev.\ Lett.\  {\bf 80}, 1582 (1998).

\bibitem{kes1}
T.~Chiba, T.~Okabe and M.~Yamaguchi,
Phys.\ Rev.\  D {\bf 62}, 023511 (2000).

\bibitem{kes2}
C.~Armendariz-Picon, V.~F.~Mukhanov and P.~J.~Steinhardt,
Phys.\ Rev.\ Lett.\  {\bf 85}, 4438 (2000).

\bibitem{Carroll98}
S.~M.~Carroll,
Phys.\ Rev.\ Lett.\  {\bf 81}, 3067 (1998).

\bibitem{Kolda}
C.~F.~Kolda and D.~H.~Lyth,
Phys.\ Lett.\  B {\bf 458}, 197 (1999).

\bibitem{fRearly1}
S.~Capozziello, Int. J. Mod. Phys. {\bf D 11}, 483, (2002).

\bibitem{fRearly2}
S.~Capozziello, V.~F.~Cardone, S.~Carloni and A.~Troisi,
Int. \ J.\ Mod.\ Phys.\ {\bf D}, 12, 1969 (2003).

\bibitem{fRearly3}
S.~M.~Carroll, V.~Duvvuri, M.~Trodden and M.~S.~Turner,
Phys.\ Rev.\  D {\bf 70}, 043528 (2004).

\bibitem{fRearly4}
S.~Nojiri and S.~D.~Odintsov,
Phys.\ Rev.\  D {\bf 68}, 123512 (2003).

\bibitem{stearly1}
L.~Amendola,
Phys.\ Rev.\  D {\bf 60}, 043501 (1999).

\bibitem{stearly2}
J.~P.~Uzan,
Phys.\ Rev.\  D {\bf 59}, 123510 (1999).

\bibitem{stearly3}
T.~Chiba,
Phys.\ Rev.\ D {\bf 60}, 083508 (1999).

\bibitem{stearly4}
N.~Bartolo and M.~Pietroni,
Phys.\ Rev.\ D {\bf 61} 023518 (2000).

\bibitem{stearly5}
F.~Perrotta, C.~Baccigalupi and S.~Matarrese,
Phys.\ Rev.\ D {\bf 61}, 023507 (2000).

\bibitem{stearly6}
A.~Riazuelo and J.~P.~Uzan,
Phys.\ Rev.\  D {\bf 66}, 023525 (2002).

\bibitem{DGP}
G.~R.~Dvali, G.~Gabadadze and M.~Porrati,
Phys.\ Lett.\  B {\bf 485}, 208 (2000).

\bibitem{Nicolis}
A.~Nicolis, R.~Rattazzi and E.~Trincherini,
Phys.\ Rev.\  D {\bf 79}, 064036 (2009).

\bibitem{NOS05}
S.~Nojiri, S.~D.~Odintsov and M.~Sasaki,
Phys.\ Rev.\  D {\bf 71}, 123509 (2005).

\bibitem{NO05}
S.~Nojiri and S.~D.~Odintsov,
Phys.\ Lett.\  B {\bf 631}, 1 (2005).

\bibitem{Soussa}
M.~E.~Soussa and R.~P.~Woodard,
Gen.\ Rel.\ Grav.\  \textbf{36}, 855 (2004).

\bibitem{Allem}
G.~Allemandi, A.~Borowiec and M.~Francaviglia, 
Phys.\ Rev.\ D \textbf{70}, 103503 (2004).

\bibitem{Easson}
D.~A.~Easson, 
Int.\ J.\ Mod.\ Phys.\ A \textbf{19}, 5343 (2004).

\bibitem{Dick04}
R.~Dick,
Gen.\ Rel.\ Grav.\  {\bf 36}, 217 (2004).

\bibitem{Carloni}
S.~Carloni, P.~K.~S.~Dunsby, S.~Capozziello and A.~Troisi, 
Class.\ Quant.\ Grav.\  \textbf{22}, 4839 (2005).

\bibitem{Brookfield}
A.~W.~Brookfield, C.~van de Bruck and L.~M.~H.~Hall,
Phys.\ Rev.\  D {\bf 74}, 064028 (2006).

\bibitem{Chiba}
T.~Chiba,
Phys.\ Lett.\ B {\bf 575}, 1 (2003).

\bibitem{Dolgov}
A.~D.~Dolgov and M.~Kawasaki,
Phys.\ Lett.\  B {\bf 573}, 1 (2003).

\bibitem{Olmo}
G.~J.~Olmo,
Phys.\ Rev.\  D {\bf 72}, 083505 (2005).

\bibitem{Navarro}
I.~Navarro and K.~Van Acoleyen,
JCAP {\bf 0702}, 022 (2007).

\bibitem{matterper1}
S.~M.~Carroll, I.~Sawicki, A.~Silvestri and M.~Trodden,
New J.\ Phys.\  \textbf{8}, 323 (2006).

\bibitem{matterper2}
R.~Bean, D.~Bernat, L.~Pogosian, A.~Silvestri and M.~Trodden,
Phys.\ Rev.\  D \textbf{75}, 064020 (2007).

\bibitem{SongHu1}
Y.~S.~Song, W.~Hu and I.~Sawicki,
Phys.\ Rev.\  D {\bf 75}, 044004 (2007).

\bibitem{SongHu2}
I.~Sawicki and W.~Hu,
Phys.\ Rev.\  D {\bf 75}, 127502 (2007).

\bibitem{APT}
L.~Amendola, D.~Polarski and S.~Tsujikawa,
Phys.\ Rev.\ Lett.\  {\bf 98}, 131302 (2007).

\bibitem{APT2}
L.~Amendola, D.~Polarski and S.~Tsujikawa,
Int.\ J.\ Mod.\ Phys.\  D {\bf 16}, 1555 (2007).

\bibitem{Teg}
T.~Faulkner, M.~Tegmark, E.~F.~Bunn and Y.~Mao,
Phys.\ Rev.\  D {\bf 76}, 063505 (2007).

\bibitem{AGPT}
L.~Amendola, R.~Gannouji, D.~Polarski and S.~Tsujikawa,
Phys.\ Rev.\  D {\bf 75}, 083504 (2007).

\bibitem{LiBarrow}
B.~Li and J.~D.~Barrow,
Phys.\ Rev.\  D {\bf 75}, 084010 (2007).

\bibitem{AmenTsuji07}
L.~Amendola and S.~Tsujikawa,
Phys.\ Lett.\  B {\bf 660}, 125 (2008).

\bibitem{Fay07}
S.~Fay, S.~Nesseris and L.~Perivolaropoulos,
Phys.\ Rev.\  D {\bf 76}, 063504 (2007).

\bibitem{Pogosian}
L.~Pogosian and A.~Silvestri,
Phys.\ Rev.\  D {\bf 77}, 023503 (2008).

\bibitem{Sendouda}
N.~Deruelle, M.~Sasaki and Y.~Sendouda,
Phys.\ Rev.\  D {\bf 77}, 124024 (2008).

\bibitem{Hu07}
W.~Hu and I.~Sawicki,
Phys.\ Rev.\  D {\bf 76}, 064004 (2007).

\bibitem{Star07}
A.~A.~Starobinsky,
JETP Lett.\  {\bf 86}, 157 (2007).

\bibitem{Appleby}
S.~A.~Appleby and R.~A.~Battye,
Phys.\ Lett.\  B {\bf 654}, 7 (2007).

\bibitem{Tsuji08}
S.~Tsujikawa,
Phys.\ Rev.\  D {\bf 77}, 023507 (2008).

\bibitem{LinderfR}
E.~V.~Linder,
Phys.\ Rev.\  D {\bf 80}, 123528 (2009).

\bibitem{BD}
C.~Brans and R.~H.~Dicke,
Phys.\ Rev.\  {\bf 124}, 925 (1961).

\bibitem{ohanlon}
J.~O'Hanlon,
Phys.\ Rev.\ Lett.\ {\bf 29}, 137 (1972).

\bibitem{coupled}
L.~Amendola,
Phys.\ Rev.\  D {\bf 62}, 043511 (2000).

\bibitem{TUMTY}
S.~Tsujikawa, K.~Uddin, S.~Mizuno, R.~Tavakol and J.~Yokoyama,
Phys.\ Rev.\  D {\bf 77}, 103009 (2008).

\bibitem{Cembranos}
J.~A.~R.~Cembranos,
Phys.\ Rev.\  D {\bf 73}, 064029 (2006).

\bibitem{CapoTsuji}
S.~Capozziello and S.~Tsujikawa,
Phys.\ Rev.\  D {\bf 77}, 107501 (2008).

\bibitem{Gannouji10}
R.~Gannouji, B.~Moraes, D.~F.~Mota, D.~Polarski, 
S.~Tsujikawa and H.~A.~Winther,
Phys.\ Rev.\  D {\bf 82}, 124006 (2010).

\bibitem{Van}
P.~Brax, C.~van de Bruck, A.~C.~Davis and D.~J.~Shaw,
Phys.\ Rev.\  D {\bf 78}, 104021 (2008).

\bibitem{KW1}
J.~Khoury and A.~Weltman,
Phys.\ Rev.\ Lett.\  {\bf 93}, 171104 (2004).

\bibitem{KW2}
J.~Khoury and A.~Weltman,
Phys.\ Rev.\  D {\bf 69}, 044026 (2004).

\bibitem{DGPnon1}
C.~Deffayet, G.~R.~Dvali, G.~Gabadadze and A.~I.~Vainshtein,
Phys.\ Rev.\  D {\bf 65}, 044026 (2002).

\bibitem{DGPnon2}
M.~Porrati,
Phys.\ Lett.\  B {\bf 534}, 209 (2002).

\bibitem{Vainshtein}
A.~I.~Vainshtein,
Phys.\ Lett.\  B {\bf 39}, 393 (1972).

\bibitem{DGPghost1}
A.~Nicolis and R.~Rattazzi,
JHEP {\bf 0406}, 059 (2004).

\bibitem{DGPghost2}
K.~Koyama and R.~Maartens,
JCAP {\bf 0601}, 016 (2006).

\bibitem{DGPghost3}
D.~Gorbunov, K.~Koyama and S.~Sibiryakov,
Phys.\ Rev.\  D {\bf 73}, 044016 (2006).

\bibitem{DGPobser1}
I.~Sawicki and S.~M.~Carroll,
arXiv:astro-ph/0510364.

\bibitem{DGPobser2}
M.~Fairbairn and A.~Goobar,
Phys.\ Lett.\  B {\bf 642} , 432 (2006).

\bibitem{DGPobser3}
R.~Maartens and E.~Majerotto,
Phys.\ Rev.\  D {\bf 74}, 023004  (2006).

\bibitem{DGPobser4}
U.~Alam and V.~Sahni,
Phys.\ Rev.\  D {\bf 73}, 084024 (2006).

\bibitem{DGPobser5}
Y.~S.~Song, I.~Sawicki and W.~Hu,
Phys.\ Rev.\  D {\bf 75}, 064003 (2007).

\bibitem{DGPobser6}
J.~Q.~Xia,
Phys.\ Rev.\  D {\bf 79}, 103527 (2009).

\bibitem{Deffayetga1}
C.~Deffayet, G.~Esposito-Farese and A.~Vikman,
Phys.\ Rev.\  D {\bf 79}, 084003 (2009).

\bibitem{Deffayetga2}
C.~Deffayet, S.~Deser and G.~Esposito-Farese,
Phys.\ Rev.\  D {\bf 80}, 064015 (2009).

\bibitem{DT2}
A.~De Felice and S.~Tsujikawa,
Phys.\ Rev.\ Lett.\  {\bf 105}, 111301 (2010).

\bibitem{DT3}
A.~De Felice and S.~Tsujikawa,
arXiv:1008.4236 [hep-th].

\bibitem{obsermo1}
C.~Schimd, J.~P.~Uzan and A.~Riazuelo,
Phys.\ Rev.\  D {\bf 71}, 083512 (2005).

\bibitem{obsermo2}
M.~Ishak, A.~Upadhye and D.~N.~Spergel,
Phys.\ Rev.\  D {\bf 74}, 043513 (2006).

\bibitem{obsermo3}
L.~Knox, Y.~S.~Song and J.~A.~Tyson,
Phys.\ Rev.\  D {\bf 74}, 023512 (2006).

\bibitem{obsermo4}
D.~Huterer and E.~V.~Linder,
Phys.\ Rev.\  D {\bf 75}, 023519 (2007).

\bibitem{obsermo5}
P.~Zhang, M.~Liguori, R.~Bean and S.~Dodelson,
Phys.\ Rev.\ Lett.\  {\bf 99}, 141302 (2007).

\bibitem{obsermo6}
S.~Wang, L.~Hui, M.~May and Z.~Haiman,
Phys.\ Rev.\  D {\bf 76}, 063503 (2007).

\bibitem{obsermo7}
B.~Jain and P.~Zhang,
Phys.\ Rev.\  D {\bf 78}, 063503 (2008).

\bibitem{obsermo8}
S.~F.~Daniel, R.~R.~Caldwell, A.~Cooray and A.~Melchiorri,
Phys.\ Rev.\  D {\bf 77}, 103513 (2008).

\bibitem{obsermo9}
E.~Bertschinger and P.~Zukin,
Phys.\ Rev.\  D {\bf 78}, 024015 (2008).

\bibitem{Carloniper}
S.~Carloni, P.~K.~S.~Dunsby and A.~Troisi,
Phys.\ Rev.\  D {\bf 77}, 024024 (2008).

\bibitem{Ananda}
K.~N.~Ananda, S.~Carloni and P.~K.~S.~Dunsby,
Class.\ Quant.\ Grav.\  {\bf 26}, 235018 (2009).

\bibitem{obsermo10}
G.~B.~Zhao, L.~Pogosian, A.~Silvestri and J.~Zylberberg,
Phys.\ Rev.\  D {\bf 79}, 083513 (2009).

\bibitem{obsermo11}
Y.~S.~Song and K.~Koyama,
JCAP {\bf 0901}, 048 (2009).

\bibitem{obsermo12}
Y.~S.~Song and O.~Dore,
JCAP {\bf 0903}, 025 (2009).

\bibitem{obsermo13}
S.~A.~Thomas, F.~B.~Abdalla and J.~Weller,
Mon.\ Not.\ Roy.\ Astron.\ Soc.\  {\bf 395}, 197 (2009).

\bibitem{Borisov08}
A.~Borisov and B.~Jain,
Phys.\ Rev.\  D {\bf 79}, 103506 (2009).

\bibitem{obsermo14}
J.~Guzik, B.~Jain and M.~Takada,
Phys.\ Rev.\  D {\bf 81}, 023503 (2010).

\bibitem{obsermo15}
P.~Brax, C.~van de Bruck, A.~C.~Davis and D.~Shaw,
JCAP {\bf 1004}, 032 (2010).

\bibitem{obsermo16}
R.~Bean and M.~Tangmatitham,
Phys.\ Rev.\  D {\bf 81}, 083534 (2010).

\bibitem{Reyes}
R.~Reyes, R.~Mandelbaum, U.~Seljak, T.~Baldauf, J.~E.~Gunn, L.~Lombriser and R.~E.~Smith,
Nature {\bf 464}, 256 (2010).

\bibitem{obsermo17}
G.~B.~Zhao {\it et al.},
Phys.\ Rev.\  D {\bf 81}, 103510 (2010).

\bibitem{obsermo18}
S.~F.~Daniel {\it et al.,}
Phys.\ Rev.\  D {\bf 81}, 123508 (2010).

\bibitem{Daniel10}
S.~F.~Daniel and E.~V.~Linder,
Phys.\ Rev.\  D {\bf 82}, 103523 (2010).

\bibitem{Girones}
Z.~Girones, A.~Marchetti, O.~Mena, C.~Pena-Garay and N.~Rius,
JCAP {\bf 1011}, 004 (2010).

\bibitem{Narikawa1}
T.~Narikawa and K.~Yamamoto,
Phys.\ Rev.\  D {\bf 81}, 043528 (2010).

\bibitem{Narikawa2}
K.~Yamamoto, G.~Nakamura, G.~Hutsi, T.~Narikawa and T.~Sato,
Phys.\ Rev.\  D {\bf 81}, 103517 (2010).

\bibitem{Lombriser}
L.~Lombriser, A.~Slosar, U.~Seljak and W.~Hu,
arXiv:1003.3009 [astro-ph.CO].

\bibitem{Song10}
Y.~S.~Song, G.~B.~Zhao, D.~Bacon, K.~Koyama, R.~C.~Nichol and L.~Pogosian,
arXiv:1011.2106 [astro-ph.CO].

\bibitem{Palatini1919}
A.~Palatini, Rend.\,Circ.\, Mat.\, Palermo 43, 203 (1919).

\bibitem{Hehl}
F.~W.~Hehl and G.~D.~Kerling, Gen.\,Rel.\,Grav.\,{\bf 9}, 
691 (1978).

\bibitem{Liberati}
T.~P.~Sotiriou and S.~Liberati,
Annals Phys.\  {\bf 322}, 935 (2007).

\bibitem{Liberati2}
T.~P.~Sotiriou and S.~Liberati,
J.\ Phys.\ Conf.\ Ser.\  {\bf 68}, 012022 (2007).

\bibitem{Capome}
S.~Capozziello, R.~Cianci, C.~Stornaiolo and S.~Vignolo,
Class.\ Quant.\ Grav.\  {\bf 24}, 6417 (2007).

\bibitem{Ferraris}
M.~Ferraris, M.~Francaviglia and I.~Volovich,
Class.\ Quant.\ Grav.\  {\bf 11}, 1505 (1994).

\bibitem{Vollick}
D.~N.~Vollick,
Phys.\ Rev.\  D {\bf 68}, 063510 (2003).

\bibitem{Vollick2}
D.~N.~Vollick,
Class.\ Quant.\ Grav.\  {\bf 21}, 3813 (2004).

\bibitem{Flanagan0}
E.~E.~Flanagan,
Class.\ Quant.\ Grav.\  {\bf 21}, 417 (2003).

\bibitem{Flanagan}
E.~E.~Flanagan,
Phys.\ Rev.\ Lett.\  {\bf 92}, 071101 (2004).

\bibitem{Flanagan2}
E.~E.~Flanagan,
Class.\ Quant.\ Grav.\  {\bf 21}, 3817 (2004).

\bibitem{Meng}
X.~Meng and P.~Wang,
Class.\ Quant.\ Grav.\ 20, 4949 (2003).

\bibitem{Meng2}
X.~Meng and P.~Wang,
Class.\ Quant.\ Grav.\  {\bf 21}, 951 (2004).

\bibitem{Meng3}
X.~H.~Meng and P.~Wang,
Phys.\ Lett.\  B {\bf 584}, 1 (2004).

\bibitem{NOpala}
S.~Nojiri and S.~D.~Odintsov,
Gen.\ Rel.\ Grav.\  {\bf 36}, 1765 (2004).

\bibitem{Sot}
T.~P.~Sotiriou,
Class.\ Quant.\ Grav.\  {\bf 23}, 1253 (2006).

\bibitem{Sotinf}
T.~P.~Sotiriou,
Phys.\ Rev.\  D {\bf 73}, 063515 (2006).

\bibitem{Motapala}
M.~Amarzguioui, O.~Elgaroy, D.~F.~Mota and T.~Multamaki,
Astron.\ Astrophys.\  {\bf 454}, 707 (2006).

\bibitem{FayTavakol}
S.~Fay, R.~Tavakol and S.~Tsujikawa,
Phys.\ Rev.\  D {\bf 75}, 063509 (2007).

\bibitem{Kaloper}
A.~Iglesias, N.~Kaloper, A.~Padilla and M.~Park,
Phys.\ Rev.\  D {\bf 76}, 104001 (2007).

\bibitem{OlmoPRL2}
G.~J.~Olmo,
Phys.\ Rev.\ Lett.\  {\bf 98}, 061101 (2007).

\bibitem{Olmo08}
G.~J.~Olmo,
Phys.\ Rev.\  D {\bf 77}, 084021 (2008).

\bibitem{Olmo09}
G.~J.~Olmo,
arXiv:0910.3734 [gr-qc].

\bibitem{Barausse1}
E.~Barausse, T.~P.~Sotiriou and J.~C.~Miller,
Class.\ Quant.\ Grav.\  {\bf 25}, 105008 (2008).

\bibitem{KoivistoPala}
T.~Koivisto and H.~Kurki-Suonio,
Class.\ Quant.\ Grav.\  {\bf 23}, 2355 (2006).

\bibitem{KoivistoPala2}
T.~Koivisto,
Phys.\ Rev.\  D {\bf 73}, 083517 (2006).

\bibitem{LiPala0}
B.~Li and M.~C.~Chu,
Phys.\ Rev.\  D {\bf 74}, 104010 (2006).

\bibitem{LiPala}
B.~Li, K.~C.~Chan and M.~C.~Chu,
Phys.\ Rev.\  D {\bf 76}, 024002 (2007).

\bibitem{TsujiUddin}
S.~Tsujikawa, K.~Uddin and R.~Tavakol,
Phys.\ Rev.\  D {\bf 77}, 043007 (2008).

\bibitem{Star80}
A.~A.~Starobinsky,
Phys.\ Lett.\  B {\bf 91}, 99 (1980).

\bibitem{Ketov1}
S.~V.~Ketov,
Phys.\ Lett.\  B {\bf 692}, 272 (2010).

\bibitem{Ketov2}
S.~V.~Ketov and A.~A.~Starobinsky,
arXiv:1011.0240 [hep-th].

\bibitem{Muller} 
V.~Muller, H.~J.~Schmidt and A.~A.~Starobinsky,
Phys.\ Lett.\  B {\bf 202}, 198 (1988).

\bibitem{Faraonista1}
V.~Faraoni,
Phys.\ Rev.\  D {\bf 72}, 061501 (2005).

\bibitem{Faraonista2}
V.~Faraoni and S.~Nadeau,
Phys.\ Rev.\  D {\bf 72}, 124005 (2005).

\bibitem{CNOT}
S.~Capozziello, S.~Nojiri, S.~D.~Odintsov and A.~Troisi,
Phys.\ Lett.\  B {\bf 639}, 135 (2006).

\bibitem{BGT09}
K.~Bamba, C.~Q.~Geng and S.~Tsujikawa,
Phys.\ Lett.\  B {\bf 688}, 101 (2010).

\bibitem{Torres}
D.~F.~Torres,
Phys.\ Rev.\  D {\bf 66}, 043522 (2002).

\bibitem{Boi00}
B.~Boisseau, G.~Esposito-Farese, D.~Polarski and A.~A.~Starobinsky,
Phys.\ Rev.\ Lett.\  {\bf 85}, 2236 (2000).

\bibitem{Espo}
G.~Esposito-Farese and D.~Polarski,
Phys.\ Rev.\  D {\bf 63}, 063504 (2001).

\bibitem{Motohashi10}
H.~Motohashi, A.~A.~Starobinsky and J.~Yokoyama,
Prog.\ Theor.\ Phys.\  {\bf 123}, 887 (2010).

\bibitem{Bamba10}
K.~Bamba, C.~Q.~Geng and C.~C.~Lee,
JCAP {\bf 1008}, 021 (2010).

\bibitem{Bamba10d}
K.~Bamba, C.~Q.~Geng and C.~C.~Lee,
JCAP {\bf 1011}, 001 (2010).

\bibitem{Bamba08}
K.~Bamba, C.~Q.~Geng, S.~Nojiri and S.~D.~Odintsov,
Phys.\ Rev.\  D {\bf 79}, 083014 (2009).

\bibitem{Bamba09}
K.~Bamba,
Open Astron.\ J.\  {\bf 3}, 13 (2010).

\bibitem{Dev}
A.~Dev, D.~Jain, S.~Jhingan, S.~Nojiri, M.~Sami and I.~Thongkool,
Phys.\ Rev.\  D {\bf 78}, 083515 (2008).

\bibitem{Mel09}
M.~Martinelli, A.~Melchiorri and L.~Amendola,
Phys.\ Rev.\  D {\bf 79}, 123516 (2009).

\bibitem{Cardone09}
V.~F.~Cardone, A.~Diaferio and S.~Camera,
arXiv:0907.4689 [astro-ph.CO].

\bibitem{Ali10}
A.~Ali, R.~Gannouji, M.~Sami and A.~A.~Sen,
Phys.\ Rev.\  D {\bf 81}, 104029 (2010).

\bibitem{Caporecon}
S.~Capozziello, V.~F.~Cardone and A.~Troisi,
Phys.\ Rev.\  D {\bf 71}, 043503 (2005).

\bibitem{Mul06}
T.~Multamaki and I.~Vilja,
Phys.\ Rev.\  D {\bf 73}, 024018 (2006).

\bibitem{Dobado06}
A.~de la Cruz-Dombriz and A.~Dobado,
Phys.\ Rev.\  D {\bf 74}, 087501 (2006).

\bibitem{Wu07}
X.~Wu and Z.~H.~Zhu,
Phys.\ Lett.\  B {\bf 660}, 293 (2008).

\bibitem{Carloni10}
S.~Carloni, R.~Goswami and P.~K.~S.~Dunsby,
arXiv:1005.1840 [gr-qc].

\bibitem{Bardeen}
J.~M.~Bardeen,
Phys.\ Rev.\  D {\bf 22}, 1882 (1980).

\bibitem{Hwang97}
J.~c.~Hwang,
Class.\ Quant.\ Grav.\  {\bf 14}, 3327 (1997).

\bibitem{DMT}
A.~De Felice, S.~Mukohyama and S.~Tsujikawa,
Phys.\ Rev.\  D {\bf 82}, 023524 (2010).

\bibitem{OlmoPRL}
G.~J.~Olmo,
Phys.\ Rev.\ Lett.\  {\bf 95}, 261102 (2005).

\bibitem{Olmo05}
G.~J.~Olmo,
Phys.\ Rev.\  D {\bf 72}, 083505 (2005).

\bibitem{Fara06}
V.~Faraoni,
Phys.\ Rev.\  D {\bf 74}, 023529 (2006).

\bibitem{Erick06}
A.~L.~Erickcek, T.~L.~Smith and M.~Kamionkowski,
Phys.\ Rev.\  D {\bf 74}, 121501 (2006).

\bibitem{Chiba07}
T.~Chiba, T.~L.~Smith and A.~L.~Erickcek,
Phys.\ Rev.\  D {\bf 75}, 124014 (2007).

\bibitem{Kainu07}
K.~Kainulainen, J.~Piilonen, V.~Reijonen and D.~Sunhede,
Phys.\ Rev.\  D {\bf 76}, 024020 (2007).

\bibitem{Kainu08}
K.~Kainulainen and D.~Sunhede,
Phys.\ Rev.\  D {\bf 78}, 063511 (2008).

\bibitem{Maeda}
K.~i.~Maeda,
Phys.\ Rev.\  D {\bf 39}, 3159 (1989).

\bibitem{TT08}
T.~Tamaki and S.~Tsujikawa,
Phys.\ Rev.\  D {\bf 78}, 084028 (2008).
  
\bibitem{Will05} 
C.~M.~Will, 
Living Rev.\ Rel.\ \textbf{9}, 3 (2005).

\bibitem{KM1}
T.~Kobayashi and K.~i.~Maeda,
Phys.\ Rev.\  D {\bf 78}, 064019 (2008).

\bibitem{KM2}
T.~Kobayashi and K.~i.~Maeda,
Phys.\ Rev.\  D {\bf 79}, 024009 (2009).

\bibitem{TTT}
S.~Tsujikawa, T.~Tamaki and R.~Tavakol,
JCAP {\bf 0905}, 020 (2009).

\bibitem{Upadhye}
A.~Upadhye and W.~Hu,
Phys.\ Rev.\  D {\bf 80}, 064002 (2009).

\bibitem{Babi1}
E.~Babichev and D.~Langlois,
Phys.\ Rev.\  D {\bf 80}, 121501 (2009).

\bibitem{Babi2}
E.~Babichev and D.~Langlois,
Phys.\ Rev.\  D {\bf 81}, 124051 (2010).

\bibitem{Cooney}
A.~Cooney, S.~DeDeo and D.~Psaltis,
Phys.\ Rev.\  D {\bf 82}, 064033 (2010).

\bibitem{Iess}
B.~Bertotti, L.~Iess and P.~Tortora,
Nature {\bf 425}, 374 (2003).

\bibitem{Peri1}
L.~Perivolaropoulos,
JCAP {\bf 0510}, 001 (2005).  

\bibitem{Gannouji06}
R.~Gannouji, D.~Polarski, A.~Ranquet and A.~A.~Starobinsky,
JCAP {\bf 0609}, 016 (2006).
  
\bibitem{Peri2}
S.~Nesseris and L.~Perivolaropoulos,
JCAP {\bf 0701}, 018 (2007).  
  
\bibitem{Martin}
J.~Martin, C.~Schimd and J.~P.~Uzan,
Phys.\ Rev.\ Lett.\  {\bf 96}, 061303 (2006).  

\bibitem{Billyard}
A.~Billyard, A.~Coley and J.~Ibanez,
Phys.\ Rev.\  D {\bf 59}, 023507 (1999). 

\bibitem{Gunzig}
E.~Gunzig, V.~Faraoni, A.~Figueiredo, T.~M.~Rocha and L.~Brenig,
Class.\ Quant.\ Grav.\  {\bf 17}, 1783 (2000).
  
\bibitem{Verde}
V.~Acquaviva and L.~Verde,
JCAP {\bf 0712}, 001 (2007).  
  
\bibitem{Agarwal}
N.~Agarwal and R.~Bean,
Class.\ Quant.\ Grav.\  {\bf 25}, 165001 (2008).  

\bibitem{Jarv}
L.~Jarv, P.~Kuusk and M.~Saal,
Phys.\ Rev.\  D {\bf 78}, 083530 (2008).
 
\bibitem{Leach}
S.~Carloni, S.~Capozziello, J.~A.~Leach and P.~K.~S.~Dunsby,
Class.\ Quant.\ Grav.\  {\bf 25}, 035008 (2008). 
 
\bibitem{Braxchame}
P.~Brax, C.~van de Bruck, A.~C.~Davis, J.~Khoury and A.~Weltman,
Phys.\ Rev.\  D {\bf 70}, 123518 (2004).

\bibitem{Kapner:2006si}
D.~J.~Kapner {\it et al., }
Phys.\ Rev.\ Lett.\  {\bf 98}, 021101 (2007).

\bibitem{Nagata}
R.~Nagata, T.~Chiba and N.~Sugiyama,
Phys.\ Rev.\  D {\bf 69}, 083512 (2004).

\bibitem{Randall1}
L.~Randall and R.~Sundrum,
Phys.\ Rev.\ Lett.\  {\bf 83}, 3370 (1999).

\bibitem{Randall2}
L.~Randall and R.~Sundrum,
Phys.\ Rev.\ Lett.\  {\bf 83}, 4690 (1999).

\bibitem{DGP2} 
G.~R.~Dvali and G.~Gabadadze, 
Phys.\ Rev.\ D \textbf{63} (2001), 065007.

\bibitem{Deffayet1} 
C.~Deffayet, 
Phys.\ Lett.\ B \textbf{502} (2001), 199.

\bibitem{Deffayet2} 
C.~Deffayet, G.~R.~Dvali and G.~Gabadadze,
Phys.\ Rev.\ D \textbf{65} (2002), 044023.

\bibitem{Shtanov}
V.~Sahni and Y.~Shtanov,
JCAP {\bf 0311}, 014 (2003).

\bibitem{Langlois1}
P.~Binetruy, C.~Deffayet and D.~Langlois,
Nucl.\ Phys.\  B {\bf 565}, 269 (2000).

\bibitem{Langlois2}
P.~Binetruy, C.~Deffayet, U.~Ellwanger and D.~Langlois,
Phys.\ Lett.\  B {\bf 477}, 285 (2000).

\bibitem{SMS}
T.~Shiromizu, K.~i.~Maeda and M.~Sasaki,
Phys.\ Rev.\  D {\bf 62}, 024012 (2000).
  
\bibitem{Riess04}
A.~G.~Riess {\it et al.}  [Supernova Search Team Collaboration],
Astrophys.\ J.\  {\bf 607}, 665 (2004).
  
\bibitem{Astier05}
P.~Astier {\it et al.}  [The SNLS Collaboration],
Astron.\ Astrophys.\  {\bf 447}, 31 (2006).  

\bibitem{WMAP3}
D.~N.~Spergel {\it et al.}  [WMAP Collaboration],
Astrophys.\ J.\ Suppl.\  {\bf 170}, 377 (2007).

\bibitem{Xia}
J.~Q.~Xia,
Phys.\ Rev.\  D {\bf 79}, 103527 (2009).

\bibitem{Gruzinov:2001hp}
A.~Gruzinov,
New Astron.\  {\bf 10}, 311 (2005).
 
\bibitem{Luty}
M.~A.~Luty, M.~Porrati and R.~Rattazzi,
JHEP {\bf 0309}, 029 (2003). 
 
\bibitem{Kolano1}
M.~Kolanovic, M.~Porrati and J.~W.~Rombouts,
Phys.\ Rev.\  D {\bf 68}, 064018 (2003).

\bibitem{Kolano2}
M.~Kolanovic,
Phys.\ Rev.\  D {\bf 67}, 106002 (2003).

\bibitem{CascoDGP}
C.~de Rham, G.~Dvali, S.~Hofmann, J.~Khoury, O.~Pujolas, M.~Redi and A.~J.~Tolley,
Phys.\ Rev.\ Lett.\  {\bf 100}, 251603 (2008).

\bibitem{CascoDGP2}
C.~de Rham, S.~Hofmann, J.~Khoury and A.~J.~Tolley,
JCAP {\bf 0802}, 011 (2008).

\bibitem{CascoDGP3}
N.~Agarwal, R.~Bean, J.~Khoury and M.~Trodden,
Phys.\ Rev.\  D {\bf 81}, 084020 (2010).

\bibitem{deRham}
C.~de Rham, J.~Khoury and A.~J.~Tolley,
Phys.\ Rev.\  D {\bf 81}, 124027 (2010).
  
\bibitem{Sami10}
R.~Gannouji and M.~Sami,
Phys.\ Rev.\  D {\bf 82}, 024011 (2010).  
  
\bibitem{JustinGal}
N.~Chow and J.~Khoury,
Phys.\ Rev.\  D {\bf 80}, 024037 (2009).

\bibitem{KazuyaGal}
F.~P.~Silva and K.~Koyama,
Phys.\ Rev.\  D {\bf 80}, 121301 (2009).

\bibitem{Kobayashi1}
T.~Kobayashi, H.~Tashiro and D.~Suzuki,
Phys.\ Rev.\  D {\bf 81}, 063513 (2010).

\bibitem{Kobayashi2}
T.~Kobayashi,
Phys.\ Rev.\  D {\bf 81}, 103533 (2010).

\bibitem{Rham10}
C.~de Rham and A.~J.~Tolley,
JCAP {\bf 1005}, 015 (2010).

\bibitem{DT1}
A.~De Felice and S.~Tsujikawa,
JCAP {\bf 1007}, 024 (2010).

\bibitem{Cremi2}
P.~Creminelli, A.~Nicolis and E.~Trincherini,
JCAP {\bf 1011}, 021 (2010).

\bibitem{Padilla}
A.~Padilla, P.~M.~Saffin and S.~Y.~Zhou,
JHEP {\bf 1012}, 031 (2010);
arXiv:1008.0745 [hep-th].

\bibitem{Deser}
C.~Deffayet, S.~Deser and G.~Esposito-Farese,
Phys.\ Rev.\  D {\bf 82}, 061501 (2010).

\bibitem{KYY}
T.~Kobayashi, M.~Yamaguchi and J.~Yokoyama,
Phys.\ Rev.\ Lett.\  {\bf 105}, 231302 (2010).

\bibitem{DPSV}
C.~Deffayet, O.~Pujolas, I.~Sawicki and A.~Vikman,
JCAP {\bf 1010}, 026 (2010).

\bibitem{Mark1}
K.~Hinterbichler, M.~Trodden and D.~Wesley,
arXiv:1008.1305 [hep-th].

\bibitem{Ali}
A.~Ali, R.~Gannouji and M.~Sami,
arXiv:1008.1588 [astro-ph.CO].

\bibitem{Mark2}
M.~Andrews, K.~Hinterbichler, J.~Khoury and M.~Trodden,
arXiv:1008.4128 [hep-th].

\bibitem{Mark3}
G.~L.~Goon, K.~Hinterbichler and M.~Trodden,
arXiv:1008.4580 [hep-th].

\bibitem{Mizuno}
S.~Mizuno and K.~Koyama,
Phys.\ Rev.\  D {\bf 82}, 103518 (2010).

\bibitem{Burrage}
C.~Burrage, C.~de Rham, D.~Seery and A.~J.~Tolley,
arXiv:1009.2497 [hep-th].

\bibitem{Babichev}
E.~Babichev,
arXiv:1009.2921 [hep-th].

\bibitem{Zhou}
S.~Y.~Zhou,
arXiv:1011.0863 [hep-th].

\bibitem{Kimura}
R.~Kimura and K.~Yamamoto,
arXiv:1011.2006 [astro-ph.CO].  
  
\bibitem{Kamada}
K.~Kamada, T.~Kobayashi, M.~Yamaguchi and J.~Yokoyama,
arXiv:1012.4238 [astro-ph.CO].  

\bibitem{Hirano}
K.~Hirano,
arXiv: 1012.5451 [astro-ph.CO].  

\bibitem{Nesseris10}
S.~Nesseris, A.~De Felice and S.~Tsujikawa,
arXiv:1010.0407 [astro-ph.CO]
(Physical Review D, to appear).

\bibitem{Faddeev}
L.~D.~Faddeev and R.~Jackiw,
Phys.\ Rev.\ Lett.\  {\bf 60}, 1692 (1988).

\bibitem{Suyama}
A.~De Felice and T.~Suyama,
JCAP {\bf 0906}, 034 (2009);
Phys.\ Rev.\  D {\bf 80}, 083523 (2009).

\bibitem{Hicken}
M.~Hicken {\it et al.},
Astrophys.\ J.\  {\bf 700}, 1097 (2009).

\bibitem{Amanullah}
R.~Amanullah {\it et al.},
Astrophys.\ J.\  {\bf 716}, 712 (2010).

\bibitem{AIC}
H.~Akaike, IEEE Trans. Auto. Control, 19, 716 (1974).

\bibitem{Liddle04}
A.~R.~Liddle,
Mon.\ Not.\ Roy.\ Astron.\ Soc.\  {\bf 351}, L49 (2004).

\bibitem{BIC}
G.~Schwarz, Annals of Statistics, 5, 461 (1978).

\bibitem{Kret}
E.~Kretschmann,
Ann.~Phys, {\bf 53}, 575-614 (1917).

\bibitem{Stelle}
K.~S.~Stelle,
Gen.\ Rel.\ Grav.\  {\bf 9}, 353 (1978).

\bibitem{Barth}
N.~H.~Barth and S.~M.~Christensen,
Phys.\ Rev.\  D {\bf 28}, 1876 (1983).

\bibitem{Gasreview}
M.~Gasperini and G.~Veneziano,
Phys.\ Rept.\  {\bf 373}, 1 (2003).

\bibitem{ANunez}
A.~Nunez and S.~Solganik, 
Phys.\ Lett.\ B {\bf 608}, 189 (2005).

\bibitem{CaDe}
S.~M.~Carroll, A.~De Felice, V.~Duvvuri, D.~A.~Easson, M.~Trodden and M.~S.~Turner,
Phys.\ Rev.\  D {\bf 71}, 063513 (2005).

\bibitem{Calcagni05}
G.~Calcagni, S.~Tsujikawa and M.~Sami,
Class.\ Quant.\ Grav.\  {\bf 22}, 3977 (2005).

\bibitem{Sami05}
M.~Sami, A.~Toporensky, P.~V.~Tretjakov and S.~Tsujikawa,
Phys.\ Lett.\  B {\bf 619}, 193 (2005).

\bibitem{Chibaghost}
T.~Chiba,
JCAP {\bf 0503}, 008 (2005).

\bibitem{Mena}
O.~Mena, J.~Santiago and J.~Weller,
Phys.\ Rev.\ Lett.\  {\bf 96}, 041103 (2006).

\bibitem{Hervik}
J.~D.~Barrow and S.~Hervik,
Phys.\ Rev.\  D {\bf 74}, 124017 (2006).

\bibitem{Cognola}
G.~Cognola, M.~Gastaldi and S.~Zerbini,
Int.\ J.\ Theor.\ Phys.\  {\bf 47}, 898 (2008).

\bibitem{Sokolowski}
L.~M.~Sokolowski,
Class.\ Quant.\ Grav.\  {\bf 24}, 3391 (2007).

\bibitem{Defelice1}
A.~De Felice, M.~Hindmarsh and M.~Trodden,
JCAP {\bf 0608}, 005 (2006).

\bibitem{Defelice2}
G.~Calcagni, B.~de Carlos and A.~De Felice,
Nucl.\ Phys.\  B {\bf 752}, 404 (2006).

\bibitem{Koivisto1}
T.~Koivisto and D.~F.~Mota,
Phys.\ Lett.\  B {\bf 644}, 104 (2007).

\bibitem{TsujiSami07}
S.~Tsujikawa and M.~Sami,
JCAP {\bf 0701}, 006 (2007).

\bibitem{Koivisto2}
T.~Koivisto and D.~F.~Mota,
Phys.\ Rev.\  D {\bf 75}, 023518 (2007).

\bibitem{Neupane06}
I.~P.~Neupane,
Class.\ Quant.\ Grav.\  {\bf 23}, 7493 (2006).

\bibitem{Neupane}
B.~M.~Leith and I.~P.~Neupane,
JCAP {\bf 0705}, 019 (2007).

\bibitem{Sanyal}
A.~K.~Sanyal,
Phys.\ Lett.\  B {\bf 645}, 1 (2007).

\bibitem{Kawai}
S.~Kawai, M.~a.~Sakagami and J.~Soda,
Phys.\ Lett.\  B {\bf 437}, 284 (1998).

\bibitem{Ohta}
Z.~K.~Guo, N.~Ohta and S.~Tsujikawa,
Phys.\ Rev.\  D {\bf 75}, 023520 (2007).

\bibitem{AmenDavis}
L.~Amendola, C.~Charmousis and S.~C.~Davis,
JCAP {\bf 0612}, 020 (2006).

\bibitem{Cognola06}
G.~Cognola, E.~Elizalde, S.~Nojiri, S.~D.~Odintsov and S.~Zerbini,
Phys.\ Rev.\  D {\bf 73}, 084007 (2006).

\bibitem{LBM}
B.~Li, J.~D.~Barrow and D.~F.~Mota,
Phys.\ Rev.\  D {\bf 76}, 044027 (2007).

\bibitem{DeTsu}
A.~De Felice and S.~Tsujikawa,
Phys.\ Lett.\  B {\bf 675}, 1 (2009).

\bibitem{Cope09}
S.~Y.~Zhou, E.~J.~Copeland and P.~M.~Saffin,
JCAP {\bf 0907}, 009 (2009).

\bibitem{Mohseni}
M.~Mohseni,
Phys.\ Lett.\  B {\bf 682}, 89 (2009).

\bibitem{Ishak}
J.~Moldenhauer, M.~Ishak, J.~Thompson and D.~A.~Easson,
Phys.\ Rev.\  D {\bf 81}, 063514 (2010).

\bibitem{DeFelicesolar}
A.~De Felice and S.~Tsujikawa,
Phys.\ Rev.\  D {\bf 80}, 063516 (2009).

\bibitem{DMT10}
A.~De Felice, D.~F.~Mota and S.~Tsujikawa,
Phys.\ Rev.\  D {\bf 81}, 023532 (2010).

\bibitem{Gerrard}
A.~De Felice, J.~M.~Gerard and T.~Suyama,
Phys.\ Rev.\  D {\bf 82}, 063526 (2010).

\bibitem{Suyama10}
A.~De Felice and T.~Suyama,
arXiv:1010.3886 [astro-ph.CO].

\bibitem{Alimo08}
M.~Alimohammadi and A.~Ghalee,
Phys.\ Rev.\  D {\bf 79}, 063006 (2009).

\bibitem{Alimo09}
M.~Alimohammadi and A.~Ghalee,
Phys.\ Rev.\  D {\bf 80}, 043006 (2009).

\bibitem{Elizalde10}
E.~Elizalde, R.~Myrzakulov, V.~V.~Obukhov and D.~Saez-Gomez,
Class.\ Quant.\ Grav.\  {\bf 27}, 095007 (2010).

\bibitem{Tanaka}
A.~De Felice and T.~Tanaka,
Prog.\ Theor.\ Phys.\  {\bf 124}, 503 (2010).

\bibitem{gcondensate}
N.~Arkani-Hamed, H.~C.~Cheng, M.~A.~Luty and S.~Mukohyama,
JHEP {\bf 0405}, 074 (2004).

\bibitem{Rubakov}
V.~A.~Rubakov,
Theor.\ Math.\ Phys.\ {\bf 149}, 1651 (2006)
[arXiv:hep-th/0604153].

\bibitem{Libanov}
M.~Libanov, V.~Rubakov, E.~Papantonopoulos, M.~Sami and S.~Tsujikawa,
JCAP {\bf 0708}, 010 (2007).

\bibitem{Rubakovreview}
V.~A.~Rubakov and P.~G.~Tinyakov,
Phys.\ Usp.\  {\bf 51}, 759 (2008).

\bibitem{Ben}
B.~M.~Gripaios,
JHEP {\bf 0410}, 069 (2004).

\bibitem{MLVR}
M.~V.~Libanov and V.~A.~Rubakov,
JHEP {\bf 0508}, 001 (2005).

\bibitem{Horava}
P.~Horava,
Phys.\ Rev.\  D {\bf 79}, 084008 (2009).

\bibitem{LVmodels1}
S.~M.~Carroll and E.~A.~Lim,
Phys.\ Rev.\  D {\bf 70}, 123525 (2004).

\bibitem{LVmodels2}
J.~M.~Cline and L.~Valcarcel,
JHEP {\bf 0403}, 032 (2004).

\bibitem{LVmodels3}
N.~Arkani-Hamed, H.~C.~Cheng, M.~Luty and J.~Thaler,
JHEP {\bf 0507}, 029 (2005).

\bibitem{LVmodels4}
H.~C.~Cheng, M.~A.~Luty, S.~Mukohyama and J.~Thaler,
JHEP {\bf 0605}, 076 (2006).

\bibitem{LVmodels5}
S.~Kanno and J.~Soda,
Phys.\ Rev.\ D {\bf 74}, 063505 (2006).

\bibitem{LVmodelsf}
P.~G.~Ferreira, B.~M.~Gripaios, R.~Saffari and T.~G.~Zlosnik,
Phys.\ Rev.\  D {\bf 75}, 044014 (2007).

\bibitem{Piazza}
F.~Piazza and S.~Tsujikawa,
JCAP {\bf 0407}, 004 (2004).

\bibitem{Sari}
E.~N.~Saridakis,
Eur.\ Phys.\ J.\  C {\bf 67}, 229 (2010).

\bibitem{Park}
M.~i.~Park,
JCAP {\bf 1001}, 001 (2010).

\bibitem{Kodama}
H.~Kodama and M.~Sasaki,
Prog.\ Theor.\ Phys.\ Suppl.\  {\bf 78}, 1 (1984).

\bibitem{Robert}
V.~F.~Mukhanov, H.~A.~Feldman and R.~H.~Brandenberger,
Phys.\ Rept.\  {\bf 215}, 203 (1992).

\bibitem{Bassett}
B.~A.~Bassett, S.~Tsujikawa and D.~Wands,
Rev.\ Mod.\ Phys.\  {\bf 78}, 537 (2006).

\bibitem{Kofman87}
L.~A.~Kofman, V.~F.~Mukhanov and D.~Y.~Pogosian,
Sov.\ Phys.\ JETP {\bf 66}, 433 (1987).

\bibitem{Hwang02}
J.~c.~Hwang and H.~r.~Noh,
Phys.\ Rev.\  D {\bf 65}, 023512 (2002).

\bibitem{Hwang}
J.~C.~Hwang and H.~Noh, 
Phys.\ Rev.\ D  {\bf 71}, 063536 (2005).

\bibitem{Star98}
A.~A.~Starobinsky,
JETP Lett.\  {\bf 68}, 757 (1998)
[Pisma Zh.\ Eksp.\ Teor.\ Fiz.\  {\bf 68}, 721 (1998)].

\bibitem{Tsujimatterper}
S.~Tsujikawa,
Phys.\ Rev.\  D {\bf 76}, 023514 (2007).

\bibitem{Hwang10}
J.~c.~Hwang, H.~Noh and C.~G.~Park,
arXiv:1012.0885 [astro-ph.CO].

\bibitem{delaCruz08}
A.~de la Cruz-Dombriz, A.~Dobado and A.~L.~Maroto,
Phys.\ Rev.\  D {\bf 77}, 123515 (2008).

\bibitem{Motohashi}
H.~Motohashi, A.~A.~Starobinsky and J.~Yokoyama,
Int.\ J.\ Mod.\ Phys.\  D {\bf 18}, 1731 (2009).

\bibitem{Appleos1}
S.~A.~Appleby and R.~A.~Battye,
JCAP {\bf 0805}, 019 (2008).

\bibitem{Frolov}
A.~V.~Frolov,
Phys.\ Rev.\ Lett.\  {\bf 101}, 061103 (2008).

\bibitem{Appleos2}
S.~A.~Appleby, R.~A.~Battye and A.~A.~Starobinsky,
JCAP {\bf 1006}, 005 (2010).

\bibitem{Miranda}
V.~Miranda, S.~E.~Joras, I.~Waga and M.~Quartin,
Phys.\ Rev.\ Lett.\  {\bf 102}, 221101 (2009).

\bibitem{Thongkool}
I.~Thongkool, M.~Sami, R.~Gannouji and S.~Jhingan,
Phys.\ Rev.\  D {\bf 80}, 043523 (2009).

\bibitem{delaCruz}
A.~de la Cruz-Dombriz, A.~Dobado and A.~L.~Maroto,
Phys.\ Rev.\ Lett.\  {\bf 103}, 179001 (2009).

\bibitem{NOuni1}
S.~Nojiri and S.~D.~Odintsov,
Phys.\ Lett.\  B {\bf 657}, 238 (2007).

\bibitem{NOuni2}
S.~Nojiri and S.~D.~Odintsov,
Phys.\ Rev.\  D {\bf 77}, 026007 (2008).

\bibitem{TsujiTate}
S.~Tsujikawa and T.~Tatekawa,
Phys.\ Lett.\  B {\bf 665}, 325 (2008).

\bibitem{Peebles}
P.~J.~E.~Peebles, {\it Large-Scale Structure of the Universe,}
Princeton University Press (1980).

\bibitem{Wang98}
L.~M.~Wang and P.~J.~Steinhardt,
Astrophys.\ J.\  {\bf 508}, 483 (1998).

\bibitem{Linder05}
E.~V.~Linder,
Phys.\ Rev.\  D {\bf 72}, 043529 (2005).

\bibitem{Moraes08}
R.~Gannouji, B.~Moraes and D.~Polarski,
JCAP {\bf 0902}, 034 (2009).

\bibitem{Moraes09}
S.~Tsujikawa, R.~Gannouji, B.~Moraes and D.~Polarski,
Phys.\ Rev.\  D {\bf 80}, 084044 (2009).

\bibitem{Oyaizu1}
H.~Oyaizu,
Phys.\ Rev.\  D {\bf 78}, 123523 (2008).

\bibitem{Oyaizu2}
H.~Oyaizu, M.~Lima and W.~Hu,
Phys.\ Rev.\  D {\bf 78}, 123524 (2008).

\bibitem{Oyaizu3}
F.~Schmidt, M.~V.~Lima, H.~Oyaizu and W.~Hu,
Phys.\ Rev.\  D {\bf 79}, 083518 (2009).

\bibitem{ZhaoLi}
G.~B.~Zhao, B.~Li and K.~Koyama,
arXiv:1011.1257 [astro-ph.CO].

\bibitem{Stabenau}
H.~F.~Stabenau and B.~Jain,
Phys.\ Rev.\  D {\bf 74}, 084007 (2006).

\bibitem{Laszlo}
I.~Laszlo and R.~Bean,
Phys.\ Rev.\  D {\bf 77}, 024048 (2008).

\bibitem{Huparametrization}
W.~Hu and I.~Sawicki,
Phys.\ Rev.\  D {\bf 76}, 104043 (2007).

\bibitem{Koyama09}
K.~Koyama, A.~Taruya and T.~Hiramatsu,
Phys.\ Rev.\  D {\bf 79}, 123512 (2009).

\bibitem{Tateskew}
T.~Tatekawa and S.~Tsujikawa,
JCAP {\bf 0809}, 009 (2008).

\bibitem{Kunz}
L.~Amendola, M.~Kunz and D.~Sapone,
JCAP {\bf 0804}, 013 (2008).

\bibitem{Peiris}
Y.~S.~Song, H.~Peiris and W.~Hu,
Phys.\ Rev.\  D {\bf 76}, 063517 (2007).
  
\bibitem{Schmidt}
F.~Schmidt,
Phys.\ Rev.\  D {\bf 78}, 043002 (2008).

\bibitem{Mc}
P.~McDonald {\it et al.,} astro-ph/0407377.

\bibitem{Viel}
M.~Viel and M.~G.~Haehnelt,
Mon.\ Not.\ Roy.\ Astron.\ Soc.\  {\bf 365}, 231 (2006).

\bibitem{Lue}
A.~Lue, R.~Scoccimarro and G.~D.~Starkman,
Phys.\ Rev.\  D {\bf 69}, 124015 (2004).

\bibitem{KoyamaSilva}
K.~Koyama and F.~P.~Silva,
Phys.\ Rev.\  D {\bf 75}, 084040 (2007).

\bibitem{Nico09}
K.~Hinterbichler, A.~Nicolis and M.~Porrati,
JHEP {\bf 0909}, 089 (2009).

\bibitem{Yamamoto}
K.~Yamamoto, D.~Parkinson, T.~Hamana, R.~C.~Nichol and Y.~Suto,
Phys.\ Rev.\  D {\bf 76}, 023504 (2007).

\bibitem{Kase}
A.~De Felice, R.~Kase and S.~Tsujikawa,
arXiv:1011.6132 [astro-ph.CO].
  

\end{thebibliography}
\end{document}